\documentclass[aps,prd,preprint,superscriptaddress,floatfix]{revtex4-1}

\usepackage{amsmath}
\usepackage{bm}
\usepackage{mathptmx}
\usepackage{color}
\usepackage{helvet}
\usepackage{courier}
\usepackage[dvips]{graphicx}

\newcommand\up{\uparrow}
\newcommand\down{\downarrow}
\newcommand\+{\dagger}
\renewcommand\k{{\bm{k}}}
\newcommand\ek{\epsilon_\k}
\newcommand\C{\mathcal{C}}
\newcommand\E{\mathcal{E}}
\newcommand\G{\mathcal{G}}
\newcommand\I{\mathcal{I}}
\newcommand\J{\mathcal{J}}
\newcommand\K{\mathcal{K}}
\newcommand\p{{\bm{p}}}
\newcommand\q{{\bm{q}}}
\newcommand\ep{\epsilon_\p}
\newcommand\eq{\epsilon_\q}
\newcommand\reg{\mathrm{reg}}
\newcommand\kF{k_\mathrm{F}}
\newcommand\eF{\epsilon_\mathrm{F}}
\DeclareMathAlphabet{\mathcal}{OMS}{cmsy}{m}{n}
\newcommand{\beq}{\begin{eqnarray}}
\newcommand{\eeq}{\end{eqnarray}}
\renewcommand\d{\partial}
\begin{document}

\preprint{RIKEN-QHP-177}

\title{Single-particle spectral density of the unitary Fermi gas: \\
Novel approach based on the operator product expansion, \\
sum rules and the maximum entropy method}


\author{Philipp Gubler}
\email[]{pgubler@riken.jp}
\affiliation{ECT*, Villa Tambosi, 38123 Villazzano (Trento), Italy}
\affiliation{RIKEN Nishina Center, Wako, Saitama 351-0198, Japan}

\author{Naoki Yamamoto}
\affiliation{Department of Physics, Keio University, Yokohama 223-8522, Japan}

\author{Tetsuo Hatsuda}
\affiliation{RIKEN Nishina Center, Wako, Saitama 351-0198, Japan}
\affiliation{RIKEN iTHES Research Group, Wako, Saitama 351-0198, Japan}

\author{Yusuke Nishida}
\affiliation{Department of Physics, Tokyo Institute of Technology, Meguro, Tokyo 152-8551, Japan}


\date{\today}

\begin{abstract}
Making use of the operator product expansion, 
we derive a general class of sum rules for the imaginary part of the single-particle self-energy of the 
unitary Fermi gas. The sum rules are analyzed numerically with the help of the maximum entropy method, which allows us 
to extract the single-particle spectral density as a function of both energy and momentum. These spectral densities 
contain basic information on the properties of the unitary Fermi gas, such as the dispersion relation and 
the superfluid pairing gap, for which we obtain reasonable agreement with the available 
results based on quantum Monte-Carlo simulations.    
\end{abstract}


\maketitle

\section*{Contents}
\vspace{0.018cm}
\hspace*{-0.39cm}I.\,\,\hspace{0.01cm}Introduction \hfill 3 \\
\vspace*{-0.33cm}\\
\hspace{0.15cm}II.\,\,\hspace{0.01cm}Formalism \hfill 7 \\
\hspace*{0.70cm}A.\,\,\hspace{0.01cm}The operator product expansion \hfill 7 \\
\hspace*{0.70cm}B.\,\,\hspace{0.01cm}The OPE of the single-particle Green's function for general values of $a$ \hfill 8 \\
\hspace*{0.70cm}C.\,\,\hspace{0.01cm}Three-body scattering amplitude \hfill 9 \\
\hspace*{0.70cm}D.\,\,\hspace{0.01cm}The OPE of the single-particle Green's function in the unitary limit \hfill 10 \\
\hspace*{0.70cm}E.\,\,\,\hspace{0.01cm}Derivation of the sum rules \hfill 12 \\
\hspace*{0.70cm}F.\,\,\,\,\hspace{0.01cm}Choice of the kernel $\mathcal{K}(\omega)$ \hfill 14 \\
\vspace*{-0.33cm}\\
\hspace{0.00cm}III.\,\,\hspace{0.01cm}MEM analysis for the spectral density \hfill 18 \\
\hspace*{0.70cm}A.\,\,\hspace{0.01cm}The Borel window and the default model \hfill 18 \\
\hspace*{0.70cm}B.\,\,\hspace{0.01cm}The single-particle spectral density \hfill 19 \\
\vspace*{-0.33cm}\\
\hspace{0.03cm}IV.\,\,\hspace{0.01cm}Summary and conclusion \hfill 23 \\
\vspace*{-0.33cm}\\
\hspace{0.00cm}Appendix A.\,\,\hspace{0.01cm}Numerical solution of $T^{\mathrm{reg}}_{\up}(k,0;k,0)$ in the unitary limit \hfill 25 \\
\hspace{0.00cm}Appendix B.\,\,\hspace{0.01cm}Derivation of the sum rules for a generic kernel \hfill 27 \\
\hspace{0.00cm}Appendix C.\,\,\hspace{0.01cm}Finite energy sum rules for the unitary Fermi gas \hfill 34 \\
\hspace{0.00cm}Appendix D.\,\,\hspace{0.01cm}The maximum entropy method \hfill 40 \\
\vspace*{-0.33cm}\\
\hspace{0.00cm}References \hfill 42

\clearpage

\section{\label{Intro} Introduction}
The unitary Fermi gas, consisting of non-relativistic fermionic particles of two species with 
equal mass, has been studied intensively during the last decade \cite{Bloch,Giorgini,Zwerger}. 
The growing interest in this system was prompted especially by the ability 
of tuning the interaction between different fermionic species in ultracold atomic gases 
through a 
Feshbach resonance by varying an external magnetic field. This technique 
allows one to bring the two-body scattering length of the two species to 
infinity and therefore makes it possible to study the unitary Fermi gas 
experimentally. 
Using photoemission spectroscopy, 
the measurement of 
the elementary excitations of ultracold atomic gases 
has in recent years become a realistic possibility \cite{Dao,Stewart}. 
Understanding these elementary excitations from a theoretical 
point of view is hence important and 
a number of studies devoted to this topic have already been 
carried out \cite{Haussmann,Magierski,Carlson2}. 
We will in this work propose a new and independent method 
for computing the single-particle spectral density of the unitary 
Fermi gas, which makes use of the operator 
product expansion (OPE). 

The OPE, which was originally proposed in the late sixties independently by Wilson, Kadanoff and Polyakov 
\cite{Wilson,Kadanoff,Polyakov}, has proven to be a powerful 
tool for analyzing processes related to QCD (Quantum Chromo Dynamics), for which simple perturbation theory 
fails in most cases. The reason for this is the ability of the OPE to incorporate 
non-perturbative effects into the analysis as expectation values of a series of 
operators, which are ordered according to their scaling dimensions. 
Perturbative effects can on the 
other hand be treated as coefficients of these operators (the ``Wilson-coefficients"). 
The OPE has specifically been used to study deep inelastic scattering processes 
\cite{Muta} and has especially played a key role in the formulation of the 
so-called QCD sum rules \cite{Shifman1,Shifman2}.

In recent years, it was noted that the OPE can also be applied to strongly 
coupled non-relativistic systems such as the unitary Fermi gas 
\cite{Braaten1,Braaten2,Braaten3,Braaten4,Braaten5,Son,Barth,Hofmann,Goldberger,Goldberger2,Nishida,Golkar,Goldberger3}. 
Initially, the OPE was used to rederive some of the Tan-relations \cite{Tan1,Tan2,Tan3} in 
a natural way \cite{Braaten1} and, for instance, to study the dynamic structure factor of 
unitary fermions in the large energy and momentum limit \cite{Son,Goldberger}. Furthermore, the OPE 
for the single-particle Green's function of the unitary Fermi gas was computed 
by one of the present authors \cite{Nishida} 
up to operators of momentum dimension 5, from which the single-particle dispersion relation 
was extracted. 
As the OPE is an expansion at small distances and times (or large momenta and energies), 
the result of such an analysis can be expected to give the correct behavior in the large momentum limit and 
is bound to become invalid at small momenta. 
The analysis of \cite{Nishida} confirmed this, but in addition somewhat surprisingly showed that 
the OPE is valid for momenta as small as the Fermi momentum $k_{\mathrm{F}}$, where the 
OPE still shows good agreement with the results obtained from quantum Monte-Carlo 
simulations \cite{Magierski}. 

\begin{figure} 
\centering 
\vspace*{1cm}
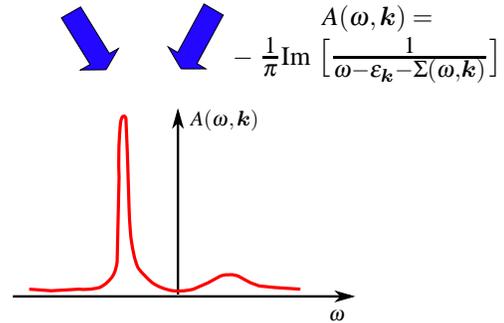 
\caption{\label{fig:steps} Steps for computing the single-particle spectral density from the OPE of the 
single-particle Green's function of a fermionic operator.} 
\end{figure}

The purpose of this paper is to extend this analysis to smaller momenta, by 
making use of the techniques of QCD sum rules, which have traditionally been employed 
to study hadronic spectra from the OPE applied to Green's functions in QCD. 
Our general strategy goes as follows: 
\begin{itemize} 
\item Step 1: Construct OPE 

At first, we need to obtain the OPE for the single-particle Green's function $\G_{\up}(k_0,\k)$ in the 
unitary limit, which can be rewritten as an expansion of the 
single-particle self-energy $\Sigma_{\up}(k_0,\k)$. 
The subscript $\up$ here represents the spin-up fermions. 
The main work of 
this step has already been carried out in \cite{Nishida}. 
$\Sigma_{\up}(k_0,\k)$ can be considered to be an analytic function on 
the complex plane of the energy variable $k_0$, with the exception of possible cuts and poles 
on the real axis. 
Considering the OPE at $T=0$, with equal densities for both fermionic species ($n_{\up} = n_{\down}$) and 
taking into account operators up to momentum dimension 5, 
the only parameters appearing in the OPE are the Bertsch parameter and 
the contact density, which are by now well known from 
both experimental measurements \cite{Ku,Zurn,Hoinka} and theoretical quantum Monte-Carlo calculations \cite{Carlson,Gandolfi}. 

\item Step 2: Derive sum rules

From the fact that the OPE is 
valid at large $|k_0|$ and the analytic properties of the self-energy, a general class of sum rules 
for $\mathrm{Im} \Sigma_{\up}(\omega,\bm{k})$ can be derived. 
In contrast to the complex $k_0$, $\omega$ here is a real parameter.  
These sum rules are relations between certain weighted integrals of 
$\mathrm{Im} \Sigma_{\up}(\omega,\bm{k})$ and corresponding analytical expressions 
that can be obtained from the OPE result (for details see Section \ref{Formalism}): 
\begin{equation}
D^{\mathrm{OPE}}_{\up}(M,\bm{k}) = \int^{\infty}_{-\infty}d\omega \mathcal{K}(\omega, M) \mathrm{Im} \Sigma_{\up}(\omega,\bm{k}). 
\label{eq:intro.sr}
\end{equation} 
The kernel $\mathcal{K}(\omega, M)$ here must be an analytic function that is real on the real axis of $\omega$ and falls off 
to zero quickly enough at $\omega \to +\infty$, while $M$ is some general parameter that characterizes the form of 
the kernel. In the practical calculations of this paper, we will use the so-called Borel kernels of the form 
$\mathcal{K}_n(\omega, M) = (\omega/M)^n e^{-\omega^2/M^2}$. 

\item Step 3: Extract $\mathrm{Im} \Sigma_{\up}(\omega,\bm{k})$ via MEM and obtain $\mathrm{Re}\Sigma_{\up}(\omega,\bm{k})$ from the Kramers-Kr$\mathrm{\ddot{o}}$nig 
relation  

As a next step, we use the maximum entropy method (MEM) to extract the most probable form of $\mathrm{Im} \Sigma_{\up}(\omega,\bm{k})$ 
from the sum rules, following an approach proposed in \cite{Gubler} for the QCD sum rule case. 

It should be mentioned here that 
this method is somewhat different from the analysis procedure most commonly employed in QCD sum rule studies, 
where the spectral function (which corresponds to $\mathrm{Im} \Sigma_{\up}$ here)  is parametrized 
using a simple functional ansatz with a small number of parameters which are then fitted to the sum rules. 
This method has traditionally worked well if some sort of prior knowledge on the spectral function is available 
and assumptions on its form can thus be justified. 
On the other hand, in cases where one does not really know what specific form the spectral function can be expected to have, 
sum rule analyses based on (potentially incorrect) assumptions on the spectral shape always involve the danger of giving ambiguous 
and even misleading results. 
MEM is therefore our method of choice, as it allows us to analyze the sum rules without making any strong assumption 
on the functional form of the spectral function and hence makes it possible to pick the most probable spectral shape among 
an infinitely large number of choices. 

Once $\mathrm{Im} \Sigma_{\up}(\omega,\bm{k})$ is obtained from the MEM analysis of the sum rules, 
it is a simple matter to compute $\mathrm{Re}\Sigma_{\up}(\omega,\bm{k})$ by 
the Kramers-Kr$\mathrm{\ddot{o}}$nig relation, 
 \begin{equation}
\mathrm{Re}\Sigma_{\up}(\omega,\bm{k}) = -\frac{1}{\pi} \mathrm{P} \int_{-\infty}^{\infty} d\omega' \frac{\mathrm{Im} \Sigma_{\up}(\omega',\bm{k})}{\omega - \omega'}. 
\label{eq:Kram.Kroe}
\end{equation}

\item Step 4: Compute single-particle spectral density

From the real and imaginary parts of the self-energy, the single-particle spectral density can then be obtained as, 
\begin{equation}
A_{\up}(\omega,\bm{k}) = -\frac{1}{\pi} \mathrm{Im} \frac{1}{\omega + i0^{+}-\epsilon_{\bm{k}}-\Sigma_{\up}(\omega+i0^{+},\bm{k})}, 
\label{eq:spe}
\end{equation}
where $\epsilon_{\bm{k}}$ is defined as $\epsilon_{\bm{k}} = \bm{k}^2/(2m)$, with $m$ being 
the fermion mass. 
\end{itemize}
The above steps are shown once more in pictorial form in Fig. \ref{fig:steps}. 

As a result of the above procedure, we find a two-peak structure in the imaginary part of the self-energy, 
the two peaks moving from the origin ($\omega=0$) 
to positive and negative directions of the energy with increasing momentum $|\bm{k}|$. Translated to the single-particle 
spectral density, this leads to a typical superfluid BCS-Bogoliubov-like dispersion relation with both hole and particle 
branches and a nonzero gap value. 

The paper is organized as follows. 
In Section \ref{Formalism}, we discuss the OPE of the single-particle Green's function and explain how it can be rewritten 
as an expansion of the single-particle self-energy. Next, we outline 
the derivation of the sum rules from 
the OPE. In Section \ref{Analysis}, the MEM analysis results of the sum rules are shown and the consequent final form of the single-particle 
spectral density and the dispersion relation are presented. The spectral density is visualized in Fig. \ref{fig:density.plot} as a density plot and 
the detailed numerical properties of the dispersion relation are described in Table \ref{tab:disp.para}. 
Section \ref{Summary} is devoted to the summary and conclusions of the paper. 
For the interested reader, we provide in the appendices detailed accounts of the relevant calculations, which were needed 
for this work. 

\section{\label{Formalism} Formalism}
\subsection{\label{OPE} The operator product expansion}
The operator product expansion (OPE) is based on the observation that a general product 
of non-local operators can be expanded as a series of local operators. This can be expressed as 
\begin{equation}
\mathcal{O}_i(x + \tfrac{1}{2}y) \mathcal{O}_j(x - \tfrac{1}{2} y) 
= \sum_{k} W_{\mathcal{O}_k}(y) \mathcal{O}_k(x). 
\label{eq:OPE.exp}
\end{equation} 
Here, we have used the abbreviations $(x)=(x_0,\,\bm{x})$ and $(y)=(y_0,\,\bm{y})$ for the 
four-dimensional vectors. 
$W_{\mathcal{O}_k}(y)$ are the Wilson-coefficients, which only depend on the relative time and distance $y$ of the 
two original operators. 
The operators on the right-hand side of Eq.(\ref{eq:OPE.exp}) 
are ordered according to their scaling dimensions $\Delta_k$, in ascending order. 
This expansion works for small time differences (or small distances), as the Wilson coefficients behave as 
$(\sqrt{|y_0|})^{\Delta_k - \Delta_i - \Delta_j}$ ($|\bm{y}|^{\Delta_k - \Delta_i - \Delta_j}$) and 
because the operators with larger scaling dimensions are thus suppressed by higher powers of $\sqrt{|y_0|}$ ($|\bm{y}|$). 
Fourier transforming Eq.(\ref{eq:OPE.exp}), the above 
statement is translated into energy-momentum space, where 
the OPE is a good approximation in the 
large energy or momentum limit as operators with larger scaling dimensions are suppressed by higher 
powers of $1/\sqrt{|k_0|}$ ($1/|\bm{k}|$). 

For the above expansion to work in the context of a non-relativistic atomic gas, certain conditions have 
to be satisfied. Firstly, it is important that the potential range $r_0$ of the atomic interaction is much smaller than 
all other length scales of the system, so that the detailed form of the interaction becomes irrelevant. 
Furthermore, the energy or momentum scale at which the system is probed needs to be much larger 
than the corresponding typical scales of the system. 
Hence, for the OPE to be a useful expansion, the following separation of scales must hold, which 
must be satisfied by either $1/\sqrt{|k_0|}$ or $1/|\k|$: 
\begin{equation}
r_0 \ll 1/\sqrt{|k_0|},\,\,1/|\k| \ll |a|,\,n^{-1/3}_{\sigma},\,\lambda_T.  
\label{eq:conditionOPE}
\end{equation}
Here, $a$ is the s-wave scattering length between spin-up and -down fermions, 
$n^{-1/3}_{\sigma}$ the mean interparticle distance of both fermionic species, and $\lambda_T \sim 1/\sqrt{mT}$ the thermal 
de Broglie wave length. In other words, $\sqrt{|k_0|}$ or $|\k|$ have to be large enough so that for example 
an expansion in $1/(a\sqrt{|k_0|})$, $n^{1/3}_{\sigma}/\sqrt{|k_0|}$ and $1/(\lambda_T \sqrt{|k_0|})$ is valid, while they should be 
still small enough not to probe the actual structure of the individual atoms. 

In practice, we will in this work take the zero-range limit $r_0 \to 0$, study the system at vanishing temperature $T=0$ and 
will in the course of the derivation of the sum rules take the unitary limit $a \to \infty$. 
Furthermore, for studying the detailed momentum dependence of the spectral-density, 
we will in the following discussion make use of an expansion in $1/\sqrt{|k_0|}$, but not in $1/|\bm{k}|$. 
$|\bm{k}|$ will instead always be kept at the order of Fermi-momentum of the studied system. 

\subsection{\label{OPEforgenerala} The OPE of the single-particle Green's function for general values of $a$}
In this paper, we will employ the OPE of the single-particle Green's function, which was computed in \cite{Nishida}. 
Let us here briefly recapitulate this result and discuss its form rewritten as an expansion of the self-energy $\Sigma_{\up}(k_0,\bm{k})$. 
The starting point of the calculation is 
\begin{equation}
 i\G_{\up}(k) \equiv \int \! dy\,e^{iky}
 \langle T[\psi_\up(x+\tfrac{y}2)\psi_\up^\+(x-\tfrac{y}2)] \rangle
 = \frac{i}{k_0-\ek-\Sigma_{\up}(k)}, 
\label{eq:Greensfunction} 
\end{equation}
where $k$ should be understood as $(k)=(k_0,\k)$. The OPE for $\G_{\up}(k)$ 
can then be carried out, as discussed in detail in \cite{Nishida}. 
If translational and rotational invariance holds, 
all sorts of currents vanish and the OPE expression (taking into account terms up to momentum dimension 5) can be simplified as follows: 
\begin{equation}
\begin{split}
\G_{\up}^{\mathrm{OPE}}(k) =&\: G(k) - G^2(k) A(k) n_{\down} 
 -\frac{\mathcal{C}}{4 \pi m a}  G^2(k) \frac{\partial}{\partial  k_0} A(k)  
 - \frac{\mathcal{C}}{m^2} G^2(k) T^{\mathrm{reg}}_{\up}(k,0;k,0) \\ 
& - G^2(k) \Big[\frac{\partial}{\partial  k_0} A(k) + \frac{m}{3} \sum_{i=1}^{3} \frac{\partial^2}{\partial  k^2_i} A(k) \Big] 
\int \frac{d \bm{q}}{(2 \pi)^3} \frac{\bm{q}^2}{2m} \Big[\rho_{\down}(\bm{q}) - \frac{\mathcal{C}}{\bm{q}^4} \Big].
\end{split}
\label{eq:GreensfunctionOPE} 
\end{equation}
Here, $G(k)$ is the free fermion propagator, 
\begin{equation}
G(k) = \frac{1}{k_0 - \epsilon_{\bm{k}}}, 
\label{eq:ferm.prop}
\end{equation}
$A(k)$ represents the two-body scattering amplitude between spin-up and -down fermions,  
\begin{equation}
A(k) = \frac{4 \pi}{m} \frac{1}{\sqrt{\tfrac{\bm{k}^2}4 - m k_0 } -1/a},  
\label{eq:scatt.amp}
\end{equation}
and $T^{\mathrm{reg}}_{\up}(k,p;k',p')$ stands for the regularized three-body scattering 
amplitude of a spin-up fermion with initial (final) momentum $k$ ($k'$) and a dimer with 
initial (final) momentum $p$ ($p'$). ``regularized" means that infrared divergences 
originally appearing in the three-body scattering amplitude have been 
subtracted (see Sections III C and III F of \cite{Nishida}): 
\begin{equation}
T^{\mathrm{reg}}_{\up}(k,0;k,0) \equiv T_{\up}(k,0;k,0) - A(k) \int \frac{d \bm{q}}{(2 \pi)^3} \frac{m^2}{\bm{q}^4}. 
\label{eq:def.reg.three.amp}
\end{equation} 
Furthermore, $\rho_{\sigma}(\q)$ is the momentum distribution function of spin-$\sigma$ fermions, 
$n_{\down}$ the density of spin-down fermions and $\mathcal{C}$ the so-called contact density \cite{Tan1,Tan2,Tan3}. 

Comparing Eq.(\ref{eq:GreensfunctionOPE}) with the definition of the self-energy of Eq.(\ref{eq:Greensfunction}), 
one can easily find an expression for $\Sigma _{\up}(k)$, which (again up to terms with momentum dimension 5) is consistent 
with the OPE of the single-particle Green's function: 
\begin{equation}
\begin{split}
\Sigma_{\up}^{\mathrm{OPE}}(k) =& - A(k) n_{\down} 
- \frac{\mathcal{C}}{4 \pi m a} \frac{\partial}{\partial  k_0} A(k)  
 - \frac{\mathcal{C}}{m^2} T^{\mathrm{reg}}_{\up}(k,0;k,0) \\ 
& -  \Big[  \frac{\partial}{\partial  k_0} A(k) + \frac{m}{3} \sum_{i=1}^{3} \frac{\partial^2}{\partial  k^2_i} A(k) \Big] 
\int \frac{d \bm{q}}{(2 \pi)^3} \frac{\bm{q}^2}{2m} \Big[\rho_{\down}(\bm{q}) - \frac{\mathcal{C}}{\bm{q}^4} \Big].
\end{split}
\label{eq:SelfenergyOPE} 
\end{equation}
Assuming the considered system to be spin symmetric [$\rho_{\up}(\q) = \rho_{\down}(\q)$], the integral 
of the momentum distribution function 
appearing in the above equation can be evaluated by one of the Tan-relations \cite{Tan1,Tan2,Tan3},  
\begin{equation}
\sum_{\sigma=\up,\down} \int \frac{d \bm{q}}{(2 \pi)^3} \frac{\bm{q}^2}{2m} \Big[\rho_{\sigma}(\bm{q}) - \frac{\mathcal{C}}{\bm{q}^4} \Big] = 
\mathcal{E} +  \frac{\mathcal{C}}{4 \pi m a}, 
\label{eq:Tan.rel.OPE} 
\end{equation}
where $\mathcal{E}$ is the energy density of the system. We hence get, 
\begin{equation}
\begin{split}
\Sigma_{\up}^{\mathrm{OPE}}(k) =& - A(k) n_{\down} 
- \frac{\mathcal{C}}{4 \pi m a} \frac{\partial}{\partial  k_0} A(k)  
 - \frac{\mathcal{C}}{m^2} T^{\mathrm{reg}}_{\up}(k,0;k,0) \\ 
& -  \frac{1}{2} \Big[ \frac{\partial}{\partial  k_0} A(k) + \frac{m}{3} \sum_{i=1}^{3} \frac{\partial^2}{\partial  k^2_i} A(k) \Big] 
\Big( \mathcal{E} +  \frac{\mathcal{C}}{4 \pi m a} \Big).
\end{split}
\label{eq:SelfenergyOPE2} 
\end{equation}
Among the various terms appearing in Eq.(\ref{eq:SelfenergyOPE2}), 
the most involved piece to evaluate is the three-body scattering amplitude $T^{\mathrm{reg}}_{\up}(k,0;k,0)$, which 
will be studied next in a separate subsection. 

\subsection{Three-body scattering amplitude} 
The difficulty in obtaining $T^{\mathrm{reg}}_{\up}(k,0;k,0)$ stems from the fact that this 
scattering amplitude by itself does not solve a closed integral equation and therefore can not 
be computed directly. We thus have to use $T_{\up}(k,0;p,k-p)$ with a more general momentum 
dependence, which will, for simplicity of notation, from now on be denoted as $T_{\up}(k;p)$. 
$T_{\up}(k;p)$ satisfies the following integral equation (note that we for the moment work with the 
non-regularized version of the amplitude): 
 \begin{equation}
 \begin{split}
 T_\up(k;p) =&\: G(-p) + i\int\!\frac{d q_0 d\q}{(2\pi)^4}\,
 T_\up(k;q)G(q)A(k-q)G(k-p-q) \\
 =& -\frac1{p_0+\ep} \\
&- \int\!\frac{d\q}{(2\pi)^3}
 \frac{4\pi}{\frac12\sqrt{3\q^2-2\q\cdot\k+\k^2-4mik_0}-\frac1a}
 \frac{T_\up(k;\eq,\q)}{\frac{(\p+\q-\k)^2}2+m p_0+\frac{\q^2}2-m k_0}.  
\end{split}
 \label{eq:scatt.amp.1}
\end{equation} 
In going to the second and third lines, the integral over $q_0$ is performed and thus $q_0$ is fixed to $\eq$. 

Next, setting $p_0 = \ep$ provides a closed equation, 
\begin{align}
 T_\up(k;\ep,\p) = -\frac{m}{\p^2} - \int\!\frac{d\q}{(2\pi)^3}
 \frac{4\pi}{\frac12\sqrt{3\q^2-2\q\cdot\k+\k^2-4m k_0}-\frac1a}
 \frac{T_\up(k;\eq,\q)}{\frac{(\p+\q-\k)^2}2+\frac{\p^2+\q^2}2-m k_0}, 
 \label{eq:scatt.amp.2}
\end{align}
which needs to be solved numerically. 
The technical details of this step are presented in Appendix \ref{ScattAmp}. 
Once the above equation is solved and $T_\up(k;\ep,\p)$ has hence been obtained, 
one can extract the desired amplitude 
$T_{\up}(k;k)$ from Eq.(\ref{eq:scatt.amp.1}) by setting $p=k$: 
\begin{equation}
\begin{split}
 T_\up(k;k) &= -\frac1{k_0+\ek} - \int\!\frac{d\q}{(2\pi)^3}
 \frac{4\pi}{\frac12\sqrt{3\q^2-2\q\cdot\k+\k^2-4m k_0}-\frac1a}
 \frac{T_\up(k;\eq,\q)}{\q^2} \\
 &= -\frac1{k_0+\ek} + \int\!\frac{d\q}{(2\pi)^3}
 \frac{4\pi}{\frac12\sqrt{3\q^2-2\q\cdot\k+\k^2-4m k_0}-\frac1a}
 \frac{m}{\q^4} \\
 &\quad - \int\!\frac{d\q}{(2\pi)^3}
 \frac{4\pi}{\frac12\sqrt{3\q^2-2\q\cdot\k+\k^2-4m k_0}-\frac1a}
 \frac{T_\up(k;\eq,\q)+\frac{m}{\q^2}}{\q^2}.
\end{split}
 \label{eq:scatt.amp.3}
\end{equation}
Finally, returning to the regularized scattering amplitude $T^{\mathrm{reg}}_{\up}(k,0;k,0) = T^{\mathrm{reg}}_{\up}(k;k)$ 
[defined in Eq.(\ref{eq:def.reg.three.amp})], we get, 
\begin{equation}
\begin{split}
&\: T_\up^\reg(k;k) \\
=&\:T_\up(k;k)
 - A(k)\int\!\frac{d\q}{(2\pi)^3}\left(\frac{m}{\q^2}\right)^2 \\
 =& -\frac1{k_0+\ek} 
+ \int\!\frac{d\q}{(2\pi)^3}
 \left[\frac{4\pi}{\frac12\sqrt{3\q^2-2\q\cdot\k+\k^2-4m k_0}-\frac1a}
 - \frac{4\pi}{\frac12\sqrt{\k^2-4m k_0}-\frac1a}\right]
 \frac{m}{\q^4} \\
 & - \int\!\frac{d\q}{(2\pi)^3}
 \frac{4\pi}{\frac12\sqrt{3\q^2-2\q\cdot\k+\k^2-4m k_0}-\frac1a}
 \frac{T_\up(k;\eq,\q)+\frac{m}{\q^2}}{\q^2}. 
\end{split}
 \label{eq:scatt.amp.4}
\end{equation}

\subsection{\label{OPE.unitary.limit} The OPE of the single-particle Green's function in the unitary limit}
So far, we have studied the OPE for arbitrary values of the s-wave scattering length $a$ between 
the two spin degrees of freedom (which should however be kept large enough for the conditions of 
a valid OPE to apply). 
One could in principle continue with these general expressions, derive sum rules for nonzero $a^{-1}$ values and 
analyze them according to our strategy outlined in the introduction. 

In order to provide a clear account of the proposed method, 
we will however not do this here but concentrate on the unitary limit ($a \to \infty$), 
which considerably simplifies many of the equations needed to derive the sum rules, 
but already exhibits all non-trivial technical difficulties that will arise in an analogous, but more involved manner 
when generalizing the calculations to nonzero $a^{-1}$.  

Firstly, looking at the unitary limit of the OPE result of Eq.(\ref{eq:SelfenergyOPE2}), the terms 
proportional to $a^{-1}$ vanish and the factor containing derivatives of $A(k)$ can be 
obtained in a simple form: 
\begin{equation}
\frac{\partial}{\partial  k_0} A(k) + \frac{m}{3} \sum_{i=1}^{3} \frac{\partial^2}{\partial  k^2_i} A(k)  
= \frac{2^{5/2} \pi}{m^{3/2}} \frac{\ek -  k_0}{(\ek - 2 k_0)^{5/2}}.
\label{eq:unitarylimit1}
\end{equation}

As for the calculation of the three-body scattering amplitude $T^{\mathrm{reg}}_{\up}(k,0;k,0)$, 
the integral equation of Eq.(\ref{eq:scatt.amp.2}) is made slightly more manageable because of a vanishing $a^{-1}$ term in 
the first denominator of the integrand on the right-hand side. 
The regularized scattering amplitude itself, given in Eq.(\ref{eq:scatt.amp.4}), also simplifies as the integral 
appearing in its second term [see Eq.(\ref{eq:scatt.amp.4})] can now be performed analytically: 
\begin{equation}
\begin{split}
\int\!\frac{d\q}{(2\pi)^3}
\left[\frac{4\pi}{\frac12\sqrt{3\q^2-2\q\cdot\k+\k^2-4m k_0}}
- \frac{4\pi}{\frac12\sqrt{\k^2-4m k_0}}\right]
\frac{m}{\q^4}\\
=\frac{1}{\pi} \Bigg[
\frac{\sqrt{3}}{2k_0 - \ek} + \frac{3k_0 - \ek}{\sqrt{\ek}(\ek - 2k_0)^{3/2}} 
\log \Bigg(\frac{1 + \sqrt{3} \sqrt{1 - 2k_0 /\ek}}{-1 + \sqrt{3} \sqrt{1 - 2k_0 / \ek}} \Bigg)
\Bigg]. 
\end{split}
\label{eq:integral1} 
\end{equation}

For a spin-symmetric system, making use of the equations of motion and Tan-relations, it is 
possible to express the expectation values of the local operators appearing in the OPE in terms 
of particle density $n_{\down}$, energy density $\E$, and contact density $\C$ [see Eq.(\ref{eq:SelfenergyOPE2})]. 
In the unitary limit, these quantities only depend on one single scale, which determines 
the properties of the system. 
Here, we define the Fermi momentum and the Fermi energy by
$n_\up=n_\down\equiv\kF^3/(6\pi^2)$ and $\eF\equiv\kF^2/(2m)$.
At infinite scattering length $a\to\infty$ (and zero temperature $T=0$), $\E$ and $\C$ are then given as 
\begin{align} 
 \E = \xi\,\frac{\kF^5}{10\pi^2m}, \qquad\qquad
 \C = \zeta\,\frac{\kF^4}{3\pi^2}. 
\end{align}
These values have by now been extracted from both theoretical quantum Monte-Carlo simulations 
and experimental measurements, which give consistent results, as shown in Table \ref{tab:input.para}. 
\begin{table}
\begin{center}
\caption{Numerical values of the Bertsch parameter $\xi$ and the dimensionless contact density $\zeta$ in the unitary limit 
at zero temperature. 
The column ``simulation" gives numbers extracted from quantum Monte-Carlo simulations, while the column ``experiment" contains 
values from ultracold-atom experiments.} 
\vspace{0.2cm}
\label{tab:input.para}
\begin{tabular}{lll} \hline
 & \hspace{1cm} simulation & experiment \\ \hline
$\xi$ & \hspace{1cm} 0.372(5) \,\,\cite{Carlson} \hspace{1.0cm}&  0.370(5)(8) \,\, \cite{Ku,Zurn} \\
$\zeta$ & \hspace{1cm} 3.40(1) \,\,\cite{Gandolfi} \hspace{1.0cm} &   3.33(7) \,\, \cite{Hoinka} \\
\hline
\end{tabular}
\end{center}
\end{table}
In the specific analyses presented in this paper, we will use the values obtained 
from quantum Monte-Carlo studies (denoted as ``simulation" in Table \ref{tab:input.para}). 

Assembling all the results and definitions of the last two subsections, we 
reach the following final form for the OPE in the unitary limit, 
\begin{equation}
\begin{split}
\Sigma^{\mathrm{OPE}}_{\up}(k_0, \bm{k}) =   
&-\frac{8}{3\pi} \eF^{3/2} \frac{1}{\sqrt{\ek - 2 k_0}} 
+\frac{4}{3\pi^2} \zeta \eF^2 \Biggl[  \frac{1}{k_0 + \ek } - \frac{\sqrt{3}}{\pi} \frac{1}{2 k_0 - \ek } \\
&- \frac{1}{\pi} \frac{3k_0 - \ek}{\sqrt{\ek}(\ek - 2k_0)^{3/2}} 
\log \Bigg(\frac{1 + \sqrt{3} \sqrt{1 - 2k_0 /\ek}}{-1 + \sqrt{3} \sqrt{1 - 2k_0 / \ek}} \Bigg) 
+\frac{1}{\ek} L\bigl(\tfrac{k_0}{\ek}\bigr) \Biggr] \\
&- \frac{8}{5\pi} \xi  \eF^{5/2} \frac{\ek -  k_0}{(\ek - 2 k_0)^{5/2}} +O(k_0^{-2}),  
\end{split}
\label{eq:OPE1}
\end{equation}
where we, for simplicity of notation, have introduced the 
function $L(x)$, which is defined as: 
\begin{equation}
L\bigl(\tfrac{k_0}{\ek}\bigr) =  \ek \int\!\frac{d\q}{(2\pi)^3}
 \frac{4\pi}{\frac12\sqrt{3\q^2-2\q\cdot\k+\k^2-4m k_0}}
 \frac{T_\up(k;\eq,\q)+\frac{m}{\q^2}}{\q^2}. 
 \label{eq:definition1}
\end{equation}
Note that we here have made use of the fact that $L(x)$ is dimensionless and hence can only depend 
on the ratio $k_0/\ek$. 
As mentioned earlier, $L(x)$ can be obtained by solving Eq.(\ref{eq:scatt.amp.2}) and substituting the 
result into the above definition. 
The detailed steps of this procedure are given in Appendix \ref{ScattAmp}. 
Here, we simply note that the imaginary part 
of $L(x)$ (which is its only piece that will play a role in the sum rules to be derived later) is a finite, but sharply peaked function, 
which is non-zero only 
in the interval: $1/3 < x < 1$ (see Fig. \ref{fig:inteq.res}). 

\subsection{Derivation of the sum rules}
\begin{figure}
\begin{center}
\includegraphics[width=10.0cm]{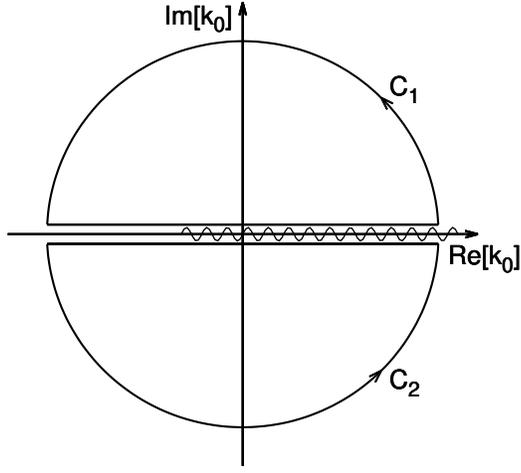}%
\vspace{-0.5cm}
\caption{\label{fig:contour} The contours $C_1$ and $C_2$ on the complex plane of $k_0$, used for deriving the sum rules. The wavy line 
on the real axis represents possible locations of non-analytic poles or cuts of $\Sigma_{\up}(k_0, \bm{k})$ and $\Sigma^{\mathrm{OPE}}_{\up}(k_0, \bm{k})$.} 
\end{center}
\end{figure}

We now derive the sum rules from the OPE of Eq.(\ref{eq:OPE1}). 
For doing this, we consider $k_0$ to be a complex variable and study the 
contour integral, 
\begin{equation}
\int_{C_1+C_2} d k_0 \Big[ \Sigma_{\up}(k_0, \bm{k}) - \Sigma^{\mathrm{OPE}}_{\up}(k_0, \bm{k})\Big]\mathcal{K}(k_0) = 0. 
\label{eq:integral}
\end{equation}
Here, $\Sigma_{\up}(k_0, \bm{k})$ is the exact (and at this moment unknown) self-energy, $\Sigma^{\mathrm{OPE}}_{\up}(k_0, \bm{k})$ is 
its approximate OPE expression of Eq.(\ref{eq:OPE1}). 
$\mathcal{K}(k_0)$ is assumed to be an analytic function on the upper and lower half of the complex plane of $k_0$ 
and to be real on the real axis, 
but is otherwise completely arbitrary. 
The contours $C_1$ and $C_2$ are shown in Fig. \ref{fig:contour}, in which 
the wavy line depicts possible non-analytic poles or cuts of $\Sigma_{\up}(k_0, \bm{k})$ and $\Sigma^{\mathrm{OPE}}_{\up}(k_0, \bm{k})$, whose actual locations depend 
on the chosen value of $|\bm{k}|$. 
The above integral vanishes because the exact self-energy $\Sigma_{\up}(k_0, \bm{k})$ and its OPE counterpart are analytic 
in the upper and lower half of the complex plane. Furthermore, we know that the OPE is valid at large 
$|k_0|$, from which follows that the integrand on the left-hand side of Eq.(\ref{eq:integral}) 
vanishes (to the order we are considering) along the large half-circles in $C_1$ and $C_2$. 
As we have assumed $\mathcal{K}(k_0)$ to be real on the real axis, 
it is noted that 
the added contour sections along the real axis leave just the imaginary parts of the self-energies, while their real parts vanish. 
Thus, we can now write down the sum rules as
\begin{equation}
\int_{-\infty}^{\infty} d \omega \mathrm{Im}\Sigma_{\up}(\omega+i0^{+}, \bm{k}) \mathcal{K}(\omega) 
= \int_{-\infty}^{\infty} d \omega \mathrm{Im}\Sigma^{\mathrm{OPE}}_{\up}(\omega+i0^{+}, \bm{k}) \mathcal{K}(\omega), 
\label{eq:sum.rule}
\end{equation}
where here and in the rest of the paper $\omega$ is understood to be a real variable. 
The right-hand side of this equation can be calculated from Eq.(\ref{eq:OPE1}), once the kernel 
$\mathcal{K}(\omega)$ is specified. 
This last step, however needs some care, as some terms of Eq.(\ref{eq:OPE1}) 
at first sight lead to divergences on the right-hand side of Eq.(\ref{eq:sum.rule}). This is for instance 
the case for the last term in Eq.(\ref{eq:OPE1}), which has an imaginary part for 
$k_0=\omega>\ek/2$ and diverges as $(\omega - \ek/2)^{-5/2}$, when $\omega$ 
approaches $\ek/2$ from above. This superficial divergence originates in our sloppiness of 
treating cuts in the above derivation and can be cured by taking into account all parts of the contours $C_1$ and 
$C_2$ which run along the cuts and their thresholds. The details of this procedure are given in Appendix \ref{OPE.detail}, where it 
is explicitly shown how all superficial divergences cancel and that hence the right-hand side of  Eq.(\ref{eq:sum.rule}) 
is indeed finite. 

All this then leads us to the following form of the sum rules: 
\begin{equation}
\begin{split}
&\:\int^{\infty}_{-\infty}d\omega \mathcal{K}(\omega) \mathrm{Im} \Sigma_{\uparrow}(\omega + i0^{+}, \k) \\
=&\:\frac{8}{3\pi} \eF^{3/2} \int^{\infty}_{\ek/2} d\omega \sqrt{2\omega - \ek} \mathcal{K}'(\omega) 
+\frac{4}{3\pi} \zeta \eF^2 \Bigl[ \frac{\sqrt{3}}{\pi}\mathcal{K}\bigl(\tfrac{\ek}{3}\bigr) - \mathcal{K}(-\ek) \Bigr] \\
&+\frac{4}{3\pi^2} \zeta \frac{\eF^2}{\sqrt{\ek}} \int_{\ek/3}^{\ek/2} d\omega \sqrt{\ek - 2\omega} 
\Bigl[ 6\mathcal{K}'(\omega) - (\ek -3\omega) \mathcal{K}''(\omega)\Bigr] \\
&+\frac{4}{3\pi^2} \zeta \frac{\eF^2}{\ek} \int^{\ek}_{\ek/3} d\omega \mathcal{K}(\omega) 
\mathrm{Im} \Bigl[ L\bigl( \tfrac{\omega}{\ek} \bigr) \Bigr] \\
&-\frac{8}{15\pi} \xi \eF^{5/2} \int_{\ek/2}^{\infty} d\omega \sqrt{2\omega - \ek} 
\Bigl[ 3\mathcal{K}''(\omega) + (\omega - \ek) \mathcal{K}'''(\omega)\Bigr]. 
\end{split}
\label{eq:sum.rule2}
\end{equation}
For deriving this expression, we have, additionally to the assumptions mentioned earlier, assumed that $\mathcal{K}(\omega)$ vanishes 
at $\omega \to \infty$ faster than $1/\sqrt{\omega}$. If one wishes to use kernels which behave differently 
(as for instance in the so-called finite energy sum rules in QCD \cite{Krasnikov}, see also Appendix \ref{finite.energy}), 
one should go back to the OPE of 
Eq.(\ref{eq:OPE1}) and rederive the corresponding sum rules. 
Our statement made above on the cancellations of superficial divergences however still holds for this case. 

Furthermore, in the limit $k_0=\omega \gg \ek$, Eq.(\ref{eq:OPE1}) takes a considerably 
simpler form, making it thus possible to derive the resultant sum rule with much less effort. 
Moreover, if one introduces certain assumptions of the functional form of the self-energy, one can even 
analytically extract some of its properties from the sum rules. 
How this can be done by making use of the finite energy sum rules, is demonstrated in Appendix \ref{finite.energy}. 
While providing a simple and qualitatively correct picture, 
this approach however has the drawback of relying rather heavily on mean-field theory for fixing the form of the 
self-energy and therefore is inferior to the MEM analysis to be presented in the following sections, which does not 
need any other input besides the sum rules themselves. 

\subsection{Choice of the kernel $\mathcal{K}(\omega)$}
\begin{figure}
\begin{center}
\includegraphics[width=7.40cm]{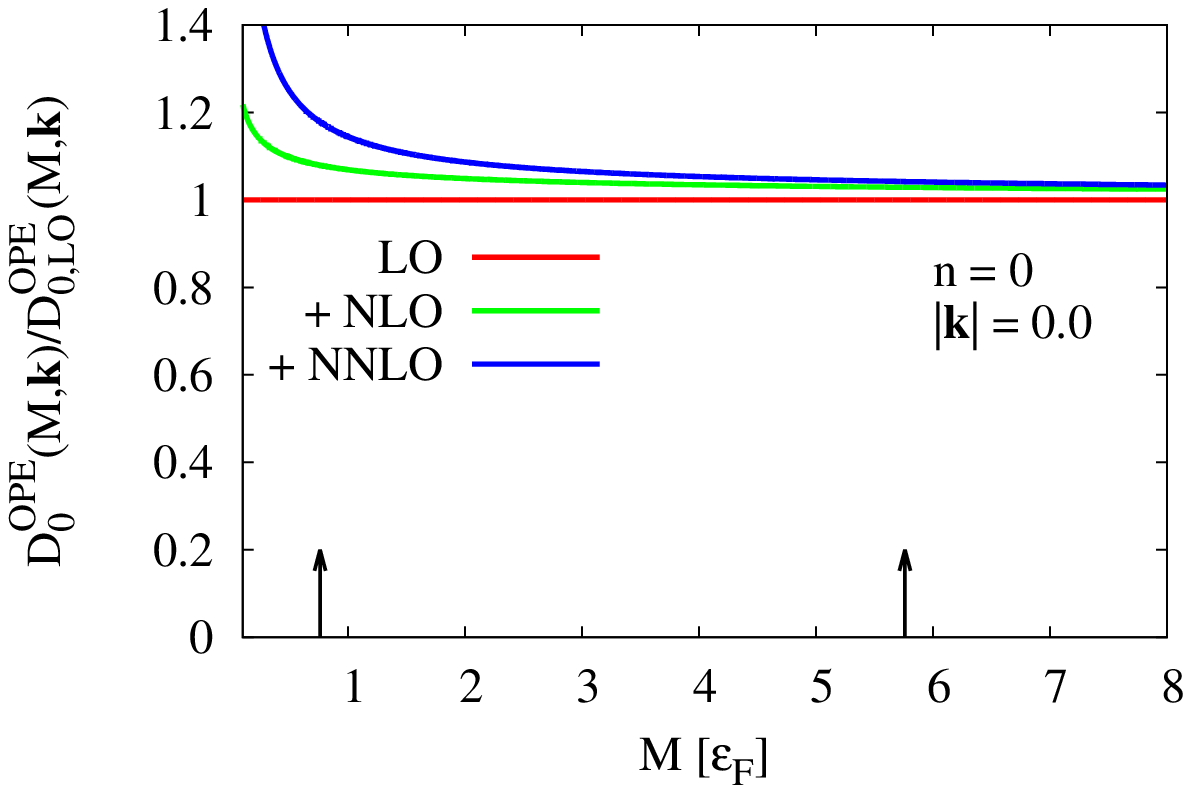} 
\includegraphics[width=7.40cm]{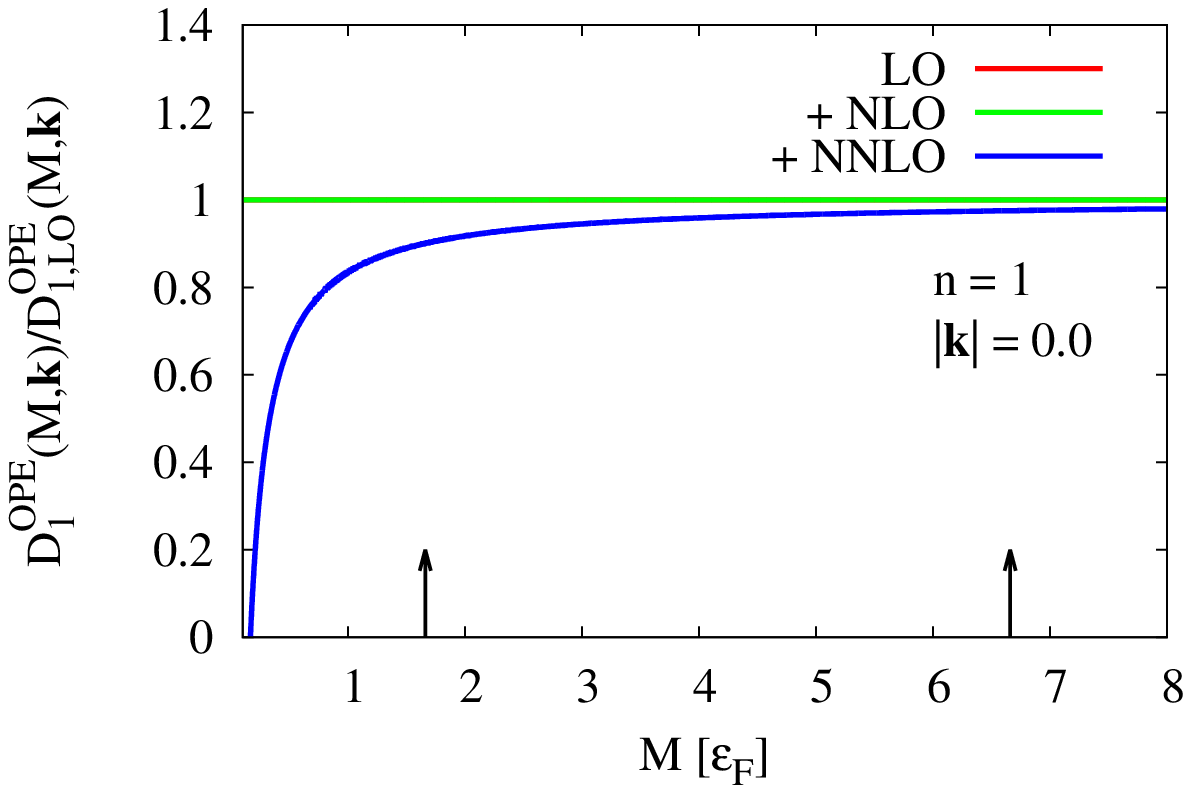}
\includegraphics[width=7.40cm]{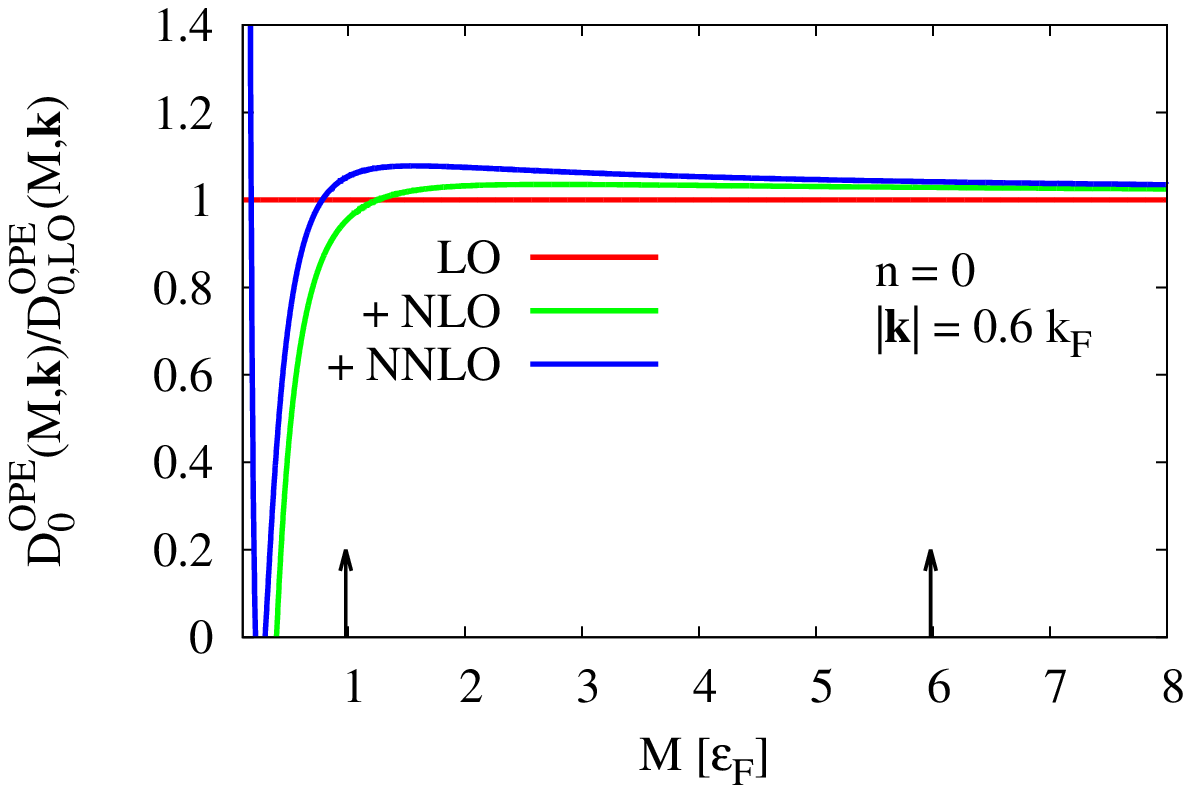} 
\includegraphics[width=7.40cm]{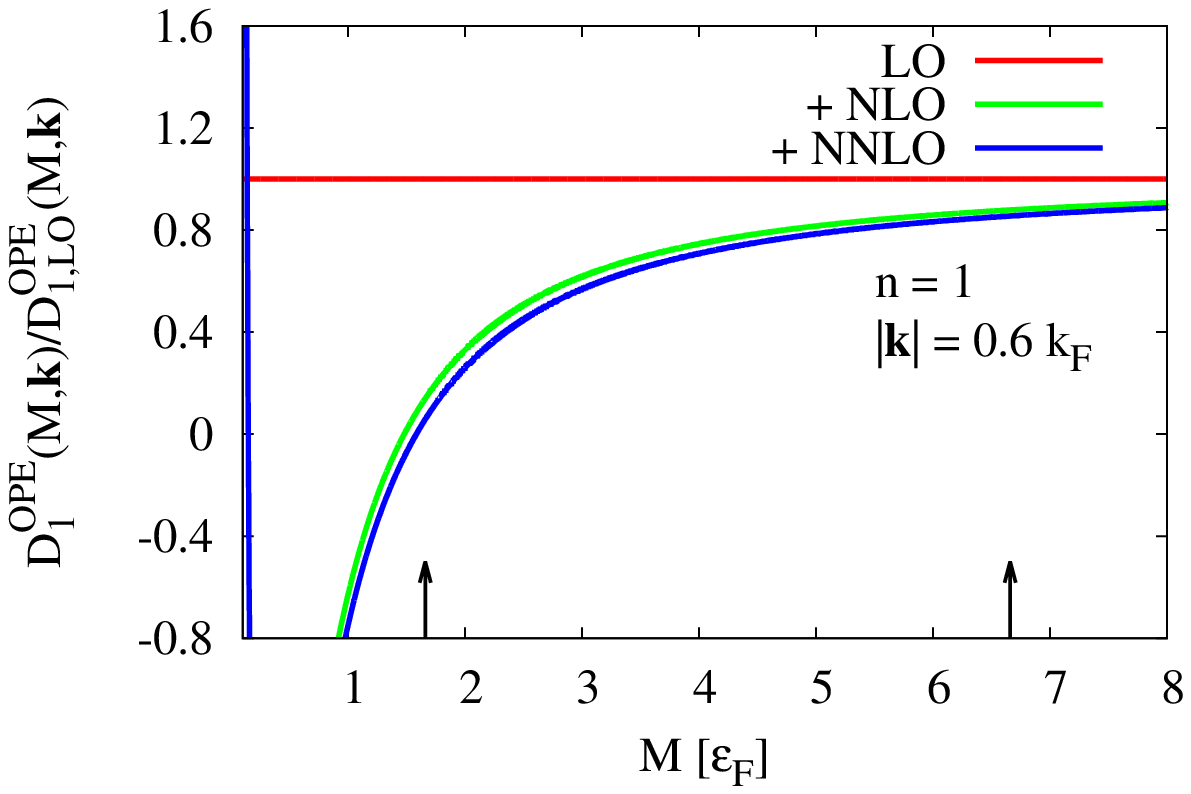}
\includegraphics[width=7.40cm]{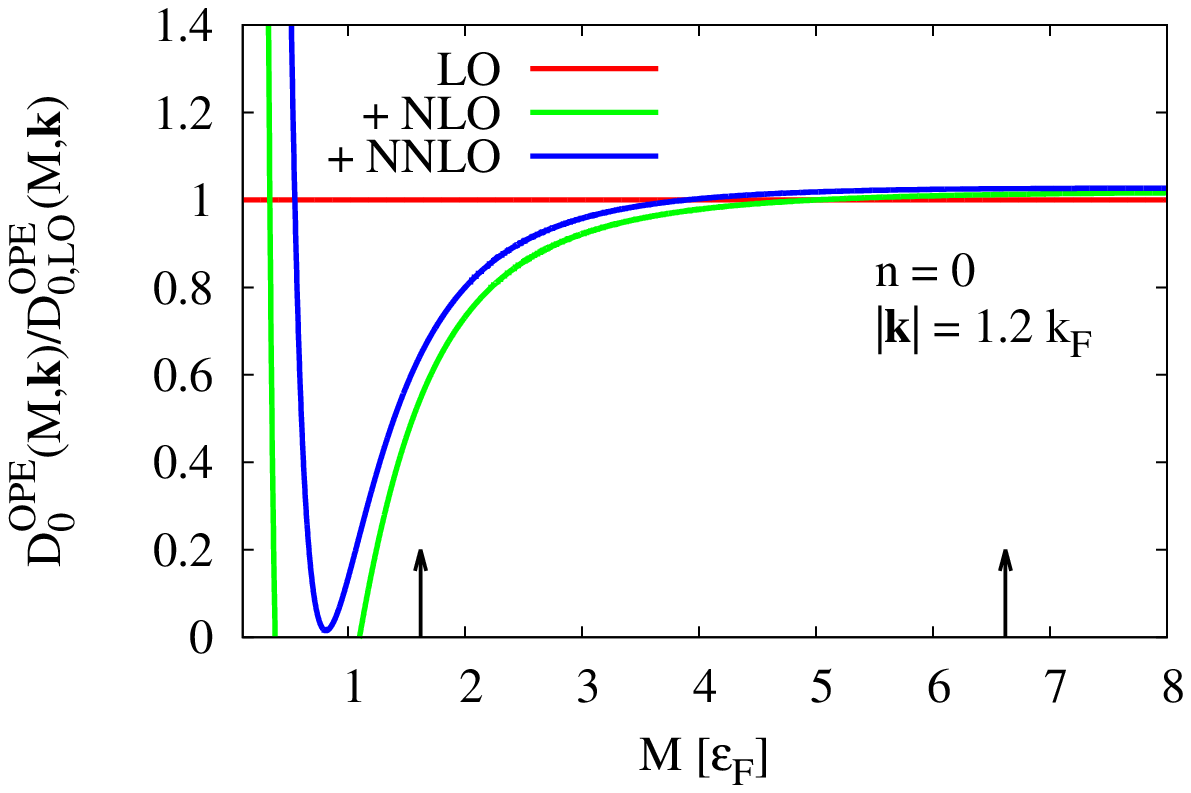} 
\includegraphics[width=7.40cm]{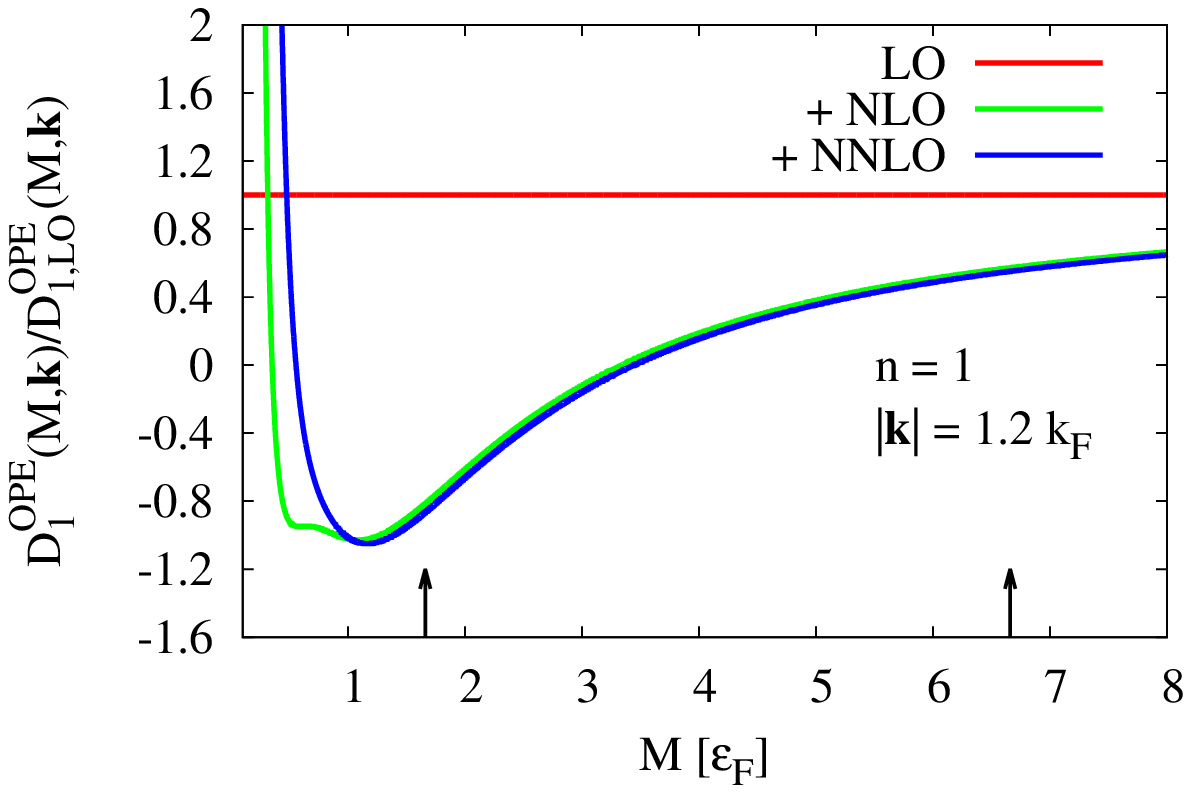}
\vspace{-0.3cm}
\caption{\label{fig:OPE} 
The right-hand sides of Eqs.(\ref{eq:sum.rule3.1}) and (\ref{eq:sum.rule3.2}), divided by their LO terms, 
as a function of the Borel mass $M$. The left and right plots 
show the cases of $n=0$ and $n=1$, respectively. Starting from the top, each line shows the OPE for momenta $|\bm{k}|/k_{\mathrm{F}}=0$, 
$0.6$ and $1.2$. 
Here, LO corresponds to the first line on the right-hand side of Eqs.(\ref{eq:sum.rule3.1}) and (\ref{eq:sum.rule3.2}), 
NLO to the second and third lines and NNLO to the fourth and fifth lines. 
The vertical arrows at the bottom of each 
plot indicate the lower and upper boundaries of the regions of $M$, which will be used in the MEM analysis of Section \ref{Analysis}.}
\end{center}
\end{figure}

As a next step, we have to fix the concrete form of the kernel $\mathcal{K}(\omega)$. 
As discussed in the previous sections, this kernel must be analytic on the complex plane of 
$\omega$ and real on the real axis. Furthermore, $\mathcal{K}(\omega)$ should vanish 
faster than $1/\sqrt{\omega}$ 
at $\omega \to \infty$ on the real axis. 
Obviously, these restrictions still give room for an infinite number of choices. 
From the experience of QCD sum rule analyses, it is however known that a simple Gaussian centered 
at the origin 
works well for extracting the lowest poles of the spectral function. We will in this paper follow a similar strategy and use 
\begin{equation}
\mathcal{K}_n(\omega, M) = \Bigl( \frac{\omega}{M}\Bigr)^n e^{-\omega^2/M^2}, \hspace{1cm} n=0,1  
\label{eq:Borel1}
\end{equation}
as our kernel. $M$ is usually referred to as the Borel mass in the QCD sum rule literature, 
which we will follow in this work, while 
in \cite{Goldberger,Goldberger2} the symbol $\omega_0$ was used for this variable. 
$M$ can in principle be freely chosen as long as the OPE converges. 
As will however be shown in Fig. \ref{fig:OPE}, the OPE convergence worsens for decreasing values of $M$, which means that there exists some lower boundary of $M$, below 
which the OPE is not a valid expansion. 

As the imaginary part of  the self-energy on the right-hand side of Eq.(\ref{eq:sum.rule2}) extends to both positive and negative values of $\omega$ and 
is in general not an even function, it is noted that using only the most simple kernel with $n=0$ does not suffice to determine 
$\mathrm{Im} \Sigma_{\up}(\omega + i0^{+}, \bm{k})$ as for this kernel all odd-function contributions automatically drop out of the sum rules. 
We hence need to introduce one more kernel which should be an odd function in $\omega$, 
for which the $n=1$ case in Eq.(\ref{eq:Borel1}) seems to be the most natural choice. 

Let us mention here that in the literature of QCD sum rules, other 
kernel choices have been proposed, such as a Gaussian with a variable center \cite{Bertlmann,Orlandini,Ohtani} or with complex Borel masses, 
which leads to an oscillating kernel \cite{Ioffe,Araki}. 
For this first study, we however prefer Eq.(\ref{eq:Borel1}) because of its simple analytic form. 

Substituting the above kernels into Eq.(\ref{eq:sum.rule2}) then gives the final form of the sum rules, 
\begin{align}
&\int^{\infty}_{-\infty}d\omega \mathcal{K}_0(\omega, M) \mathrm{Im} \Sigma_{\up}(\omega, \bm{k}) 
= D^{\mathrm{OPE}}_{\up,\,0}(M,\bm{k}) = \nonumber \\
& -\frac{2\sqrt{2}}{3\pi} \eF^{3/2} \sqrt{\ek} e^{-\frac{\ek^2}{8M^2}} K_{\frac{1}{4}}\Big(\frac{\ek^2}{8M^2}\Big) \nonumber \\
&+\frac{4}{3\pi} \zeta \eF^2 \Bigl( \frac{\sqrt{3}}{\pi}e^{-\frac{\ek^2}{9M^2}}  - e^{-\frac{\ek^2}{4M^2}}  \Bigr) \nonumber \\
&-\frac{8}{3\pi^2} \zeta \eF^2 \Bigl(\frac{M}{\sqrt{\ek}}\Bigr)^{1/2} G^1_0\Big(\frac{\ek}{M}\Big)
+ \frac{4}{3\pi^2} \zeta \eF^2 \frac{M}{\ek} G^2_0 \Big(\frac{\ek}{M}\Big) \nonumber \\
&-\frac{1}{30} \xi \eF^{5/2} \frac{1}{\sqrt{\ek}} e^{-\frac{\ek^2}{8M^2}} 
\Biggl\{ \Big(12 + 3\frac{\ek^2}{M^2} - \frac{\ek^4}{M^4}\Big) I_{\frac{1}{4}}\Big(\frac{\ek^2}{8M^2} \Big)
+\frac{\ek^2}{M^2} \Big(1 + \frac{\ek^2}{M^2}\Big) I_{-\frac{1}{4}}\Big(\frac{\ek^2}{8M^2} \Big) \nonumber \\
&- \frac{\ek^2}{M^2}\Big(3 + \frac{\ek^2}{M^2}\Big) \Biggl[ I_{\frac{3}{4}}\Big(\frac{\ek^2}{8M^2} \Big) -  I_{\frac{5}{4}}\Big(\frac{\ek^2}{8M^2} \Big)   \Biggr]
 \Biggr\} 
\label{eq:sum.rule3.1}
\end{align}
and
\begin{align}
&\int^{\infty}_{-\infty}d\omega \mathcal{K}_1(\omega, M) \mathrm{Im} \Sigma_{\up}(\omega, \bm{k}) 
= D^{\mathrm{OPE}}_{\up,\,1}(M,\bm{k}) = \nonumber \\
& -\frac{1}{6} \eF^{3/2}  \frac{M}{\sqrt{\ek}}  e^{-\frac{\ek^2}{8M^2}} 
\Biggl\{ \Big(4 - \frac{\ek^2}{M^2} \Big) I_{\frac{1}{4}}\Big(\frac{\ek^2}{8M^2} \Big) 
+\frac{\ek^2}{M^2} \Biggr[ I_{-\frac{1}{4}}\Big(\frac{\ek^2}{8M^2} \Big)
-I_{\frac{3}{4}}\Big(\frac{\ek^2}{8M^2} \Big) +  I_{\frac{5}{4}}\Big(\frac{\ek^2}{8M^2} \Big) \Biggr]  \Biggr\} \nonumber \\
&+\frac{4}{3\pi} \zeta \eF^2 \frac{\ek}{M} \Bigl( \frac{\sqrt{3}}{3\pi}e^{-\frac{\ek^2}{9M^2}}  + e^{-\frac{\ek^2}{4M^2}}  \Bigr) \nonumber \\
&+\frac{4}{3\pi^2} \zeta \eF^2  \frac{\sqrt{M}}{\sqrt{\ek}} G^1_1(\ek/M)
+ \frac{4}{3\pi^2} \zeta \eF^2 \frac{M}{\ek} G^2_1(\ek/M) \nonumber \\
&+\frac{1}{60}  \xi  \eF^{5/2} \frac{\sqrt{\ek}}{M} e^{-\frac{\ek^2}{8M^2}}  
\Biggl\{\Big(6 + 2\frac{\ek^2}{M^2} - \frac{\ek^4}{M^4}\Big) I_{-\frac{1}{4}}\Big(\frac{\ek^2}{8M^2} \Big) \nonumber \\
&-\Big(6 + 6\frac{\ek^2}{M^2} -\frac{\ek^4}{M^4}\Big) I_{\frac{1}{4}}\Big(\frac{\ek^2}{8M^2} \Big) 
+ \frac{\ek^4}{M^4} \Biggl[ I_{\frac{3}{4}}\Big(\frac{\ek^2}{8M^2} \Big) -  I_{\frac{5}{4}}\Big(\frac{\ek^2}{8M^2} \Big)   \Biggr]
 \Biggr\}, 
\label{eq:sum.rule3.2}
\end{align}
where $I_{\nu}(y)$ and $K_{\nu}(y)$ are the modified Bessel functions of the first and second kind, respectively. Furthermore, 
the functions $G^i_n(y)$ have been defined as follows: 
\begin{equation}
\begin{split}
G^1_0(y) =&  \int_{y/3}^{y/2} dx  \sqrt{y - 2x} \Bigl[ 6x 
- (y - 3x) (1 - 2x^2) \Bigr] e^{-x^2}, \\
G^1_1(y) =& \int_{y/3}^{y/2} dx  \sqrt{y - 2x} \Bigl[ 6(1 -2x^2) 
+ 2x(y - 3x)(3 - 2x^2) \Bigr] e^{-x^2}, \\
G^2_0(y) =& \int^{y}_{y/3} dx  
\mathrm{Im} \Bigl[ L\bigl( \tfrac{x}{y} \bigr) \Bigr] e^{-x^2}, \\
G^2_1(y) =& \int^{y}_{y/3} dx x  
\mathrm{Im} \Bigl[ L\bigl( \tfrac{x}{y} \bigr) \Bigr] e^{-x^2}. 
\end{split}
\label{eq:defs}
\end{equation}
The ratios of the 
right-hand sides of Eqs.(\ref{eq:sum.rule3.1}-\ref{eq:sum.rule3.2}) and their respective leading order terms are shown in Fig. \ref{fig:OPE} 
as functions of the Borel mass $M$ for 
three typical values of the momentum $|\bm{k}|$. 

The sum rules of Eqs.(\ref{eq:sum.rule3.1}) and (\ref{eq:sum.rule3.2}) look quite cumbersome, 
but their analytic structure becomes clearer if one takes the small momentum limit 
($\ek \to 0$). 
Using the kernel of Eq.(\ref{eq:Borel1}) with general values of $n$, one can show that in this limit 
the LO term behaves as $M^{1/2+n}$ and the NNLO term as $M^{-1/2+n}$. 
The NLO term on the other hand can be shown to be proportional to $M^0=1$ for $n=0$, 
while it vanishes for all other positive $n$ values. 
The results for $n=0$ and $n=1$ are given by  
\begin{align}
&\int^{\infty}_{-\infty}d\omega \mathcal{K}_0(\omega, M) \mathrm{Im} \Sigma_{\up}(\omega, \bm{k}) \nonumber \\
=& -\frac{2 \sqrt{2}}{3 \pi} \Gamma(1/4) \eF^{3/2} M^{1/2} - 
\frac{0.207498}{3 \pi} \eF^2 \zeta -\frac{4}{5} \frac{1}{\Gamma(1/4)} \eF^{5/2} \Bigl( \xi - \frac{5}{3} \frac{\ek}{\eF} \Bigr) \frac{1}{M^{1/2}} 
\label{eq:sum.rule4.1}
\end{align}
and 
\begin{align}
&\int^{\infty}_{-\infty}d\omega \mathcal{K}_1(\omega, M) \mathrm{Im} \Sigma_{\up} (\omega, \bm{k}) \nonumber \\
=& -\frac{4}{3} \frac{1}{\Gamma(1/4)} \eF^{3/2} M^{3/2} + \frac{\sqrt{2}}{10 \pi} \Gamma(1/4) \eF^{5/2} \Bigl( \xi - \frac{5}{3} \frac{\ek}{\eF} \Bigr) M^{1/2}. 
\label{eq:sum.rule4.2}
\end{align}
Here, the term proportional to $\ek$ in the last term comes from Taylor expanding the leading order terms 
of the first lines of Eqs.(\ref{eq:sum.rule3.1}) and (\ref{eq:sum.rule3.2}) in $\ek/M$. 
The above equations 
should give the reader an idea on the behavior of the OPE at least for small $|\bm{k}|$. 
In the actual analysis of the next section, we will however use the full result of 
Eqs.(\ref{eq:sum.rule3.1}) and (\ref{eq:sum.rule3.2}). 

\section{\label{Analysis} MEM analysis for the spectral density}
Next, we discuss the imaginary parts of the self-energies, which we have extracted numerically from the 
sum rules by using the maximum entropy method (MEM). This sort of approach for 
analyzing sum rules, was recently applied to QCD in a similar way \cite{Gubler} and has 
during the last few years been used to study hadrons in various environments \cite{Ohtani,Gubler3,Gubler2,Suzuki,Ohtani2,Gubler4}. 
For the technical details of this analysis, we refer the reader to Appendix \ref{MEM} and the references cited 
therein. 

\subsection{The Borel window and the default model}
Before discussing our results, let us here at first briefly explain how to determine 
the lower and upper boundaries of the Borel mass $M$ used in the 
analysis. For fixing the lower boundary $M_{\mathrm{min}}$, we demand that the 
highest order (NNLO) OPE term, which is proportional to $\xi$, 
should be smaller than 10\% of the leading order term. 
Note, that this condition generally leads to a value of $M_{\mathrm{min}}$, which depends on the momentum $|\bm{k}|$. 
We will here first fix $M_{\mathrm{min}}$ at $|\bm{k}|=0$ and 
take this momentum dependence into account only if it leads 
to an increasing value of $M_{\mathrm{min}}$. This keeps the momentum dependence of $M_{\mathrm{min}}$ to a minimum and 
at the same time ensures that for any value of $|\bm{k}|$, only Borel mass ranges with a satisfactory OPE convergence are 
used as input for the MEM analysis. 

For fixing the upper boundary $M_{\mathrm{max}}$, we do 
not have such a clear-cut criterion and therefore can in principle choose it freely as long as it lies above $M_{\mathrm{min}}$. 
For the analysis presented in this paper, 
we will set it as $M_{\mathrm{max}} = M_{\mathrm{min}} + x$, with $x = 5\,\epsilon_F$. 
We have checked that 
our results do not much depend on this choice and the exact value of $x$ hence does not play any important role in the present analysis. 
The specific values of $M_{\mathrm{min}}$ and $M_{\mathrm{max}}$ for some typical momentum 
values are indicated in Fig. \ref{fig:OPE} as vertical arrows at the bottom of each plot. 

As for the default model $m(\omega)$, which is an input of  the MEM algorithm (see Appendix \ref{MEM} for details), 
we will use 
\begin{equation}
m(\omega) = -\frac{4\sqrt{2}}{3\pi} \eF^{3/2} \frac{1}{(\omega^2 + y)^{1/4}}, 
\label{eq:def.model}
\end{equation}
with $y=\epsilon_{\mathrm{F}}^2$. 
As can be understood from Eq.(\ref{eq:OPE1}), 
the above default model approaches the correct asymptotic limit 
of $\mathrm{Im} \Sigma_{\up}(\omega,\bm{k}) \simeq -(4\sqrt{2} \eF^{3/2})/(3\pi \sqrt{\omega})$, 
as $\omega \gg  \ek$ and is therefore a suitable choice for the present analysis. 
For avoiding singularities at $\omega=0$, we have introduced the parameter 
$y$ for smoothing out the function around the origin. 
We have tested different choices for $y$ and found that this 
affects our analysis results only very weakly. 

\subsection{The single-particle spectral density}
\begin{figure}
\begin{center}
\includegraphics[width=5.2cm]{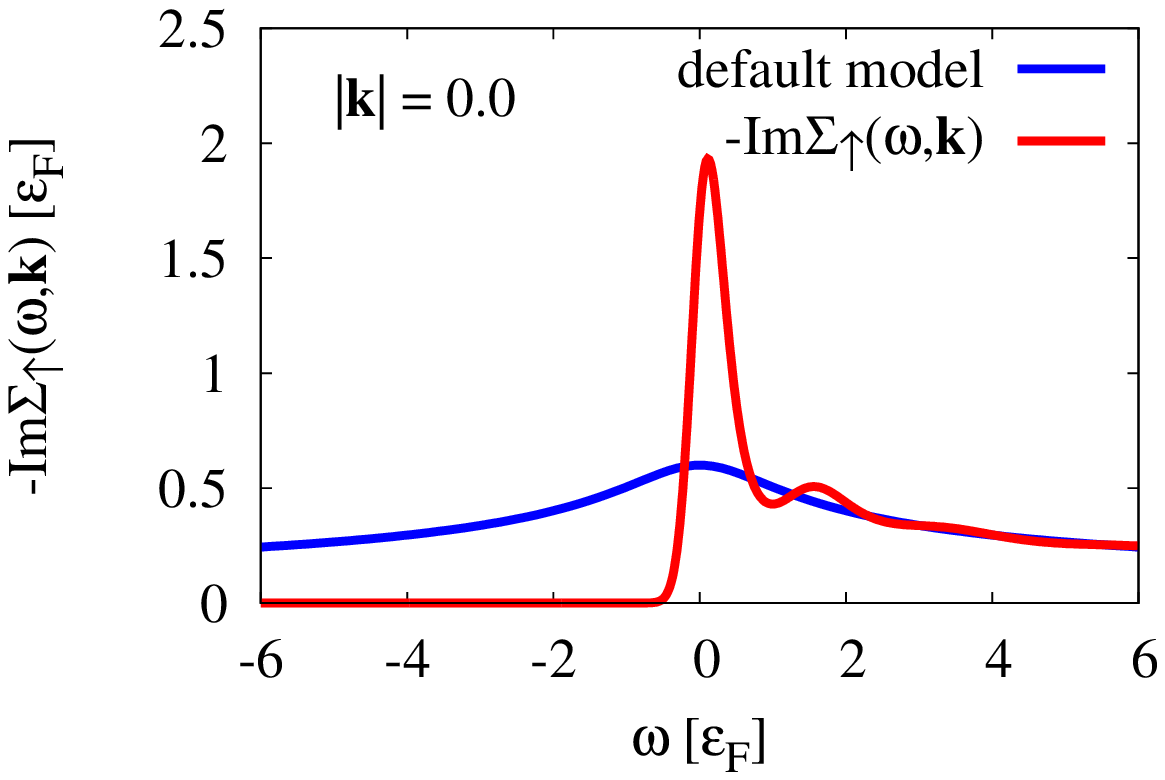} 
\includegraphics[width=5.2cm]{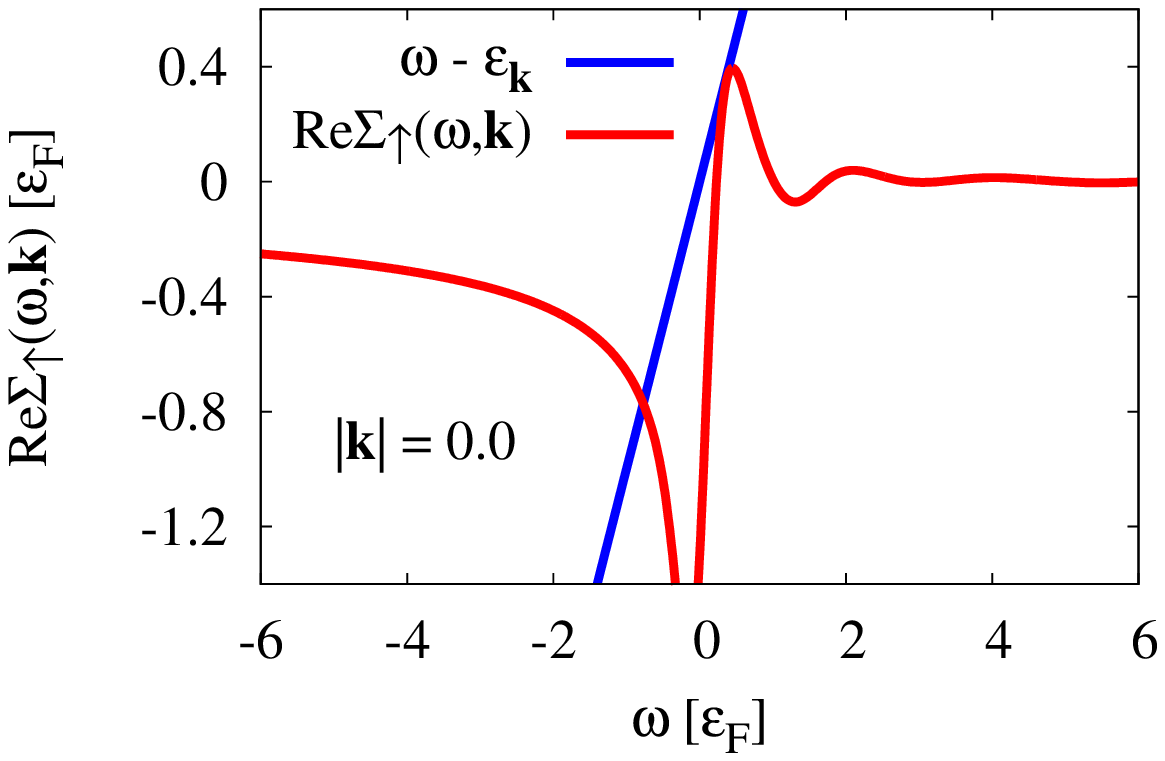}
\includegraphics[width=5.2cm]{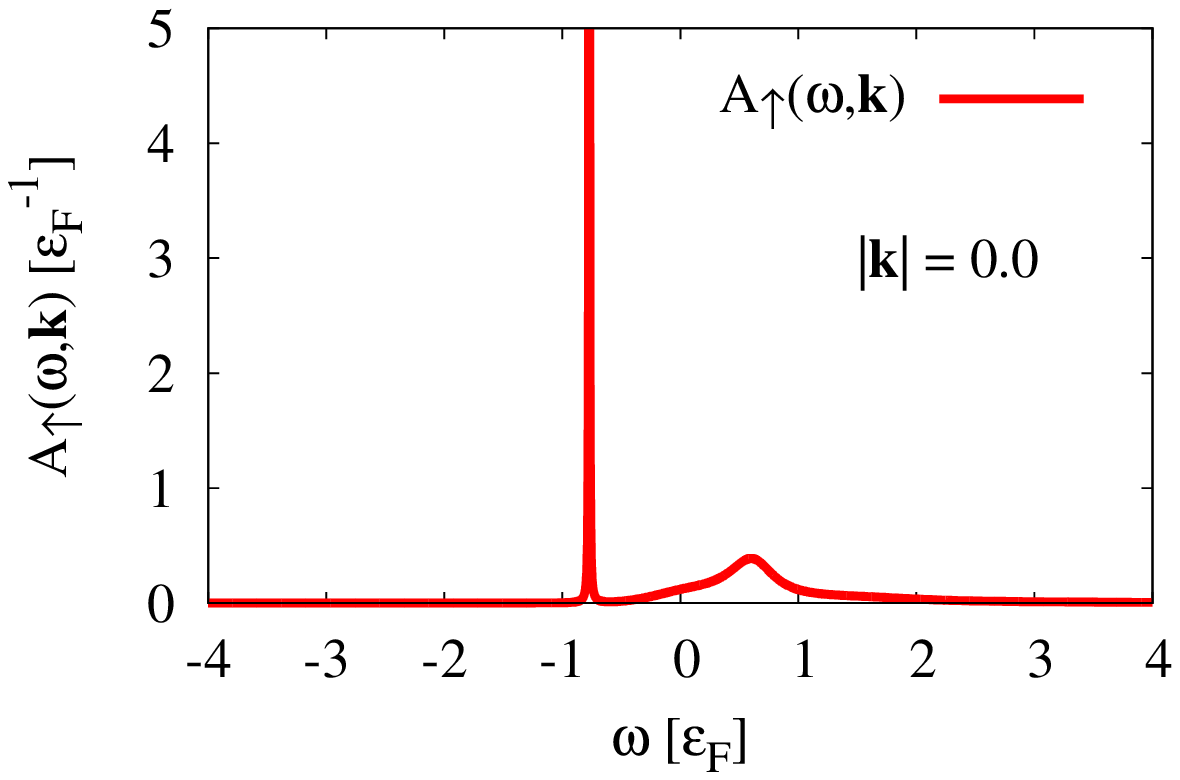}
\includegraphics[width=5.2cm]{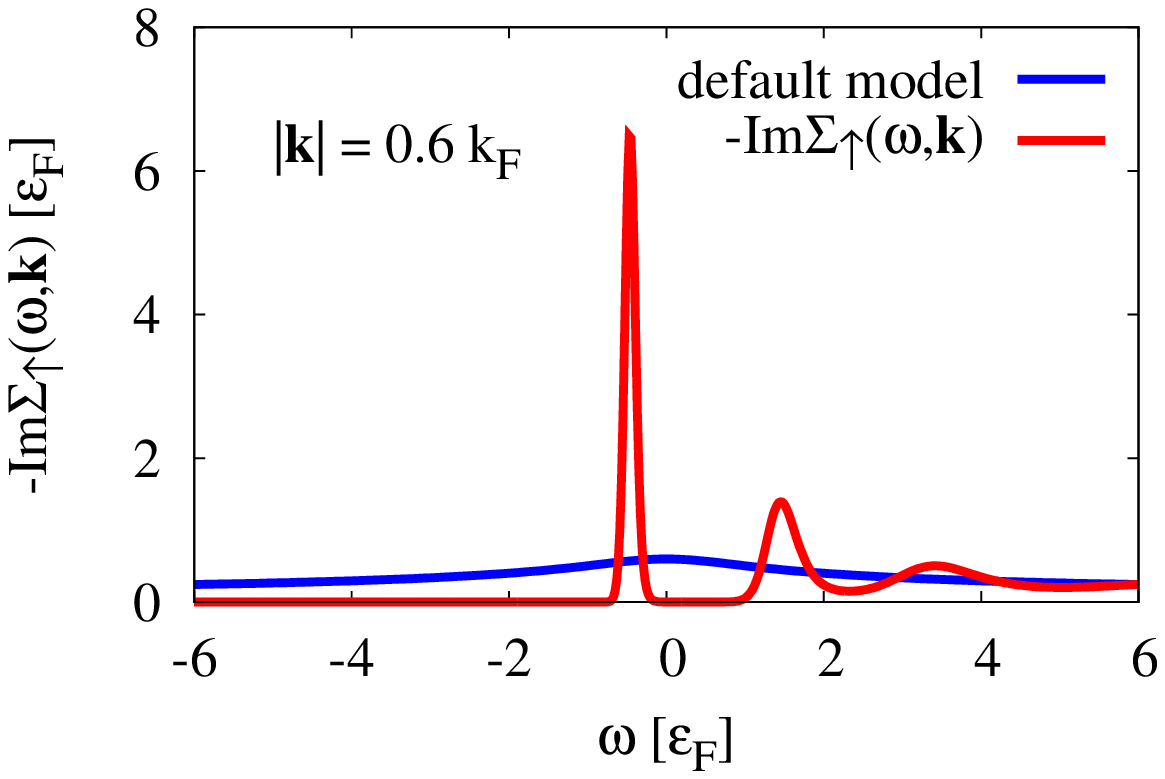} 
\includegraphics[width=5.2cm]{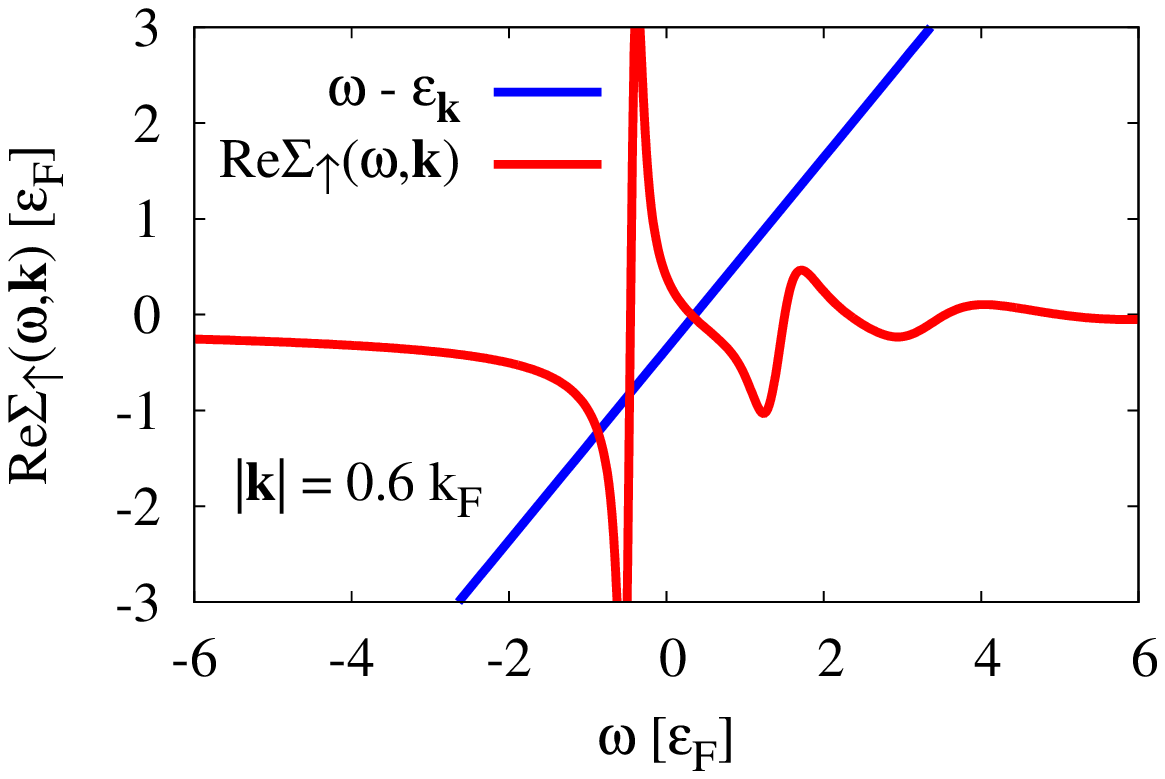}
\includegraphics[width=5.2cm]{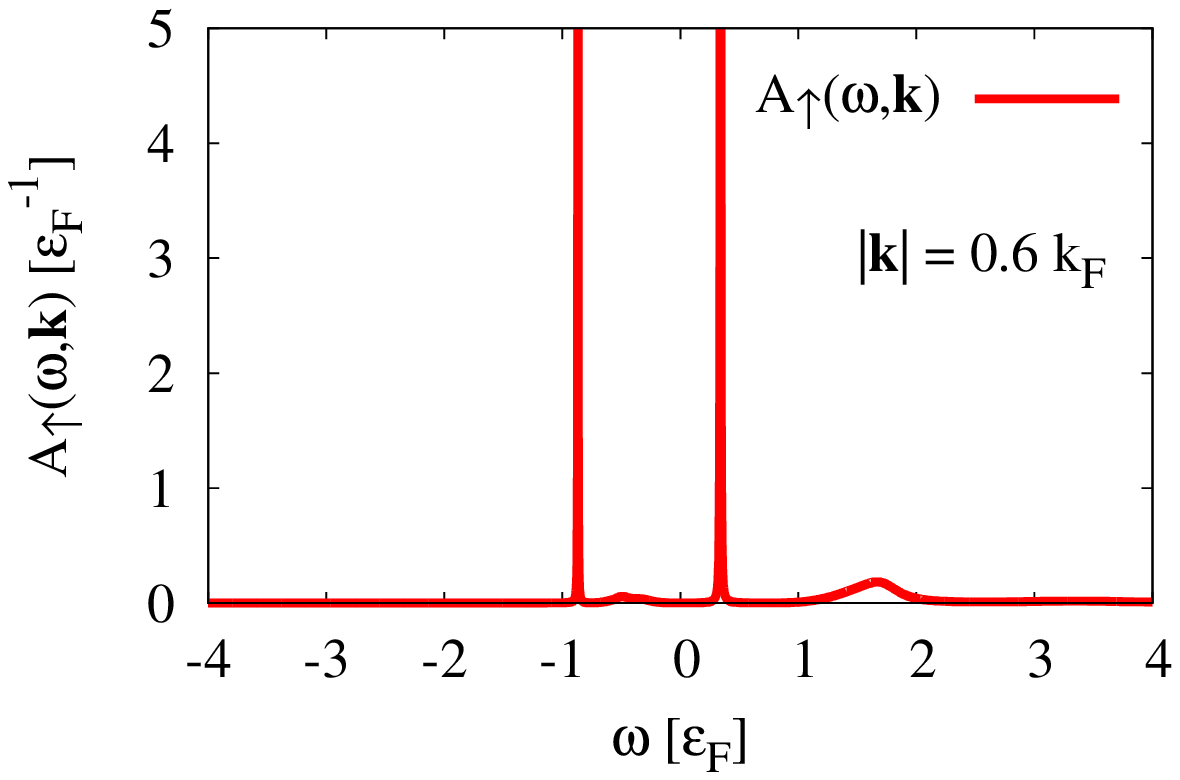}
\includegraphics[width=5.2cm]{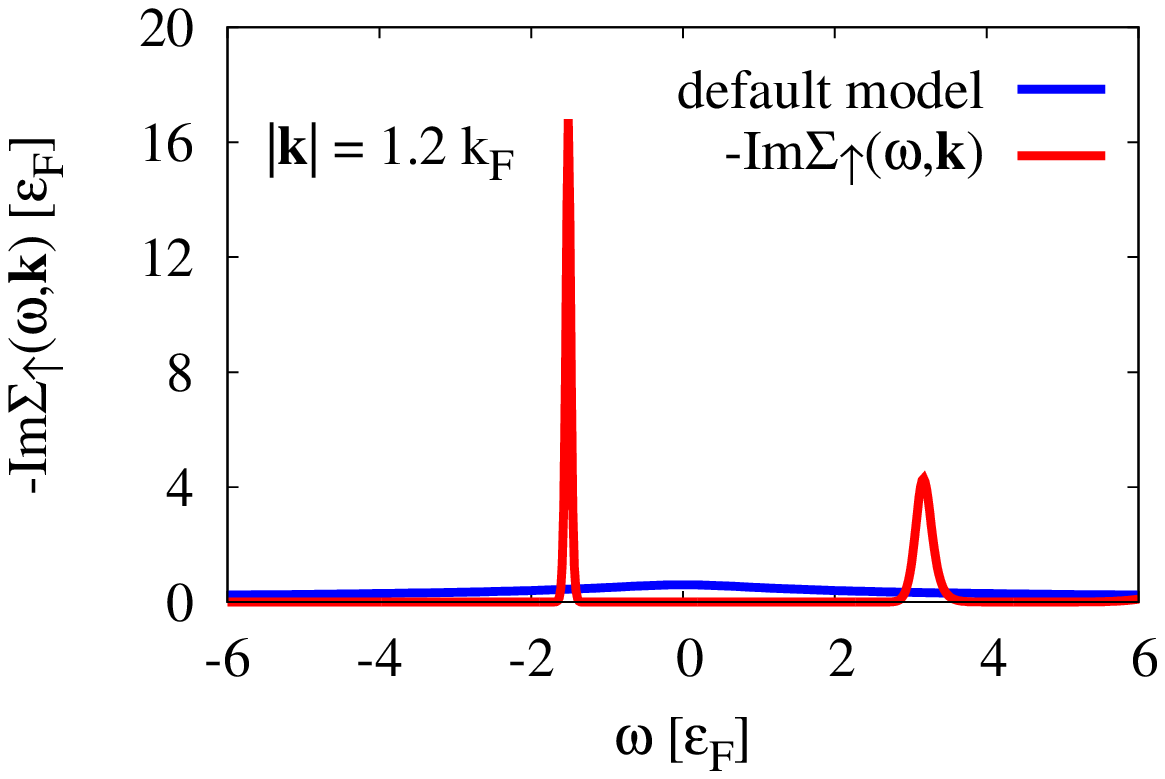} 
\includegraphics[width=5.2cm]{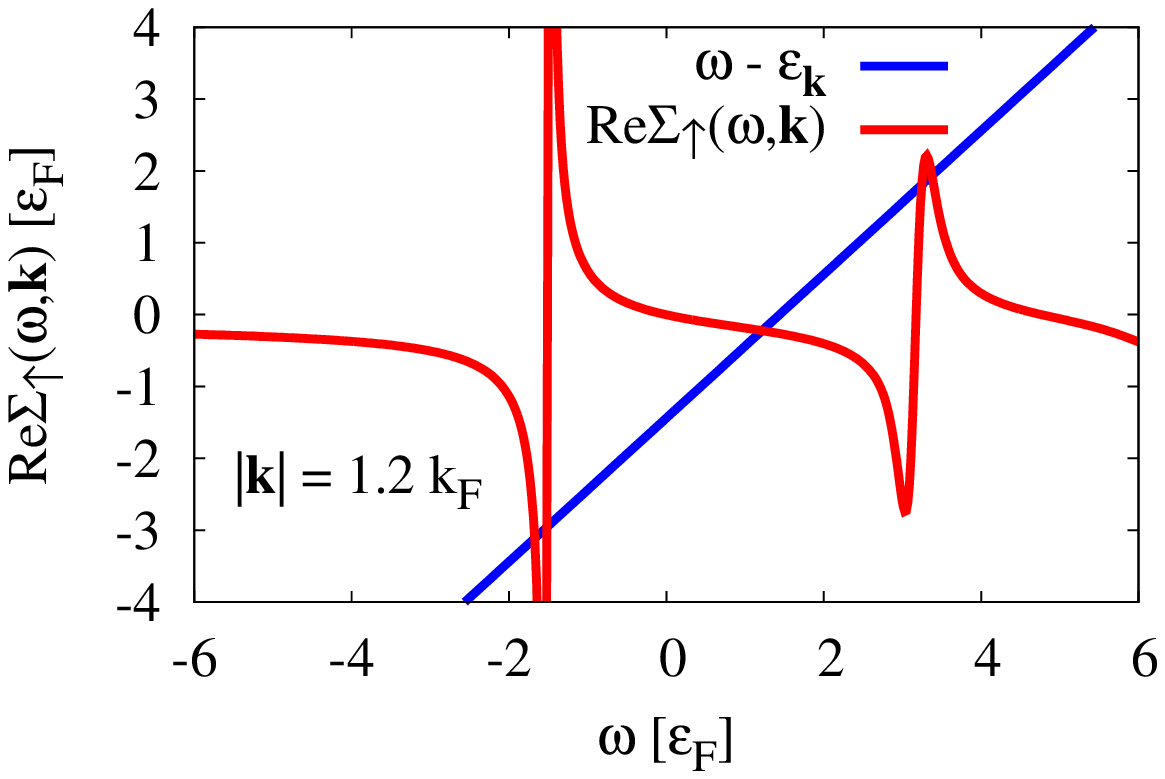}
\includegraphics[width=5.2cm]{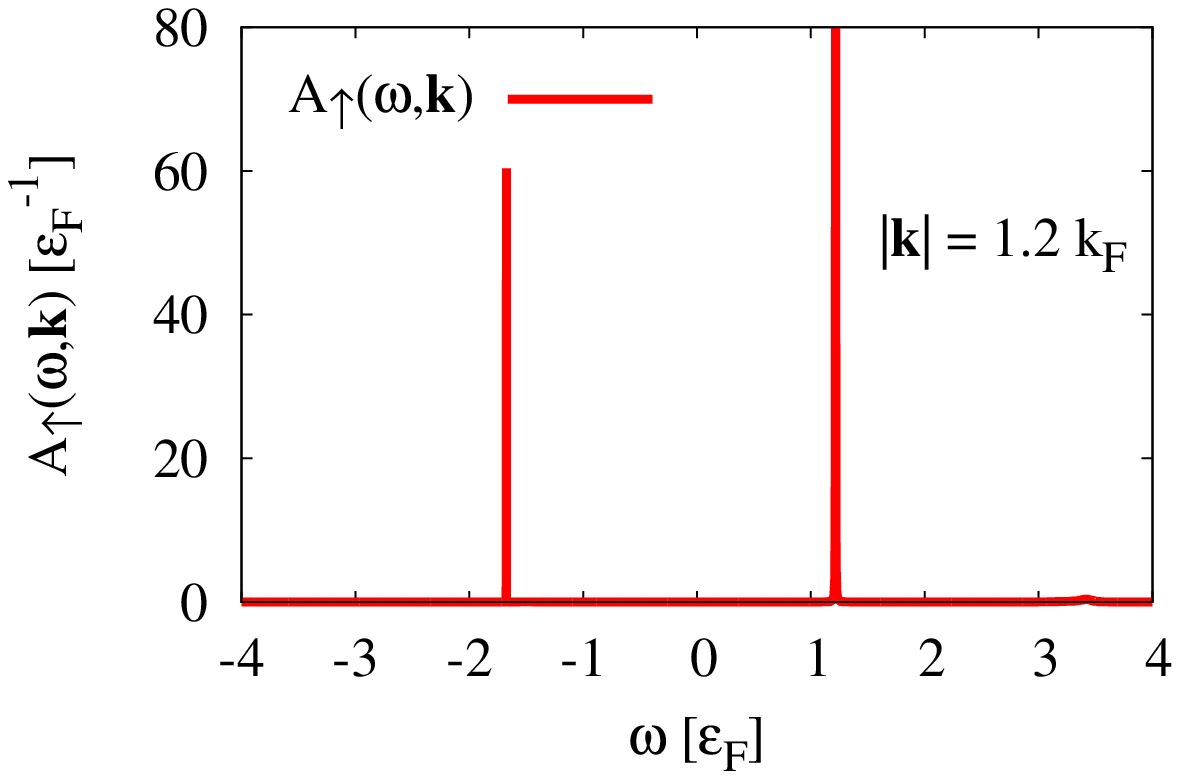}
\vspace{-0.5cm}
\caption{\label{fig:spec.func} Left column: Results of the MEM analysis of Eqs.(\ref{eq:sum.rule3.1}) and (\ref{eq:sum.rule3.2}) 
are shown as red lines, while the used default model [see Eq.(\ref{eq:def.model})] is indicated in blue. Middle column: The real parts of 
the self-energies obtained from Eq.(\ref{eq:disp}) and $\mathrm{Im} \Sigma_{\up}(\omega,\k)$ are plotted as red lines, and 
the function $\omega - \epsilon_{\bm{k}}$ is given in blue. Right column: The spectral density 
$A_{\up}(\omega,\bm{k})$, as computed from the results of the two columns on the left and Eq.(\ref{eq:spe}). 
As in Fig. \ref{fig:OPE}, each row from top to bottom corresponds to momenta $|\bm{k}|/k_{\mathrm{F}}=0.0$, 
$0.6$ and $1.2$, respectively.}
\end{center}
\end{figure}

After these preparations, we can now finally proceed to our analysis results. 
First, we show the imaginary part of the self-energy, for three representative 
momenta in the left column of Fig.~\ref{fig:spec.func}. 
For illustration, we show in these plots also the used default model of Eq.(\ref{eq:def.model}). 
It is seen that for zero momentum, 
the spectral function is composed of one single peak around $\omega=0$ and 
a continuum behaving as $\sim 1/\sqrt{\omega}$ in the positive 
energy region. As the momentum increases, the initial peak separates into 
two distinct peaks which start to move into opposite directions. The continuum 
also recedes into the positive $\omega$ region with increasing momentum, 
leaving a growing region around the origin without any strength at all. 

With the extracted 
$\mathrm{Im} \Sigma_{\up}(\omega,\bm{k})$, 
we next compute the real part of the self-energy 
by using the Kramers-Kr$\mathrm{\ddot{o}}$nig relation 
\begin{equation}
\mathrm{Re}\Sigma_{\up}(\omega,\bm{k}) = 
- \frac{1}{\pi} \mathrm{P} \int_{-\infty}^{\infty} d\omega' \frac{\mathrm{Im} \Sigma_{\up}(\omega',\bm{k})}{\omega - \omega'},
\label{eq:disp}
\end{equation}
and executing the principal value integral numerically. The result of this evaluation is given in the 
middle column of Fig. \ref{fig:spec.func}, where we also show the curve $\omega - \epsilon_{\bm{k}}$, which appears in the 
denominator of the 
right-hand side of Eq.(\ref{eq:spe}). It is clear from this equation that if the imaginary part of the self-energy 
happens to be small, the single-particle spectral density will have a narrow peak wherever 
$\mathrm{Re}\Sigma_{\up}(\omega,\bm{k})$ coincides with $\omega - \epsilon_{\bm{k}}$. 

\begin{center}
\begin{figure}
\vspace{-1.0cm}
\hspace*{-0.5cm}
\includegraphics[width=15cm,bb=0 0 360 252]{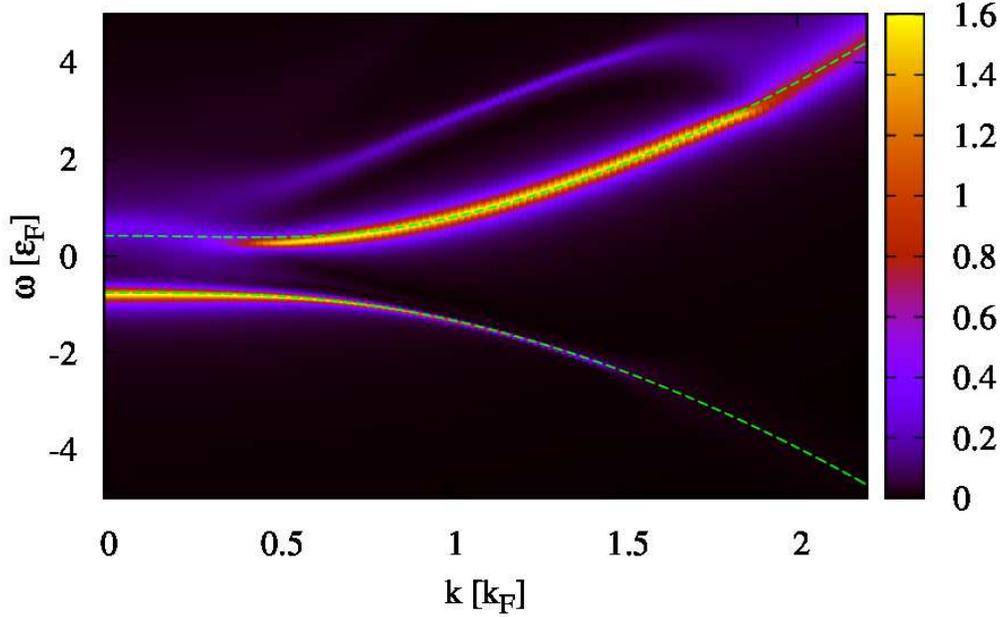}
\vspace{-1.0cm}
\caption{\label{fig:density.plot} Density plot of the spectral density $A_{\up}(\omega,\bm{k})$ shown as a function of energy $\omega$ and 
momentum $|\bm{k}|$. The green dashed lines indicate the results of a fit of the particle and hole peak-maxima to Eq.(\ref{eq:fit.func}).}
\end{figure}
\end{center}

As a last step, we simply plug the real and imaginary parts of the self-energy into 
\begin{equation}
A_{\up}(\omega,\bm{k}) = -\frac{1}{\pi} \mathrm{Im} \frac{1}{\omega + i0^{+}-\epsilon_{\bm{k}}-\Sigma_{\up}(\omega+i0^{+},\bm{k})}, 
\label{eq:spe.second.time}
\end{equation} 
to obtain the single-particle spectral density $A_{\up}(\omega,\bm{k})$. The resulting functions are given in the right column of 
Fig. \ref{fig:spec.func}. It can be seen there, that for small momenta $|\bm{k}|$, the spectral density is dominated by the narrow 
hole-branch in the negative energy region, while the particle-branch consists of only a relatively broad bump. This changes 
at around $|\bm{k}| \sim 0.5\,k_{\mathrm{F}}$, where the main strength of the spectral density switches over to the particle branch, 
which, as the momentum is further increased, starts to move into the positive energy direction. 
On the other hand, the hole-branch bends back into the negative energy region, while gradually losing its strength. 
To give the reader a better visual grasp of the spectral density as a whole and especially on the behavior of the particle and 
hole branches, we show $A_{\up}(\omega,\bm{k})$ in a density plot as a function of both energy $\omega$ and 
momentum $|\bm{k}|$ in Fig. \ref{fig:density.plot}. To improve the visibility of this plot without changing its essential features, we have 
artificially increased the imaginary part of $\Sigma_{\up}(\omega,\bm{k})$ in Eq.(\ref{eq:spe.second.time}) by an amount of $0.2\,\eF$. 

In this figure, the typical BCS-like dispersion of the particle and hole branches clearly manifest themselves. 
Qualitatively, this result agrees with the spectral densities extracted from both quantum Monte-Carlo calculations 
\cite{Magierski} and a Luttinger-Ward approach \cite{Haussmann}. 
In order to make a quantitative comparison with other methods, we fit the peak maxima to a dispersion relation parametrized 
as 
\begin{equation}
E_{\bm{k}}^{\pm} = \mu \pm \sqrt{\left( \frac{m}{m^{\pm}} \epsilon_{\bm{k}} + U^{\pm} - \mu \right)^2 + \Delta^2}, 
\label{eq:fit.func}
\end{equation}
which we have adopted from \cite{Haussmann}. 
The resultant curves are shown in Fig. \ref{fig:density.plot} as green dashed lines, while 
the corresponding values of $\mu$, $\Delta$, $m^{\pm}$ and $U^{\pm}$ are given in Table \ref{tab:disp.para}. 
\begin{table}
\begin{center}
\caption{Fit results of the particle and hole branches shown in Fig. \ref{fig:density.plot} to a dispersion relation 
parametrized as in Eq.(\ref{eq:fit.func}).} 
\label{tab:disp.para}
\begin{tabular}{ccccccc} \hline
& & &\multicolumn{2}{c}{Particle} & \multicolumn{2}{c}{Hole} \\ \cline{4-5} \cline{6-7} 
& $\mu/\epsilon_{\mathrm{F}}$ & $\Delta /\epsilon_{\mathrm{F}}$ & $m^{+}/m$ & $U^{+}/\epsilon_{\mathrm{F}}$ & $m^{-}/m$ & $U^{-}/\epsilon_{\mathrm{F}}$ \\ \hline
this work & \hspace*{0.5cm}-0.18\hspace*{0.5cm} & \hspace*{0.5cm}0.57\hspace*{0.5cm} & \hspace*{0.5cm}1.02\hspace*{0.5cm} 
&\hspace*{0.5cm} -0.37 \hspace*{0.5cm} &\hspace*{0.5cm} 1.09\hspace*{0.5cm} & \hspace*{0.5cm}-0.12\hspace*{0.5cm} \\
\cite{Haussmann} & \hspace*{0.5cm}0.36\hspace*{0.5cm} & \hspace*{0.5cm}0.46\hspace*{0.5cm} & \hspace*{0.5cm}1.00\hspace*{0.5cm} 
&\hspace*{0.5cm} -0.50 \hspace*{0.5cm} &\hspace*{0.5cm} 1.19\hspace*{0.5cm} & \hspace*{0.5cm}-0.35\hspace*{0.5cm} \\
\hline
\end{tabular}
\end{center}
\end{table} 
It is seen in Figure \ref{fig:density.plot} that the fit is able to reproduce our dispersion relation fairly well, with 
the exception of the low momentum region of the particle branch, whose curvature can not be 
captured fully by the simple formula of Eq.(\ref{eq:fit.func}). Note that this leads to a slight overestimation of the gap $\Delta$. If we simply 
read it of from the point at which the particle and hole branches are closest, we get a value of 
$\Delta/\eF=0.54$ instead of the one given in Table \ref{tab:disp.para}. 
 
Comparing the values of this work with those of \cite{Haussmann}, it is seen that 
the two approaches give comparable results for the gap parameter $\Delta$, 
effective masses $m^{\pm}$ and Hartree shifts $U^{\pm}$ for both the particle and hole branches. 
On the other hand, 
the chemical potential $\mu$ 
deviates significantly from \cite{Haussmann}, even giving a different sign. 
The reason for this discrepancy apparently originates in the low sensitivity of the sum rules to 
the absolute position of the $\omega$ axis. 
This can be understood by inspecting the OPE of Eq.(\ref{eq:OPE1}). 
After setting $k_0 = \omega$ and making a change 
of variables $\omega \to \omega'$ as $\omega = \omega' + \omega_0$, 
with $\omega_0$ of the order of $\epsilon_{\mathrm{F}}$ and 
expanding the resulting expression in $\omega_0/\omega'$, 
one notes that only the NNLO term of the OPE will be modified, which must be 
kept small 
due to the convergence condition of the OPE. 
Therefore, we can expect that such a change of variables will introduce no 
qualitative modification of the OPE, while the spectral density experiences a 
parallel shift of $\omega_0$. 

It is in principle possible to choose $\omega_0$ such that 
the fitted value of $\mu$ approaches the correct value of around $0.36\,\eF$. 
Due to the convergence criterion of the OPE, such a choice however leads to 
a significantly larger value of $M_{\mathrm{min}}$ and therefore to a rather poor 
resolution of the MEM extraction of $\mathrm{Im}\Sigma_{\up}(\omega,\bm{k})$. 
We have thus not explored this possibility any further and 
simply note that at the present stage, 
the absolute positions of the structures appearing in the spectral 
density should not be taken too seriously. 

\begin{center}
\begin{figure}
\includegraphics[width=12cm]{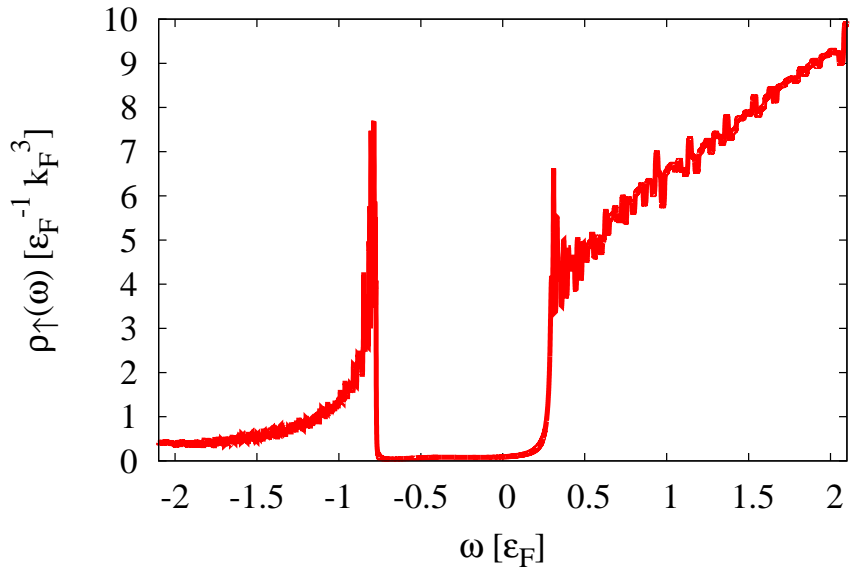}
\vspace{-0.7cm}
\caption{\label{fig:int.k.plot} 
The density of states, $\rho_{\up}(\omega)$, obtained by integrating 
the spectral density $A_{\up}(\omega,\bm{k})$ over the momentum $\bm{k}$ as shown in Eq.(\ref{eq:int.k}).}
\end{figure}
\end{center}

As a final point, we study the density of states of the single 
argument $\omega$, $\rho_{\up}(\omega)$, which is obtained by 
integrating the spectral density over the momentum $|\bm{k}|$: 
\begin{equation}
\rho_{\up}(\omega) = \int \frac{d^3k}{(2\pi)^3} A_{\up}(\omega,\bm{k}).    
\label{eq:int.k}
\end{equation} 
This function is shown in Fig. \ref{fig:int.k.plot}, from which 
one can immediately read off the approximate gap value, which can be regarded as 
half the width of the region where $\rho_{\up}(\omega)$ loses almost all of its strength. 
To draw Fig. \ref{fig:int.k.plot}, 
we have added an constant amount of $0.002\,\eF$ to the imaginary part of 
$\Sigma_{\up}(\omega,\bm{k})$, which reduces artificial effects caused by evaluating the integral numerically from 
a discrete number of data points, but does not change the gap structure of this plot. 

\section{\label{Summary} Summary and conclusion}
The work presented in this paper was carried out with two essential goals in mind. 
As the introduced techniques are new and have not been applied to cold atom systems so far, we first 
needed to test 
to what extent the sum rules and MEM are able 
to extract the single-particle spectral density 
from the result of the OPE. This is by no means a trivial test, 
because the OPE considered in momentum space does not converge for momenta 
below the Fermi momentum \cite{Nishida}, as we have already discussed in the introduction. 
It was therefore at the beginning not clear 
to what degree the 
sum rules can extend the 
applicability of the OPE to lower momenta or energies. 
As it however turns out, even at zero momentum $|\bm{k}|$ and small $\omega$, the sum rules 
of Eqs.(\ref{eq:sum.rule3.1}) and (\ref{eq:sum.rule3.2}) lead a fairly reasonable behavior for the spectral density, 
which suggests that our approach is indeed useful for extracting the spectral density at any momentum and energy. 

Once the proposed method is proven to work well, our second goal was to provide an independent 
framework for evaluating the superfluid pairing gap $\Delta$ of the unitary Fermi gas. Our obtained value is given 
in Table \ref{tab:disp.para} and can be inferred from Fig. \ref{fig:density.plot}. 
We wish to emphasize here that even though we have only taken into account the first few terms of 
the OPE, in which the Bertsch parameter and the contact density are the only input values, our 
numerical result shows reasonable agreement with 
other theoretical approaches \cite{Haussmann,Carlson2}. 
Specifically, we obtain $\Delta/\eF=0.54$, when extracting the gap from point of smallest distance between the particle and 
hole branches and $\Delta/\eF=0.57$ from an overall fit of our dispersion relation to Eq.(\ref{eq:fit.func}), while \cite{Haussmann} and 
\cite{Carlson2} get 
$\Delta/\eF=0.46$ and $\Delta/\eF=0.50(3)$, respectively. 
For confirming these results in the future, it will be necessary to consider 
still higher order terms in the OPE, evaluate the size of their contributions 
and examine their impact on the spectral density. 

Using the method proposed in this work, we have so far 
only studied the fermionic single-particle channel at zero temperature. 
As long as the conditions for its applicability (that is $r_0 \ll 1/\sqrt{|k_0|} \ll  |a|,\,n^{-1/3},\,\lambda_T$) are satisfied, the 
OPE technique is fairly general and can in principle be applied to 
any kind of bosonic or fermionic systems with one or more constituents. 
One can therefore envisage various future applications of this approach. 
For instance, in \cite{Goldberger} the OPE for the retarded correlator of the density 
operator has already been worked out, and one in principle just needs to apply 
MEM or some other sort of fitting method to extract information on the dynamic 
structure factor from the OPE expression. 
Another interesting direction of research could be the generalization of this approach 
to finite temperature. For being able to do this, one however needs information 
on the finite temperature behavior of the operator expectation values which 
appear in the OPE of the channel of interest. 
For the system considered in this paper, this would correspond to the 
finite temperature values of the Bertsch parameter and the contact density, 
which are calculable using quantum Monte-Carlo simulations \cite{Drut}. 

\section*{Acknowledgments}
This work was supported by RIKEN Foreign Postdoctoral
Researcher Program, the RIKEN iTHES Project 
and JSPS KAKENHI Grant Numbers 20192700, 25887020, 26887032 and \break 30130876. 

\appendix
\section{\label{ScattAmp} Numerical solution of $T^{\mathrm{reg}}_{\up}(k,0;k,0)$ in the unitary limit}
As discussed in Section \ref{OPEforgenerala}, we need to solve the integral equation given in Eq.(\ref{eq:scatt.amp.2}) 
numerically, which is then substituted in Eq.(\ref{eq:scatt.amp.4}) to obtain the desired  scattering 
amplitude $T^{\mathrm{reg}}_{\up}(k,0;k,0)$. 
The technical details necessary for this task will be outlined in this Appendix, in which 
we generalize the discussion given in 
Appendix B of \cite{Nishida}, where $k_0$ was set to $\ek$, while we here have to keep it as 
an independent variable and will specifically consider $k_0=\omega + i0^{+}$, with $\omega$ being 
real. 

Firstly, it is noticed that the dimensionless function 
\begin{equation}
s_{\up}(k;\ep,\p) = \frac{\k^2}{m} T_{\up}(k;\ep,\p + \tfrac{1}3 \k)
\label{eq:dimlessfuncs}
\end{equation}
satisfies a simpler integral equation, which is given as 
\begin{equation}
\begin{split}
s_{\up}(k;\ep,\p) =& -\frac{\k^2}{(\p + \tfrac{1}3 \k)^2} \\
&- \int\!\frac{d\q}{(2\pi)^3}
\frac{16\pi}{\sqrt{3\q^2+\tfrac{2}3 \k^2-4m k_0}} 
\frac{s_{\up}(k;\eq,\q)}{2\p^2 + 2\q^2 +2\p \cdot \q + \tfrac{1}3 \k^2 - 2m k_0} \\
\equiv &  - \I(\k;\p) - \int\!\frac{d\q}{(2\pi)^3} \J(k;\p,\q) s_{\up}(k;\eq,\q).
\end{split}
\label{eq:simp.integral.eq}
\end{equation}
The important point here is that the Kernel $\J(k;\p,\q)$ now depends only on 
the angle between $\q$ and $\p$, which will permit a partial wave expansion 
of the above integral equation. 

Next, we expand $s_{\up}(k;\ep,\p)$ into its partial waves, 
which depend on the angle $\theta$ between $\k$ and $\p$ as 
\begin{equation}
s_{\up}(k;\ep,\p) = \sum_{l=0}^{\infty} s_{\up}^{(l)} \Big(\tfrac{k_0}{\ek},\tfrac{\ep}{\ek} \Big) P_{l}(\cos \theta),   
\label{eq:partialwave.exp}
\end{equation}
where $P_l(x)$ are the Legendre polynomials. We have made use of the fact that 
$s_{\up}^{(l)}$ is a dimensionless function, which can hence only depend on the ratios 
$k_0/\ek$ and $\ep/\ek$. 
It can be shown that each term  
$s_{\up}^{(l)}(k_0/\ek,\ep/\ek)$ in the sum of Eq.(\ref{eq:partialwave.exp}) satisfies a 
closed integral equation,  
\begin{equation}
s_{\up}^{(l)}\Big( \tfrac{k_0}{\ek},\tfrac{\ep}{\ek} \Big) = - \I^{(l)}\Big( \tfrac{\ep}{\ek} \Big) - 
\int_0^{\infty} d \tfrac{|\q|}{|\k|}  \J^{(l)} \Big(\tfrac{k_0}{\ek},\tfrac{\ep}{\ek},\tfrac{\eq}{\ek} \Big) s_{\up}^{(l)}\Big( \tfrac{k_0}{\ek},\tfrac{\eq}{\ek}\Big).  
\label{eq:sl.integraleq}
\end{equation}
Here, the function $\I^{(l)}$ is defined as 
\begin{equation}
\I^{(l)}\Big( \tfrac{\ep}{\ek} \Big) \equiv \frac{2l+1}{2} \int_{-1}^{1} d \cos \theta  P_{l}(\cos \theta) \I(\k;\p), 
\label{eq:I.function.def}
\end{equation}
which can be rewritten with the help of the Gaussian hypergeometric function $_{2}F_1(a,b;c;y)$: 
\begin{equation}
\I^{(l)}(x) = \frac{l!}{(2l-1)!!} \frac{9}{1+9x} \Big(-\frac{6\sqrt{x}}{1+9x} \Big)^{l} \,
_{2}F_1 \Bigg[\frac{l+1}{2},\frac{l+2}{2};l+\frac{3}{2}; \frac{36 x}{(1+9x)^2} \Bigg]. 
\label{eq:I.function}
\end{equation} 
Furthermore, we have defined $\J^{(l)}$ as shown below:   
\begin{equation}
\J^{(l)}\Big(\tfrac{k_0}{\ek},\tfrac{\ep}{\ek},\tfrac{\eq}{\ek} \Big) \equiv \frac{|\k| |\q|^2}{4 \pi^2} \int_{-1}^{1} d \cos \theta  P_{l}(\cos \theta) \J(k;\p,\q).  
\label{eq:J.function.def}
\end{equation}
Using again the Gaussian hypergeometric function this gives,  
 \begin{equation}
 \begin{split}
\J^{(l)}(x,y,z) =&\: \frac{8}{\pi} \frac{l!}{(2l+1)!!} \frac{z}{\sqrt{3z + \tfrac{2}{3} - 2x}}
\frac{1}{2y + 2z+ \tfrac{1}{3} - x} \Big( \frac{2\sqrt{yz}}{2y+2z + \tfrac{1}{3}-x}\Big)^l \\
&\times _{2}F_1 \Bigg[\frac{l+1}{2},\frac{l+2}{2};l+\frac{3}{2};  \frac{4yz}{(2y+2z + \tfrac{1}{3}-x)^2} \Bigg]. 
\end{split}
\label{eq:J.function}
\end{equation}

As a next step, we need to solve Eq.(\ref{eq:sl.integraleq}) numerically for general values of $l$. In practice, 
we however will not deal with this equation directly, but first define 
\begin{equation}
\delta s_{\up}^{(l)}\Big( \tfrac{k_0}{\ek},\tfrac{\ep}{\ek} \Big) = 
s_{\up}^{(l)}\Big( \tfrac{k_0}{\ek},\tfrac{\ep}{\ek} \Big) + \I^{(l)}\Big( \tfrac{\ep}{\ek} \Big), 
\label{eq:define.delta.s}
\end{equation} 
which satisfies 
\begin{equation}
\begin{split}
\delta s_{\up}^{(l)}\Big( \tfrac{k_0}{\ek},\tfrac{\ep}{\ek} \Big) =& 
\int_0^{\infty} d \tfrac{|\q|}{|\k|}  \J^{(l)} \Big(\tfrac{k_0}{\ek},\tfrac{\ep}{\ek},\tfrac{\eq}{\ek} \Big) \I^{(l)}\Big( \tfrac{\eq}{\ek} \Big) \\
&- \int_0^{\infty} d \tfrac{|\q|}{|\k|}  \J^{(l)} \Big(\tfrac{k_0}{\ek},\tfrac{\ep}{\ek},\tfrac{\eq}{\ek} \Big) 
\delta s_{\up}^{(l)}\Big( \tfrac{k_0}{\ek},\tfrac{\eq}{\ek}\Big), 
\end{split}
\label{eq:delta.s.int.eq}
\end{equation} 
and then solve this equation for $\delta s_{\up}^{(l)}(k_0/\ek,\ep/\ek )$. 
This is done in order to avoid (or at least to weaken) the singularities that appear in $s_{\up}^{(l)}(k_0/\ek, \ep/\ek )$ 
for certain values of $k_0/\ek$ and $\ep/\ek$ and use instead the better behaved 
$\delta s_{\up}^{(l)}(k_0/\ek,\ep/\ek)$. 
Once this is done, the result is substituted into Eq.(\ref{eq:definition1}), which, by making use of the above definitions, 
can be rephrased as 
\begin{align}
\begin{split}
L\Big( \frac{k_0}{\ek}\Big) =&\: \ek \int\!\frac{d\q}{(2\pi)^3}
 \frac{8\pi}{\sqrt{3\q^2-2\q\cdot\k+\k^2-4m k_0}}
 \frac{T_\up(k;\eq,\q)+\frac{m}{\q^2}}{\q^2} \\
=&\: 
\frac{\sqrt{2}}{\pi} \sum_{l=0}^{\infty} \frac{1}{2l +1} 
\int_0^{\infty} d \tfrac{|\q|}{|\k|} 
\frac{\tfrac{\eq}{\ek}}{\sqrt{\tfrac{3}{2}\tfrac{\eq}{\ek} + \tfrac{1}{3} - \tfrac{k_0}{\ek}}} 
\I^{(l)}\Big( \tfrac{\eq}{\ek} \Big) \delta s_{\up}^{(l)}\Big( \tfrac{k_0}{\ek},\tfrac{\eq}{\ek}\Big).
\end{split}
\label{eq:substitute.delta.s}
\end{align}

After obtaining $\delta s_{\up}^{(l)}(k_0/\ek,\ep/\ek)$ for each value of $l$ 
individually, the corresponding contributions are added in Eq.(\ref{eq:substitute.delta.s}), which then gives 
the final form of $L(k_0/\ek)$. It suffices to evaluate the function $L(k_0/\ek)$ 
for one specific value of $\ek$, as its form for general $\ek$ can be obtained by a simple rescaling 
of its argument. 

\begin{figure}
\centering
\includegraphics[width=8.0cm]{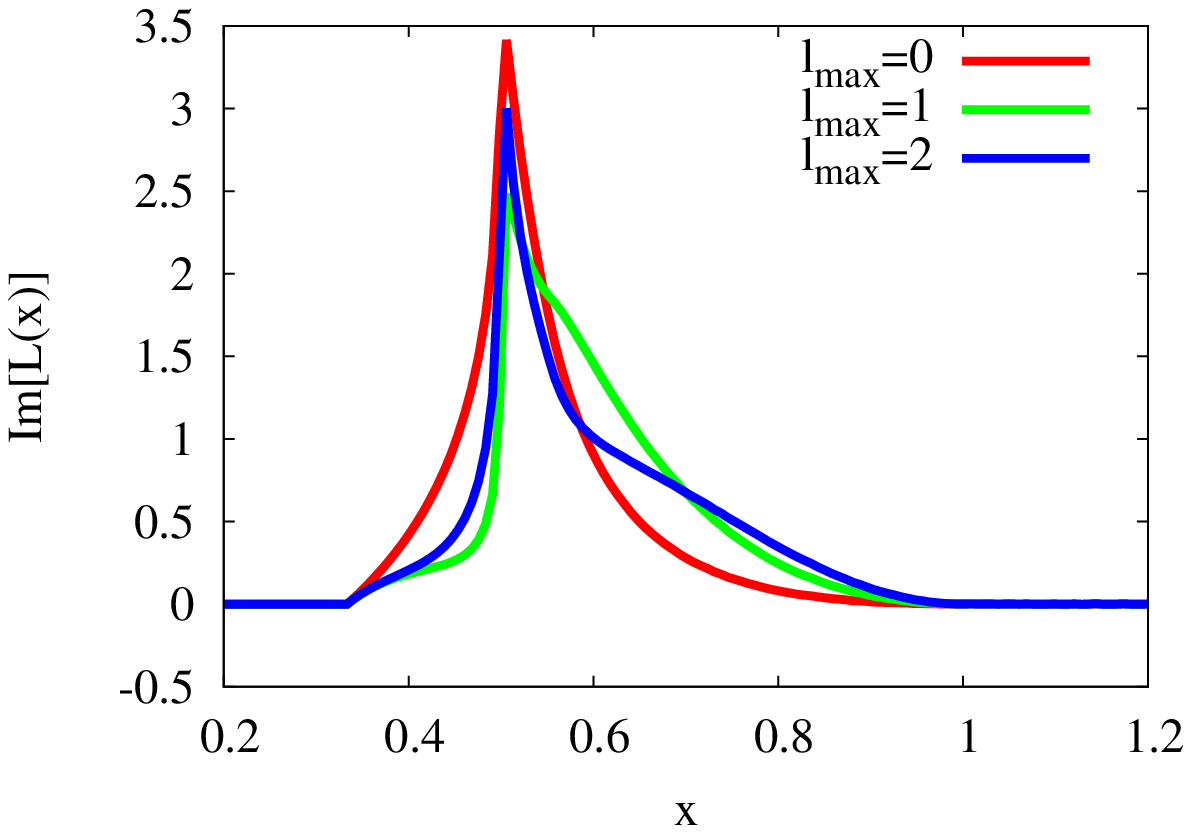}
\includegraphics[width=8.0cm]{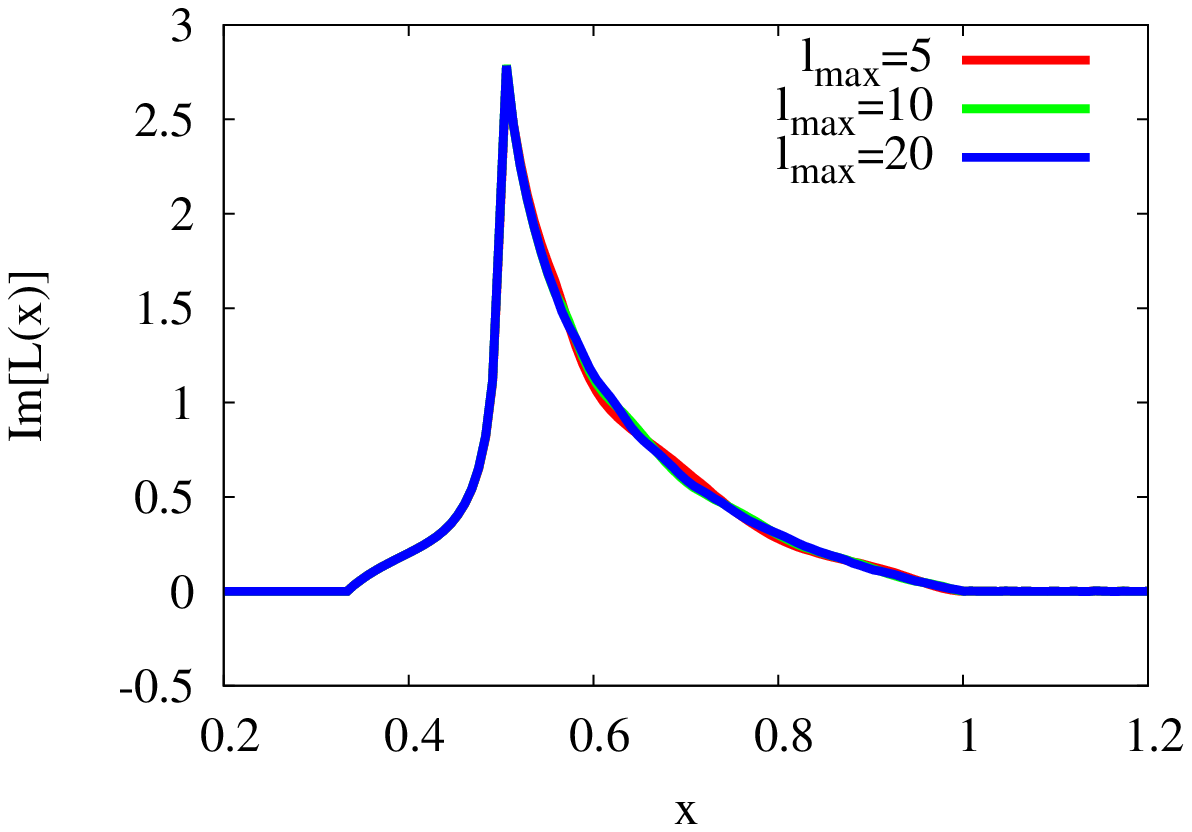}
\vspace{-0.5cm}
\caption{$\mathrm{Im}\bigl[L(x)\bigr]$, obtained by solving Eq.(\ref{eq:delta.s.int.eq}) and adding up the 
results in Eq.(\ref{eq:substitute.delta.s}) to various maximum values of $l$.}
\label{fig:inteq.res}
\end{figure}

In Fig. \ref{fig:inteq.res}, we show the final results for $\mathrm{Im} \bigl[ L(x)\bigr]$ 
for various maximum values of $l$ in the sum of Eq.(\ref{eq:substitute.delta.s}). 
(We show only the imaginary part because this is the only piece that will be needed for constructing the sum rules.) 
It is seen in these plots that $\mathrm{Im}\bigl[ L(x) \bigr]$ 
is essentially determined by the first 5 terms in the sum over $l$ 
and that the expansion converges quickly for values beyond $l \sim 10$. 
In this work, we will take into account the terms up to $l=20$. 

Furthermore, it can be seen in Fig. \ref{fig:inteq.res} that $\mathrm{Im} \bigl[ L(x) \bigr]$ takes non-zero values 
only in the interval $\tfrac{1}{3} < x < 1$, where it peaks sharply at around $x \sim 0.5$. 

\section{\label{OPE.detail} Derivation of the sum rules for a generic kernel} 
To derive the general form of the sum rule, given in Eq.(\ref{eq:sum.rule2}), we need to compute the 
right-hand side of Eq.(\ref{eq:sum.rule}), or, to be more precise, need to evaluate the contour integrals 
of $\K(k_0) \Sigma^{\mathrm{OPE}}(k_0,\k)$ along the sections of the contours
$C_1$ and $C_2$, which run above and below the real axis. 
The OPE expression for the self-energy is given in Eq.(\ref{eq:OPE1}) of Section \ref{OPE.unitary.limit} and 
is reproduced here once more: 
\begin{equation}
\begin{split}
&\Sigma^{\mathrm{OPE}}_{\up}(k_0, \bm{k}) = \\ 
&-\frac{8}{3\pi} \eF^{3/2} \frac{1}{\sqrt{\ek - 2 k_0}} 
+\frac{4}{3\pi^2} \zeta \eF^2 \Biggl[  \frac{1}{k_0 + \ek } - \frac{\sqrt{3}}{\pi} \frac{1}{2 k_0 - \ek } \\
&- \frac{1}{\pi} \frac{3k_0 - \ek}{\sqrt{\ek}(\ek - 2k_0)^{3/2}} 
\log \Bigg(\frac{1 + \sqrt{3} \sqrt{1 - 2k_0 /\ek}}{-1 + \sqrt{3} \sqrt{1 - 2k_0 / \ek}} \Bigg) 
+\frac{1}{\ek} L\bigl(\tfrac{k_0}{\ek}\bigr) \Biggr] \\
&- \frac{8}{5\pi} \xi  \eF^{5/2} \frac{\ek -  k_0}{(\ek - 2 k_0)^{5/2}} +O(k_0^{-2}).  
\end{split}
\label{eq:OPE1.appendix}
\end{equation}
The kernel $\K(k_0)$ is 
assumed to be analytic in the whole complex $k_0$ plane and to vanish 
faster that $1/\sqrt{k_0}$ at $k_0 \to \infty$ on the positive real axis. 

As some of the
derivations are somewhat involved, we will consider each term of the OPE individually. 
As in the main text, we here consider $k_0$ to be a complex variable, while $\omega$ is 
understood to be purely real.  

\subsection{Leading order (LO)}
Using
\begin{equation}
\mathrm{Im}\Bigl[ \frac{1}{\sqrt{\ek - 2\omega - i0^+}}\Bigr] = \theta(2\omega - \ek) \frac{1}{\sqrt{2\omega - \ek}}, 
\label{eq:imag1}
\end{equation}
we immediately get,  
\begin{equation}
\begin{split}
&\int^{\infty}_{-\infty}d\omega \mathcal{K}(\omega) \mathrm{Im} \Sigma^{\mathrm{LO}}_{\uparrow}(\omega + i0^{+}, \k) \\
&= -\frac{8}{3\pi} \eF^{3/2} \int^{\infty}_{\ek/2}d\omega  \frac{1}{\sqrt{2\omega - \ek}} \mathcal{K}(\omega) \\
&= \frac{8}{3\pi} \eF^{3/2} \int^{\infty}_{\ek/2} d\omega \sqrt{2\omega - \ek} \mathcal{K}'(\omega).
\end{split}
\label{eq:LO}
\end{equation} 
Note that for the above integrals to converge, 
the assumption of 
$\mathcal{K}(\omega)$ to approach 0 quicker than $1/\sqrt{\omega}$ at $\omega \to \infty$ is needed here. 

\subsection{Next-to-leading order (NLO)}
Being proportional to the contact density parameter $\zeta$, the NLO expression consists of two pole terms, one $\log$-term 
and one term containing the function $L(k_0/\ek)$. 
The pole terms are easily treated using 
\begin{equation}
\mathrm{Im}\Bigl[ \frac{1}{\omega - x + i0^+}\Bigr] = -\pi \delta(\omega - x),  
\label{eq:imag2}
\end{equation}
which gives 
\begin{equation}
\begin{split}
&\int^{\infty}_{-\infty}d\omega \mathcal{K}(\omega) \mathrm{Im} \Sigma^{\mathrm{NLO,\,\mathrm{pole}}}_{\uparrow}(\omega + i0^{+}, \k) \\
&= -\frac{4}{3\pi} \zeta \eF^2 \Bigl[\mathcal{K}(-\ek) - \frac{\sqrt{3}}{2\pi} \mathcal{K}\big(\tfrac{\ek}{2}\big) \Bigr].
\end{split}
\label{eq:NLO1}
\end{equation} 

Next, we consider the $\log$-term, which needs a somewhat more careful treatment of the contour integral, because 
simply taking its imaginary part leads to a divergence at $\omega = \ek / 2$. 
Before doing this, we note that 
\begin{equation}
\frac{1 + \sqrt{3} \sqrt{1 - 2\omega /\ek}}{-1 + \sqrt{3} \sqrt{1 - 2\omega / \ek}},  
\label{eq:expression1}
\end{equation}
which is the argument of the $\log$ in Eq.(\ref{eq:OPE1}),  is positive and 
real for $\omega < \ek / 3$ and negative and real for 
$\ek / 3 < \omega < \ek / 2$, where the $\log$ therefore has a cut. On the other hand, for 
$\omega > \ek / 2$, the above expression can be rewritten as follows:
\begin{equation}
\begin{split}
\frac{1 + \sqrt{3} \sqrt{1 - 2\omega /\ek - i0^+}}{-1 + \sqrt{3} \sqrt{1 - 2\omega / \ek - i0^+}} 
=&\: \frac{1 - i\sqrt{3}\sqrt{2\omega/\ek -1}}{-1 - i\sqrt{3}\sqrt{2\omega/\ek - 1}} \\
=&\: \frac{1}{3\omega - \ek}\Bigl(3\omega - 2\ek - i\sqrt{3\ek}\sqrt{2\omega - \ek}\Bigr) \\
=&\: e^{i\theta},
\end{split}
\end{equation}
where $\theta$ is given as
\begin{equation}
\theta = \tan^{-1}\Bigl( \frac{\sqrt{3 \ek}\sqrt{2\omega - \ek}}{3\omega - 2\ek} \Bigr).
\end{equation}
Therefore, the $\log$ of Eq.(\ref{eq:expression1}) is purely imaginary for $\omega > \ek / 2$. 
In this region, the root 
in front of the $\log$ in Eq.(\ref{eq:OPE1}) is also purely imaginary, which means that the term as a whole is real and 
that there is no cut for $\omega > \ek / 2$. 

\begin{figure}
\hspace*{1.0cm}
\includegraphics[width=10cm]{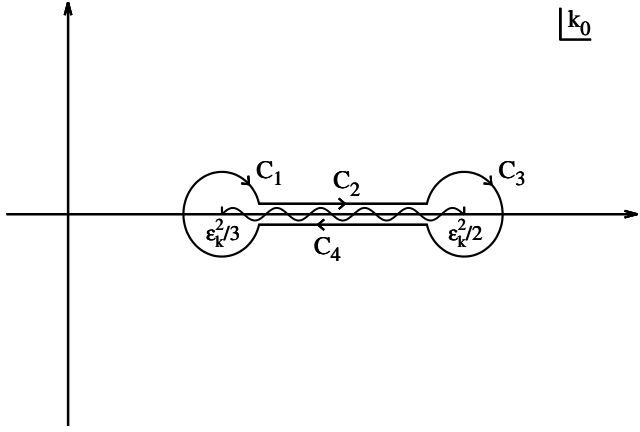} 
\vspace{-1.0cm}
\caption{The contour integral on the complex plane of the variable $k_0$ needed to calculate the NLO $\log$-term contribution to 
the sum rule.}
\label{fig:contour1}
\end{figure}

Hence, it is understood that we just have to evaluate the contour shown in Fig. \ref{fig:contour1}. 
The corresponding analytical formula is 
\begin{equation}
\begin{split}
&\frac{1}{2i} \oint_{C_1\mathrm{ - }C_4}d k_0 \mathcal{K}(k_0) \Sigma^{\mathrm{NLO,\,\mathrm{log}}}_{\uparrow}(k_0, \k) \\
&=-\frac{4}{3\pi^3} \zeta \frac{\eF^2}{\sqrt{\ek}} \frac{1}{2i} \oint_{C_1\mathrm{ - }C_4}d k_0 \mathcal{K}(k_0) \frac{3 k_0 - \ek}{(\ek - 2k_0)^{3/2}}
\log \Biggl( \frac{1 + \sqrt{3}\sqrt{1 - 2k_0/\ek}}{-1 + \sqrt{3}\sqrt{1 - 2k_0/\ek}} \Biggr),
\end{split}
\label{eq:NLO2}
\end{equation}
for which we below calculate the parts $C_1$ - $C_4$ separately. 

Firstly, it is seen that the integrand is not singular at $k_0 = \ek / 3$. Thus, the contour $C_1$ circling 
around this point vanishes as its radius approaches zero:
\begin{equation}
\begin{split}
&\frac{1}{2i} \oint_{C_1}d k_0 \mathcal{K}(k_0) \Sigma^{\mathrm{NLO,\,\mathrm{log}}}_{\uparrow}(k_0, \k) = 0. 
\end{split}
\label{eq:NLO3}
\end{equation}
Next, the contour segments $C_2$ and $C_4$ are considered. They have a finite value due to the cut of the $\log$, which 
can be evaluated as follows: 
\begin{equation}
\begin{split}
&\frac{1}{2i} \oint_{C_2 + C_4}d k_0 \mathcal{K}(k_0) \Sigma^{\mathrm{NLO,\,\mathrm{log}}}_{\uparrow}(k_0, \k) \\
=&-\frac{4}{3\pi^2} \zeta \frac{\eF^2}{\sqrt{\ek}} \int_{\ek/3}^{\ek/2 - \epsilon} d\omega \mathcal{K}(\omega) 
\frac{3\omega - y^2}{(y^2 - 2\omega)^{3/2}} \\
=& -\frac{4}{3\pi^2} \zeta \eF^2 \Biggl\{
 \frac{\sqrt{\ek}}{2\sqrt{2}}
\mathcal{K}\bigl(\tfrac{\ek}{2}\bigr) \frac{1}{\sqrt{\epsilon}} -  \sqrt{3} \mathcal{K}\bigl(\tfrac{\ek}{3}\bigr) \\
&-\frac{1}{\sqrt{\ek}}  \int_{\ek/3}^{\ek/2} d\omega \sqrt{\ek - 2\omega} \Bigl[ 6\mathcal{K}'(\omega) - (\ek -3\omega) 
\mathcal{K}''(\omega)\Bigr] \Biggr\}.
\end{split}
\label{eq:NLO4}
\end{equation}
Here, $\epsilon$ stands for the radius of the circle around $k_0=\ek / 2$ of the contour $C_3$. 
The last contribution comes from $C_3$, which, after a change of variables from $k_0$ to $\theta$ ($k_0 = \ek/2 + \epsilon e^{i\theta}$), 
is divided into two parts: 
\begin{equation}
\begin{split}
&\: \frac{1}{2i} \oint_{C_3}d\omega \mathcal{K}(k_0) \Sigma^{\mathrm{NLO,\,\mathrm{log}}}_{\uparrow}(k_0, \k) \\
=&\: \frac{\epsilon}{2} \int_{\pi}^{0} d\theta e^{i\theta} \mathcal{K}(\ek/2 + \epsilon e^{i\theta}) 
\Sigma^{\mathrm{NLO,\,\mathrm{log}}}_{\uparrow}(\ek/2 + \epsilon e^{i\theta}, \k) \\
&+ \frac{\epsilon}{2} \int_{0}^{-\pi} d\theta e^{i\theta} \mathcal{K}(\ek/2 + \epsilon e^{i\theta}) 
\Sigma^{\mathrm{NLO,\,\mathrm{log}}}_{\uparrow}(\ek/2 + \epsilon e^{i\theta}, \k).
\end{split}
\label{eq:NLO5}
\end{equation}
The first part is evaluated as
\begin{equation}
\begin{split}
&\frac{\epsilon}{2} \int_{\pi}^{0} d\theta e^{i\theta} \mathcal{K}(\ek/2 + \epsilon e^{i\theta}) 
\Sigma^{\mathrm{NLO,\,\mathrm{log}}}_{\uparrow}(\ek/2 + \epsilon e^{i\theta}, \k) \\
=&-\frac{2}{3\pi^3}\zeta \frac{\eF^2}{\sqrt{\ek}} \epsilon \int_{\pi}^0 d\theta e^{i\theta} 
\mathcal{K}(\ek/2 + \epsilon e^{i\theta}) \frac{3\bigl(\ek/2 + \epsilon e^{i\theta} \bigr) - \ek}{\bigl[ 2\epsilon e^{i(\theta-\pi)}\bigr]^{3/2}} \\
&\times \log \Biggl( \frac{\sqrt{\ek} + \sqrt{3}\sqrt{2\epsilon e^{i(\theta-\pi)}}}{-\sqrt{\ek} + \sqrt{3}\sqrt{ 2\epsilon e^{i(\theta-\pi)}}} \Biggr) \\
=&\: \frac{\sqrt{2}}{6\pi^2} \zeta \eF^2 \sqrt{\ek}  \mathcal{K}\bigl(\tfrac{\ek}{2}\bigr) (1 - i) \frac{1}{\sqrt{\epsilon}} 
- \frac{1}{\sqrt{3}\pi^2} \zeta \eF^2 \mathcal{K}\bigl(\tfrac{\ek}{2}\bigr), 
\end{split}
\label{eq:NLO6}
\end{equation}
while the second one gives 
\begin{equation}
\begin{split}
&\frac{\epsilon}{2} \int_{0}^{-\pi} d\theta e^{i\theta} \mathcal{K}(\ek/2 + \epsilon e^{i\theta}) 
\Sigma^{\mathrm{NLO,\,\mathrm{log}}}_{\uparrow}(\ek/2 + \epsilon e^{i\theta}, \k) \\
=&-\frac{2}{3 \pi^3}\zeta \frac{\eF^2}{\sqrt{\ek}} \epsilon \int_{\pi}^0 d\theta e^{i\theta}  
\mathcal{K}(\ek/2 + \epsilon e^{i\theta}) \frac{3\bigl(\ek/2 + \epsilon e^{i\theta} \bigr) - \ek}{\bigl[ 2\epsilon e^{i(\theta+\pi)}\bigr]^{3/2}} \\
&\times \log \Biggl( \frac{\sqrt{\ek} + \sqrt{3}\sqrt{ 2\epsilon e^{i(\theta+\pi)}}}{-\sqrt{\ek} + \sqrt{3}\sqrt{ 2\epsilon e^{i(\theta+\pi)}}} \Biggr) \\
=&\: \frac{\sqrt{2}}{6\pi^2} \zeta \eF^2 \sqrt{\ek} \mathcal{K}\bigl(\tfrac{\ek}{2}\bigr) (1 + i) \frac{1}{\sqrt{\epsilon}} 
- \frac{1}{\sqrt{3}\pi^2} \zeta \eF^2 \mathcal{K}\bigl(\tfrac{\ek}{2}\bigr). 
\end{split}
\label{eq:NLO7}
\end{equation}
Adding the two results from above, we finally get 
\begin{equation}
\frac{1}{2i} \oint_{C_3}d k_0 \mathcal{K}(k_0) \Sigma^{\mathrm{NLO,\,\mathrm{log}}}_{\uparrow}(k_0, \k) 
=\frac{\sqrt{2}}{3\pi^2} \zeta \eF^2 \sqrt{\ek} \mathcal{K}\bigl(\tfrac{\ek}{2}\bigr) \frac{1}{\sqrt{\epsilon}} 
- \frac{2}{\sqrt{3}\pi^2} \zeta \eF^2 \mathcal{K} \bigl(\tfrac{\ek}{2}\bigr). 
\label{eq:NLO8}
\end{equation}

Thus, assembling all the contributions, we can obtain the result for the whole contour of Fig.~\ref{fig:contour1}: 
\begin{equation}
\begin{split}
& \frac{1}{2i} \oint_{C_1\mathrm{ - }C_4}d k_0 \mathcal{K}(k_0) \Sigma^{\mathrm{NLO,\,\mathrm{log}}}_{\uparrow}(k_0, \k) \\
=&\:\frac{2}{\sqrt{3}\pi^2} \zeta \eF^2 \Bigl[ 2\mathcal{K}\bigl(\tfrac{\ek}{3}\bigr) - \mathcal{K}\bigl(\tfrac{\ek}{2}\bigr) \Bigr] \\
&+\frac{4}{3\pi^2} \zeta \frac{\eF^2}{\sqrt{\ek}} \int_{\ek/3}^{\ek/2} d\omega \sqrt{\ek - 2\omega} \Bigl[ 6\mathcal{K}'(\omega) - (\ek -3\omega) 
\mathcal{K}''(\omega)\Bigr]. 
\end{split}
\label{eq:NLO9}
\end{equation}
Note 
that all divergences have vanished in this final expression.  

The last term that has to be considered contains the function $L(k_0/\ek)$. 
As its imaginary part has no divergences, it is straightforward to 
evaluate the corresponding contribution, we just need to take the imaginary part of $L(k_0/\ek)$ 
(shown in Fig. \ref{fig:inteq.res}), multiply the kernel and numerically perform the integral over $\omega$: 
\begin{equation}
\begin{split}
\int^{\infty}_{-\infty}d\omega \mathcal{K}(\omega) \mathrm{Im} \Sigma^{\mathrm{NLO,}\,L(k_0/\ek)}_{\uparrow}(\omega + i0^{+}, \k) 
=&\: \frac{4}{3\pi^2} \zeta \frac{\eF^2}{\ek} \int^{\infty}_{-\infty}d\omega \mathcal{K}(\omega) \mathrm{Im} \Bigl[ L\bigl( \tfrac{\omega}{\ek} \bigr) \Bigr] \\
=&\: \frac{4}{3\pi^2} \zeta \frac{\eF^2}{\ek} \int^{\ek}_{\ek/3} d\omega \mathcal{K}(\omega) \mathrm{Im} \Bigl[ L\bigl( \tfrac{\omega}{\ek} \bigr) \Bigr]. 
\end{split}
\label{eq:NLO11}
\end{equation}
In the last line we made use of the fact that $\mathrm{Im} \bigl[ L(x) \bigr]$ only has non-zero values in the region of 
$\frac{1}{3} < x < 1$. 

Together with the pole- and $\log$-terms, we hence can collect the NLO results as follows: 
\begin{equation}
\begin{split}
&\frac{1}{2i} \oint d k_0 \mathcal{K}(k_0) \Sigma^{\mathrm{NLO}}_{\uparrow}(k_0, \k) \\
=&\: \frac{4}{3\pi} \zeta \eF^2 \Bigl[\tfrac{\sqrt{3}}{\pi}\mathcal{K}\bigl(\tfrac{\ek}{3}\bigr) - \mathcal{K}(-\ek) \Bigr] \\
&+\frac{4}{3\pi^2} \zeta \frac{\eF^2}{\sqrt{\ek}} \int_{\ek/3}^{\ek/2} d\omega \sqrt{\ek - 2\omega} \Bigl[ 6\mathcal{K}'(\omega) - (\ek -3\omega) 
\mathcal{K}''(\omega)\Bigr] \\
&+\frac{4}{3\pi^2} \zeta \frac{\eF^2}{\ek} \int^{\ek}_{\ek/3} d\omega \mathcal{K}(\omega) \mathrm{Im} \Bigl[ L\bigl( \tfrac{\omega}{\ek} \bigr) \Bigr].
\end{split}
\label{eq:NLO10}
\end{equation} 
Let us here briefly draw the attention of the reader to the fact that 
the term containing $\mathcal{K}(\ek/2)$, which appears 
in both the pole term result of Eq.(\ref{eq:NLO1}) and the expression of Eq.(\ref{eq:NLO9}), 
happens to exactly cancel and does therefore not show up in Eq.(\ref{eq:NLO10}). 

\subsection{Next-to-next-to-leading order (NNLO)}
\begin{figure}
\hspace*{1.0cm}
\includegraphics[width=10cm]{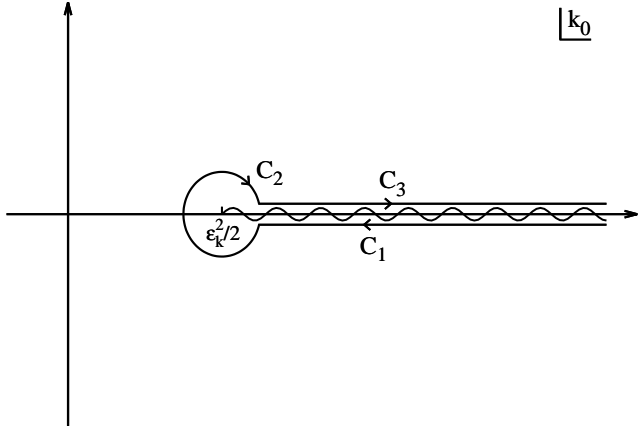}
\vspace{-1.0cm}
\caption{The contour integral on the complex plane of the variable $k_0$ needed to calculate the NNLO contribution to 
the sum rule.}
\label{fig:contour2}
\end{figure}

As in the last section, we here again have to compute a contour integral in the complex plane 
of $k_0$. This contour is shown in Fig. \ref{fig:contour2}. 
We hence have to calculate the following integral: 
\begin{equation}
\begin{split}
&\frac{1}{2i} \oint_{C_1\mathrm{ - }C_3} d k_0 \mathcal{K}(k_0) \Sigma^{\mathrm{NNLO}}_{\uparrow}(k_0, \k) \\
=&-\frac{\sqrt{2}}{5\pi} \xi \eF^{5/2} \frac{1}{2i} \oint_{C_1\mathrm{ - }C_3}d\omega \mathcal{K}(\omega) \frac{\ek - \omega}{(\ek/2 - \omega)^{5/2}}.
\end{split}
\label{eq:NNLO1}
\end{equation}

First, the contours of $C_1$ and $C_3$ are considered. Added together, they receive only contributions from the cut of 
the root-function. This then leads to 
\begin{equation}
\begin{split}
&\frac{1}{2i} \oint_{C_1 + C_3}d k_0 \mathcal{K}(k_0) \Sigma^{\mathrm{NNLO}}_{\uparrow}(k_0, \k) \\
=& -\frac{\sqrt{2}}{5\pi} \xi \eF^{5/2} \int_{\ek/2 + \epsilon}^{\infty} d\omega \mathcal{K}(\omega) \frac{\ek - \omega}{(\omega - \ek/2)^{5/2}} \\
=&-\frac{\sqrt{2}}{15\pi} \xi \eF^{5/2} \ek \mathcal{K}\bigl(\tfrac{\ek}{2}\bigr) \frac{1}{(\sqrt{\epsilon})^3} 
+ \frac{\sqrt{2}}{5\pi} \xi \eF^{5/2} \Bigl[2\mathcal{K}\bigl(\tfrac{\ek}{2}\bigr) - \ek \mathcal{K}'\bigl(\tfrac{\ek}{2}\bigr) \Bigr]  \frac{1}{\sqrt{\epsilon}} \\
&-\frac{8}{15\pi} \xi \eF^{5/2} \int_{\ek/2}^{\infty} d\omega \sqrt{2\omega - \ek} \Bigl[ 3\mathcal{K}''(\omega) + (\omega - \ek) 
\mathcal{K}'''(\omega)\Bigr], 
\end{split}
\label{eq:NNLO2}
\end{equation}
where, in similarity to the last subsection, $\epsilon$ stands for the radius of the circle around $\omega=\ek/2$ of the contour $C_2$. 

Next, the contribution of $C_2$ is calculated, leading to 
\begin{equation}
\begin{split}
&\frac{1}{2i} \oint_{C_2} d k_0 \mathcal{K}(k_0) \Sigma^{\mathrm{NNLO}}_{\uparrow}(k_0, \k) \\
=&-\frac{\sqrt{2}}{10\pi} \xi \eF^{5/2} \epsilon \int_{0}^{-2\pi} d\theta e^{i\theta} 
\mathcal{K}\bigl( \tfrac{\ek}{2} + \epsilon e^{i\theta}\bigr)  \frac{\ek/2 - \epsilon e^{i\theta}}{(\epsilon e^{i(\theta + \pi)})^{5/2}} \\
=&\:\frac{\sqrt{2}}{15\pi} \xi \eF^{5/2} \ek \mathcal{K}\bigl(\tfrac{\ek}{2}\bigr) \frac{1}{(\sqrt{\epsilon})^3} - 
\frac{\sqrt{2}}{5\pi} \xi \eF^{5/2} \Bigl[2\mathcal{K}\bigl(\tfrac{\ek}{2}\bigr) - \ek \mathcal{K}'\bigl(\tfrac{\ek}{2}\bigr) \Bigr]  \frac{1}{\sqrt{\epsilon}}.
\end{split}
\label{eq:NNLO3}
\end{equation}
Therefore, the final form of the NNLO contribution to the sum rule is found to be 
\begin{equation}
\begin{split}
&\frac{1}{2i} \oint_{C_1\mathrm{ - }C_3}d k_0 \mathcal{K}(k_0) \Sigma^{\mathrm{NNLO}}_{\uparrow}(k_0, \k) \\
=&-\frac{8}{15\pi} \xi \eF^{5/2}  \int_{\ek/2}^{\infty} d\omega \sqrt{2\omega - \ek} \Bigl[ 3\mathcal{K}''(\omega) + (\omega - \ek) 
\mathcal{K}'''(\omega)\Bigr], 
\end{split}
\label{eq:NNLO4}
\end{equation}
where, as before, only the finite term remains, while all other divergent contributions at $\epsilon \to 0$ cancel. 

\subsection{Summary}
Collecting all the terms of the last few subsections, 
we get the following form of the sum rule for a kernel $\mathcal{K}(\omega)$, which 
at $\omega \to \infty$ has to approach zero faster than $1/\sqrt{\omega}$: 
\begin{equation}
\begin{split}
&\int^{\infty}_{-\infty}d\omega \mathcal{K}(\omega) \mathrm{Im} \Sigma_{\uparrow}(\omega + i0^{+}, \k) \\
=&\: \frac{8}{3\pi} \eF^{3/2} \int^{\infty}_{\ek/2} d\omega \sqrt{2\omega - \ek} \mathcal{K}'(\omega) 
+\frac{4}{3\pi} \zeta \eF^2 \Bigl[ \frac{\sqrt{3}}{\pi}\mathcal{K}\bigl(\tfrac{\ek}{3}\bigr) - \mathcal{K}(-\ek) \Bigr] \\
&+\frac{4}{3\pi^2} \zeta \frac{\eF^2}{\sqrt{\ek}} \int_{\ek/3}^{\ek/2} d\omega \sqrt{\ek - 2\omega} 
\Bigl[ 6\mathcal{K}'(\omega) - (\ek -3\omega) \mathcal{K}''(\omega)\Bigr] \\
&+\frac{4}{3\pi^2} \zeta \frac{\eF^2}{\ek} \int^{\ek}_{\ek/3} d\omega \mathcal{K}(\omega) 
\mathrm{Im} \Bigl[ L\bigl( \tfrac{\omega}{\ek} \bigr) \Bigr] \\
&-\frac{8}{15\pi} \xi \eF^{5/2} \int_{\ek/2}^{\infty} d\omega \sqrt{2\omega - \ek} 
\Bigl[ 3\mathcal{K}''(\omega) + (\omega - \ek) \mathcal{K}'''(\omega)\Bigr]. 
\end{split}
\label{eq:sumrule2}
\end{equation}
This results corresponds to Eq.(\ref{eq:sum.rule2}) of the main text. 

\section{\label{finite.energy} Finite energy sum rules for the unitary Fermi gas} 
In this Appendix, we will demonstrate how to apply the finite energy (FE) sum rule approach \cite{Krasnikov} to the 
$k_0=\omega \gg \ek$ limit of Eq.(\ref{eq:OPE1}). This limit will considerably simplify the 
analysis of the sum rules and, after introducing certain assumptions on the functional  form 
of $\mathrm{Im} \Sigma_{\uparrow}(\omega,\bm{k})$, will even allow us to study them 
analytically. 
 
\subsection{Large frequency limit} 
To take the $k_0=\omega \gg \ek$ limit in Eq.(\ref{eq:OPE1}) and expanding the result in $\ek/\omega$ is 
mostly straightforward, the only exception being the $ L(k_0/\ek)$ term, which is 
related to the three-body scattering amplitude. 
For evaluating this term, we need to solve Eq.(\ref{eq:scatt.amp.2}) at $\bm{k} = 0$ (and $a^{-1}=0$). 
This integral equation can be rewritten as 
\begin{equation}
\begin{split}
& T_\up(k_0,0;\ep,\p) \\ 
=& -\frac{m}{\p^2} - \int\!\frac{d\q}{(2\pi)^3}
 \frac{4\pi}{\frac12\sqrt{3\q^2-4mk_0}}
 \frac{T_\up(k_0,0;\eq,\q)}{\p^2+\q^2+\p\cdot\q-mk_0} \\
 =& -\frac{m}{\p^2} - \frac2\pi\int_0^\infty d|\q|\,
 \frac{|\q|}{|\p|}\frac1{\sqrt{3\q^2-4m k_0}}
 \log\!\left(\frac{\p^2+\q^2+|\p||\q|-m k_0}{\p^2+\q^2-|\p||\q|-m k_0}\right)
 T_\up(k_0,0;\eq,\q).
\end{split}
\end{equation}
It can be understood from the last line above that $T_\up(k_0,0;\ep,\p)$ can only depend 
on $|\p|$. We hence define the dimensionless function: 
\begin{align}
 T_\up(k_0,0;\ep,\p) \equiv \frac1{k_0}t_\up(|\p|)
\end{align}
and rescale the momentum in units of $\sqrt{mk_0}$.  The integral
equation thus becomes
\begin{align}
 t_\up(|\p|) &= -\frac{1}{\p^2} - \frac2\pi\int_0^\infty\!d|\q|\,
 \frac{|\q|}{|\p|}\frac1{\sqrt{3\q^2-4}}
 \log\!\left(\frac{\p^2+\q^2+|\p||\q|-1}{\p^2+\q^2-|\p||\q|-1}\right)
 t_\up(|\q|),
\end{align}
which numerically determines $t_\up(|\q|)$. 

The term containing the function $L(x)$ of Eq.(\ref{eq:definition1}) can then be given by 
\begin{equation}
\begin{split}
\frac{1}{\ek} L\bigl(\tfrac{k_0}{\ek}\bigr) \xrightarrow{\bm{k=0}}&
 \int\!\frac{d\q}{(2\pi)^3}
 \frac{4\pi}{\frac12\sqrt{3\q^2-4m k_0}}
 \frac{T_\up(k_0,0;\eq,\q)+\frac{m}{\q^2}}{\q^2} \\
 =&\: \frac4\pi\int_0^\infty\!d|\q|\,\frac1{\sqrt{3\q^2-4m k_0}}
 \left[T_\up(k_0,0;\eq,\q)+\frac{m}{\q^2}\right] \\
 =&\: \frac1{k_0} \frac4\pi\int_0^\infty\!d|\q|\,\frac1{\sqrt{3\q^2-4}}
 \left[t_\up(|\q|)+\frac{1}{\q^2}\right]. 
\end{split}
\end{equation}
By using the numerically obtained $t_\up(|\q|)$, we find
\begin{align}
\frac{1}{\ek} L\bigl(\tfrac{k_0}{\ek}\bigr) \xrightarrow{\bm{k}=0} -  \frac{0.396797}{k_0}. 
\end{align}

Together with the other terms, we hence reach the desired limit: 
\begin{equation}
\begin{split}
\Sigma^{\mathrm{OPE}}_{\up}(k_0 = \omega, \bm{k}) \xrightarrow{\omega \gg \ek}
&-\frac{8}{3\pi} \eF^{3/2} \Bigl[ \frac{1}{\sqrt{-2 \omega}} -  \frac{\ek}{2} \frac{1}{(\sqrt{-2 \omega})^3} \Bigr] \\
&+\frac{4}{3\pi^2} \zeta \eF^2 \frac{1}{\omega} \Bigl(1 - \frac{\sqrt{3}}{\pi}- 0.396797 \Bigr) \\
&- \frac{4}{5\pi} \xi  \eF^{5/2} \frac{1}{(\sqrt{-2 \omega})^3} +O(\omega^{-2}) \\
=& -\frac{4\sqrt{2}}{3\pi} \eF^{3/2} \frac{1}{\sqrt{-\omega}} + \frac{0.20750}{3\pi^2} \zeta \eF^2 \frac{1}{\omega} \\
&- \frac{\sqrt{2}}{5\pi} \eF^{5/2} \Bigl( \xi - \frac{5}{3} \frac{\ek}{\eF} \Bigr) \frac{1}{(\sqrt{-\omega})^3} + O(\omega^{-2}). 
\end{split}
\label{eq:OPE.limit}
\end{equation}

\subsection{Ansatz of self-energy spectral function}
For streamlining the notation, we first rewrite the result of Eq.(\ref{eq:OPE.limit}) as follows,  
\beq
\label{eq:OPE.limit2}
\Sigma_{\uparrow}(\omega,\bm{k}) 
= \frac{\pi}{2} C_1 \frac{1}{\sqrt{-\omega}}
- C_2 \frac{1}{\omega}
+ \frac{\pi}{2}C_3 \frac{1}{(\sqrt{-\omega})^3},
\eeq 
where we have defined 
\beq
\label{eq:C}
C_1 = - \frac{8\sqrt{2}}{3\pi^2} \eF^{3/2}, \qquad 
C_2= - \frac{0.207498}{3\pi^2}\zeta \eF^2, \qquad 
C_3 = - \frac{2\sqrt{2}}{15\pi^2}\left(3\xi - 5 \frac{\ek}{\eF}\right) \eF^{5/2}. 
\eeq 
As Eq.(\ref{eq:OPE.limit2}) is valid at large $\omega$, we can immediately read off 
the asymptotic behavior of  
$\mathrm{Im} \Sigma_{\uparrow}(\omega+i0^+,\bm{k})$ in this region as 
\beq
\mathrm{Im} \Sigma_{\uparrow}(\omega+i0^+,\bm{k}) 
\sim \frac{\pi}{2} C_1 \frac{1}{\sqrt{\omega}}
- \frac{\pi}{2} C_3 \frac{1}{(\sqrt{\omega})^3}.
\eeq
Here we used
\begin{align}
{\rm Im}\left[\frac{1}{\sqrt{-\omega-i0^+}} \right]&=\theta(\omega)\frac{1}{\sqrt{\omega}},
\nonumber \\
{\rm Im}\left[\frac{1}{\omega + i0^+} \right]& = - \pi \delta(\omega),
\\
{\rm Im}\left[\frac{1}{(\sqrt{-\omega-i0^+})^3} \right]&
=-\theta(\omega)\frac{1}{(\sqrt{\omega})^3} \nonumber.
\end{align}
One can take the simplest ansatz for 
$\mathrm{Im} \Sigma_{\uparrow}(\omega+i0^+,\bm{k})$ satisfying the above behavior as
\beq
\label{eq:continuum}
\mathrm{Im} \Sigma_{\uparrow}(\omega,\bm{k}) = 
\theta(\omega - s_{\rm thr})\left[\frac{\pi}{2} C_1 \frac{1}{\sqrt{\omega}}
- \frac{\pi}{2} C_3 \frac{1}{(\sqrt{\omega})^3} \right],
\eeq
with some parameter $s_{\rm thr}$.
(We here assume $s_{\rm thr}>0$.)

However, it turns out that the finite energy sum rules for this
ansatz does not have a physical solution for 
$\ek/\eF < 3\xi/5$. 
Also we already know from the BCS theory in the mean-field 
approximation (MFA) that $\mathrm{Im} \Sigma_{\uparrow}(\omega,\bm{k})$
has a peak at negative $\omega$, which is absent in Eq.(\ref{eq:continuum}). 
We are thus tempted to take the modified ansatz, given by a naive
summation of the continuum (\ref{eq:continuum}) and the peak
in the MFA, 
\beq
\label{eq:ansatz.app}
\mathrm{Im} \Sigma_{\uparrow}(\omega,\bm{k}) =  - \pi C_4 \delta(\omega + \ek - 2\xi)
+ \theta(\omega - s_{\rm thr})\left[\frac{\pi}{2} C_1 \frac{1}{\sqrt{\omega}}
- \frac{\pi}{2} C_3 \frac{1}{(\sqrt{\omega})^3} \right],
\eeq
where $C_4= \Delta^2$ is the result in the MFA, which is also 
expressed using the contact density ${\cal C}$ as
\beq
C_4 = \frac{{\cal C}}{m^2} = \frac{4 \zeta}{3 \pi^2}
\eeq 
in our units. 

\subsection{Derivation of finite energy sum rule}
From the imaginary part of the self-energy $\mathrm{Im} \Sigma_{\uparrow}(\omega,\bm{k})$, 
we can obtain its real part through the Kramers-Kr$\mathrm{\ddot{o}}$nig relation  
\beq
\label{eq:dispersion.app}
{\rm Re} \Sigma_{\uparrow}(\omega, \bm{k}) = -\frac{1}{\pi} {\rm P}\int_{-\infty}^{\infty}
d \omega' \frac{\mathrm{Im} \Sigma_{\uparrow}(\omega'+i0^+,\bm{k})}{\omega-\omega'}. 
\eeq
Using the ansatz of Eq.(\ref{eq:ansatz.app}), the integral in the right-hand side above 
reduces to
\begin{align}
{\rm Re}  \Sigma_{\uparrow}(\omega, \bm{k}) 
&= \frac{C_4}{\omega + \ek - 2\xi} + 
{\rm P}\int_{s_{\rm thr}}^{\infty} \frac{d \omega'}{\omega-\omega'} 
\left[-\frac{C_1}{2}\frac{1}{\sqrt{\omega'}} + 
\frac{C_3}{2}\frac{1}{(\sqrt{\omega'})^3}  \right] 
\nonumber \\
&=  \frac{C_4}{\omega + \ek - 2\xi} +
{\rm P}\int_{\sqrt{s_{\rm thr}}}^{\infty} \frac{d t}{t^2 - \omega} 
\left({C_1} - \frac{C_3}{t^2}  \right)
\nonumber \\
& = \frac{C_4}{\omega + \ek - 2\xi} +
\frac{C_3}{\sqrt{s_{\rm thr}}} \frac{1}{\omega}
+ \left( C_1 - \frac{C_3}{\omega} \right)
{\rm P}\int_{\sqrt{s_{\rm thr}}}^{\infty} \frac{d t}{t^2 - \omega} 
\end{align}
where, in the second line, we set $t = \sqrt{\omega'}$. 
The integral can be performed for $\omega>0$ and $\omega<0$, 
respectively, as
\begin{align}
{\rm Re}  \Sigma_{\uparrow}(\omega, \bm{k}) =
\begin{cases} 
\displaystyle{ \frac{C_4}{\omega + \ek - 2\xi} +
\frac{C_3}{\sqrt{s_{\rm thr}}} \frac{1}{\omega}
- \left( C_1 - \frac{C_3}{\omega} \right)
\frac{1}{2\sqrt{\omega}} \log \left|\frac{\sqrt{\omega}-\sqrt{s_{\rm thr}}}
{\sqrt{\omega} + \sqrt{s_{\rm thr}}} \right|}
\quad  & (\omega > 0)
\\
\displaystyle{\frac{C_4}{\omega + \ek - 2\xi} +
\frac{C_3}{\sqrt{s_{\rm thr}}} \frac{1}{\omega}
+ \left( C_1 - \frac{C_3}{\omega} \right) \frac{1}{\sqrt{-\omega}}
\left(\frac{\pi}{2} - \tan^{-1} \sqrt{\frac{s_{\rm thr}}{-\omega}} \right)}
\quad  & (\omega < 0).
\end{cases}
\label{eq:Re}
\end{align}

For sufficiently large $\omega \gg s_{\rm thr}$, using
\beq
\log \left|\frac{1-x}{1+x} \right| = -2x\left(1 + \frac{x^2}{3} + \frac{x^4}{5} + \cdots \right)
\eeq
with $x=\sqrt{s_{\rm thr}/\omega} \ll 1$,
the right-hand side of Eq.(\ref{eq:Re}) can be expanded as
\beq
\frac{C_4}{\omega}\left(1 - \frac{\ek - 2\xi}{\omega} + \cdots \right) +
\frac{C_3}{\sqrt{s_{\rm thr}}} \frac{1}{\omega}
+ \left( C_1 - \frac{C_3}{\omega} \right)
\left[\frac{\sqrt{s_{\rm thr}}}{\omega} + \frac{(\sqrt{s_{\rm thr}})^3}{3\omega^2} 
+ \frac{(\sqrt{s_{\rm thr}})^5}{5\omega^3} \cdots \right]. 
\eeq
By comparing the coefficient of $1/\omega$ with that
in Eq.(\ref{eq:OPE.limit2}), we arrive at a constraint for $\sqrt{s_{\rm thr}}>0$:
\beq
\label{eq:FESR}
C_1 (\sqrt{s_{\rm thr}})^2 + C \sqrt{s_{\rm thr}} + C_3 = 0,
\eeq
where
\beq
C \equiv C_2 + C_4 = \frac{3.792502}{3 \pi^2} \zeta >0.
\eeq

\subsection{Solution of finite energy sum rule}
For Eq.(\ref{eq:FESR}) to have a real solution for $\sqrt{s_{\rm thr}}$, 
the following condition is necessary:
\beq
\label{eq:D}
D \equiv C^2 - 4C_1 C_3 > 0.
\eeq
For $\xi=0.372$ and $\zeta=3.40$,
one can check that this condition is satisfied for any $\ek$.
Then Eq.(\ref{eq:FESR}) can be solved as 
\begin{align}
\label{eq:s}
\sqrt{s_{\rm thr}}= 
\begin{cases}
\displaystyle{ \frac{- C \pm \sqrt{C^2 - 4C_1 C_3}}{2C_1} }
\qquad & (\frac{C^2}{4C_1} < C_3 <0)
\\
\displaystyle{ \frac{- C - \sqrt{C^2 - 4C_1 C_3}}{2C_1} }
\qquad & (C_3 > 0)
\end{cases}
\end{align}
where the signs are chosen such that $\sqrt{s_{\rm thr}}$ is positive.
For the smoothness of the solution of $s_{\rm thr}$ around $C_3=0$,
we assume to take
\beq
\sqrt{s_{\rm thr}} = \frac{- C - \sqrt{C^2 - 4C_1 C_3}}{2C_1} 
\eeq
for any $\ek$. 

In summary, we find
\begin{align}
\label{eq:Sigma}
& \Sigma_{\uparrow} (\omega+i0^{+},\bm{k}) \nonumber \\
=& 
\begin{cases}
\displaystyle{\frac{C_4}{\omega + \ek - 2\xi} +
\frac{C_3}{\sqrt{s_{\rm thr}}} \frac{1}{\omega} 
- \left( C_1 - \frac{C_3}{\omega} \right) 
\frac{1}{2\sqrt{\omega}} 
\left[ \log \left|\frac{\sqrt{\omega}-\sqrt{s_{\rm thr}}}
{\sqrt{\omega} + \sqrt{s_{\rm thr}}} \right|
- i \pi \theta(\omega - s_{\rm thr})\right]} 
\\
\displaystyle{\frac{C_4}{\omega + \ek - 2\xi} +
\frac{C_3}{\sqrt{s_{\rm thr}}} \frac{1}{\omega}
+ \left( C_1 - \frac{C_3}{\omega} \right) \frac{1}{\sqrt{-\omega}}
\left(\frac{\pi}{2} - \tan^{-1} \sqrt{\frac{s_{\rm thr}}{-\omega}} \right)}
\end{cases}
\end{align}
for $\omega>0$ and $\omega<0$, respectively.

\subsection{Single-particle spectral function}
Now we compute the single-particle spectral function defined by
\beq
{\cal A}_{\uparrow}(\omega, \bm{k})
= -\frac{1}{\pi} {\rm Im} \left[\frac{1}{\omega + i0^+ - \ek
- \Sigma_{\uparrow} (\omega+i0^{+},\bm{k})} \right].
\eeq
From the expression of $\Sigma_{\uparrow} (\omega+i0^{+},\bm{k})$ in 
Eq.(\ref{eq:Sigma}), ${\cal A}_{\uparrow}(\omega, \bm{k})$ reads 
\begin{align}
\label{eq:A}
{\cal A}_{\uparrow}(\omega, \bm{k}) =
\begin{cases}
\displaystyle{\sum_{n} F_n \delta \left(\omega- \omega_n \right)}
\qquad &(\omega < s_{\rm thr})
\\
\displaystyle{ - \frac{1}{\pi} \frac{{\rm Im}\Sigma_{\uparrow} (\omega+i0^{+},\bm{k}) }
{\left[\omega - \ek - {\rm Re} \Sigma_{\uparrow} (\omega+i0^{+},\bm{k})  \right]^2
+ \left[{\rm Im} \Sigma_{\uparrow} (\omega+i0^{+},\bm{k})  \right]^2}}
\qquad &(\omega > s_{\rm thr}).
\end{cases}
\end{align}
\begin{figure}
\begin{center}
\includegraphics[width=7.5cm]{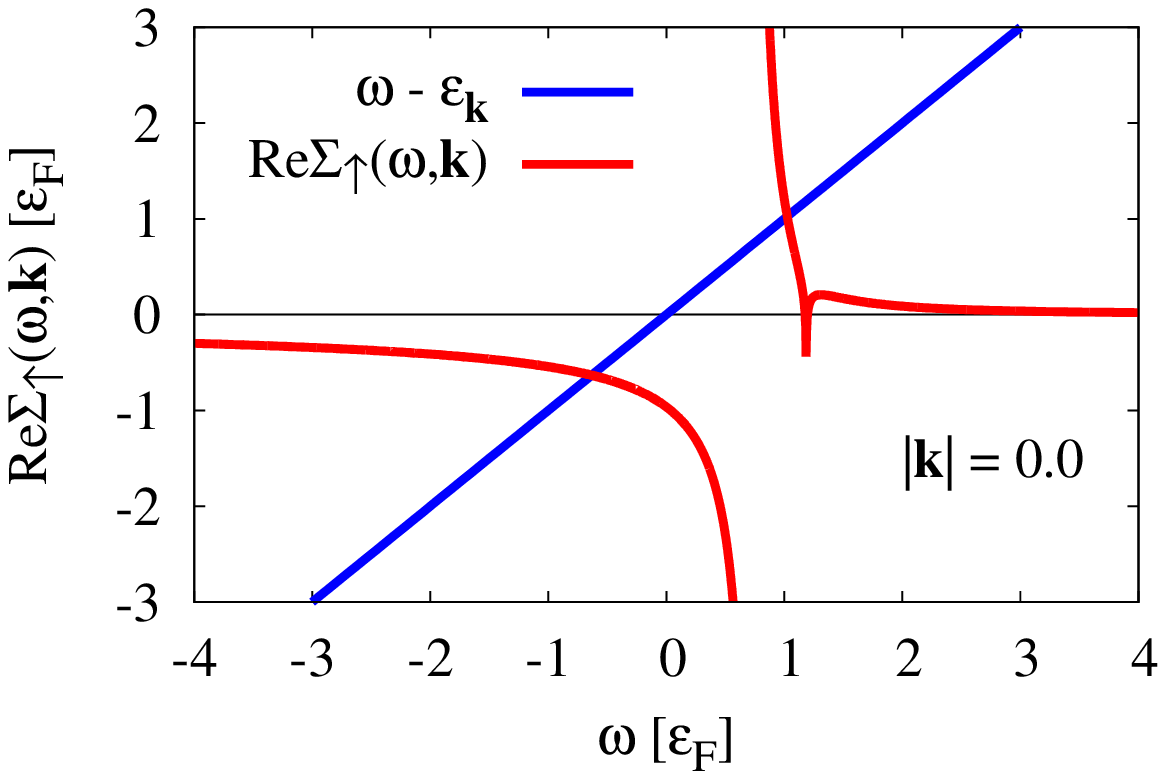} 
\includegraphics[width=7.5cm]{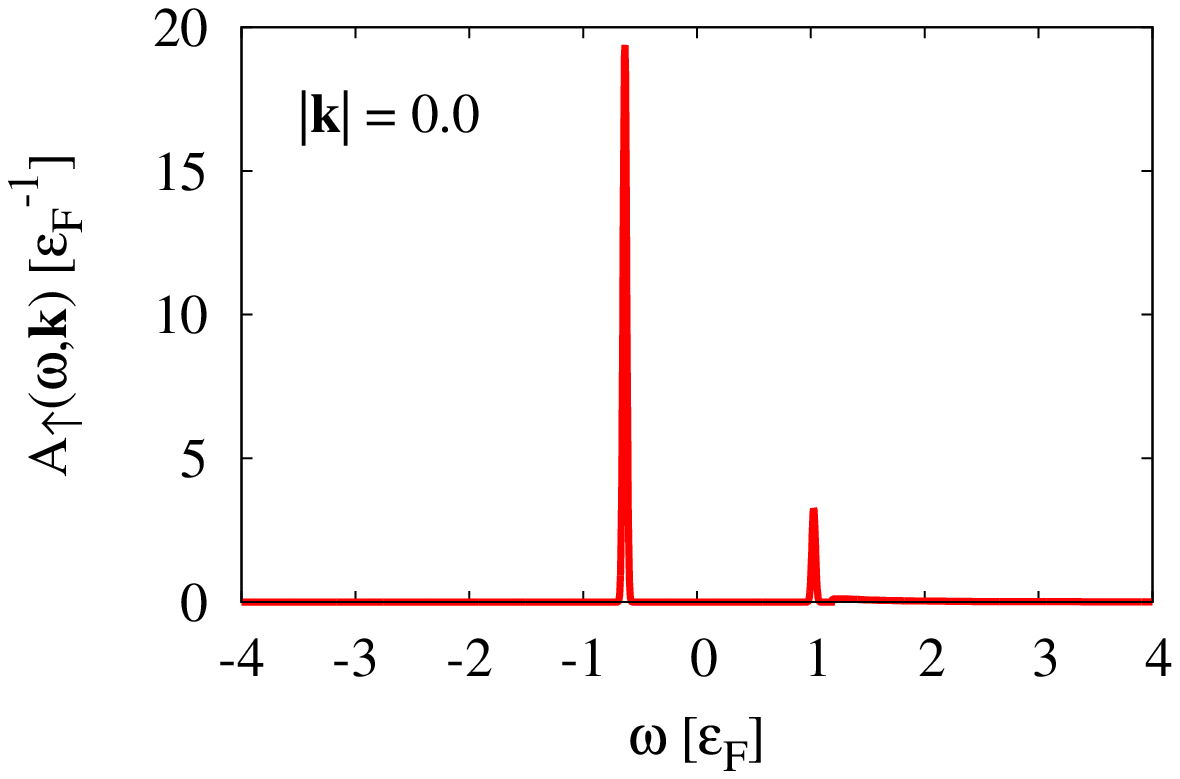}
\includegraphics[width=7.5cm]{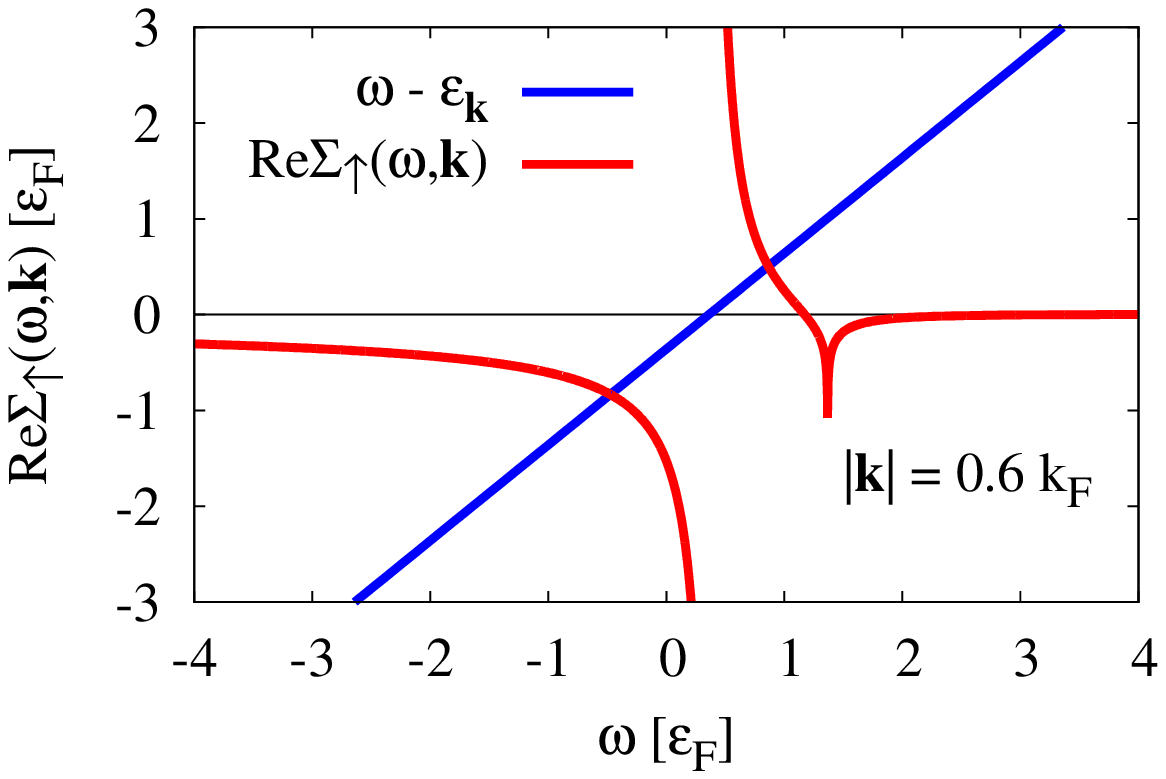} 
\includegraphics[width=7.5cm]{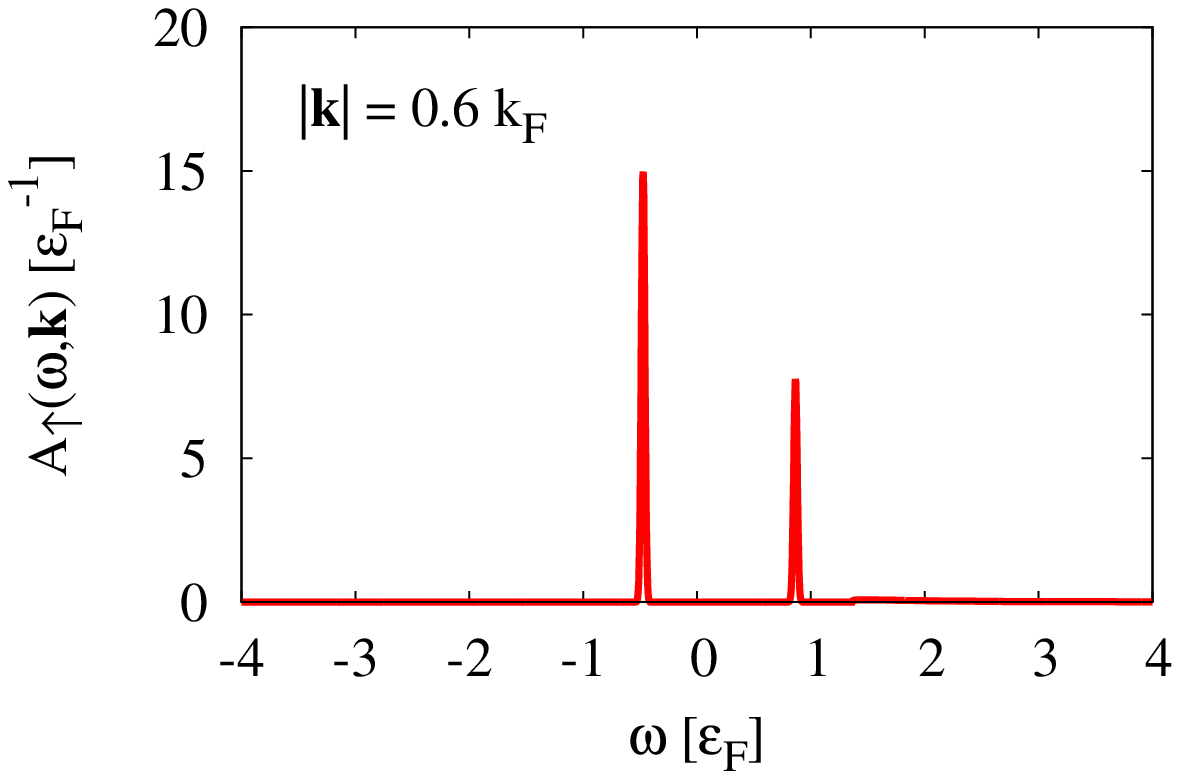}
\includegraphics[width=7.5cm]{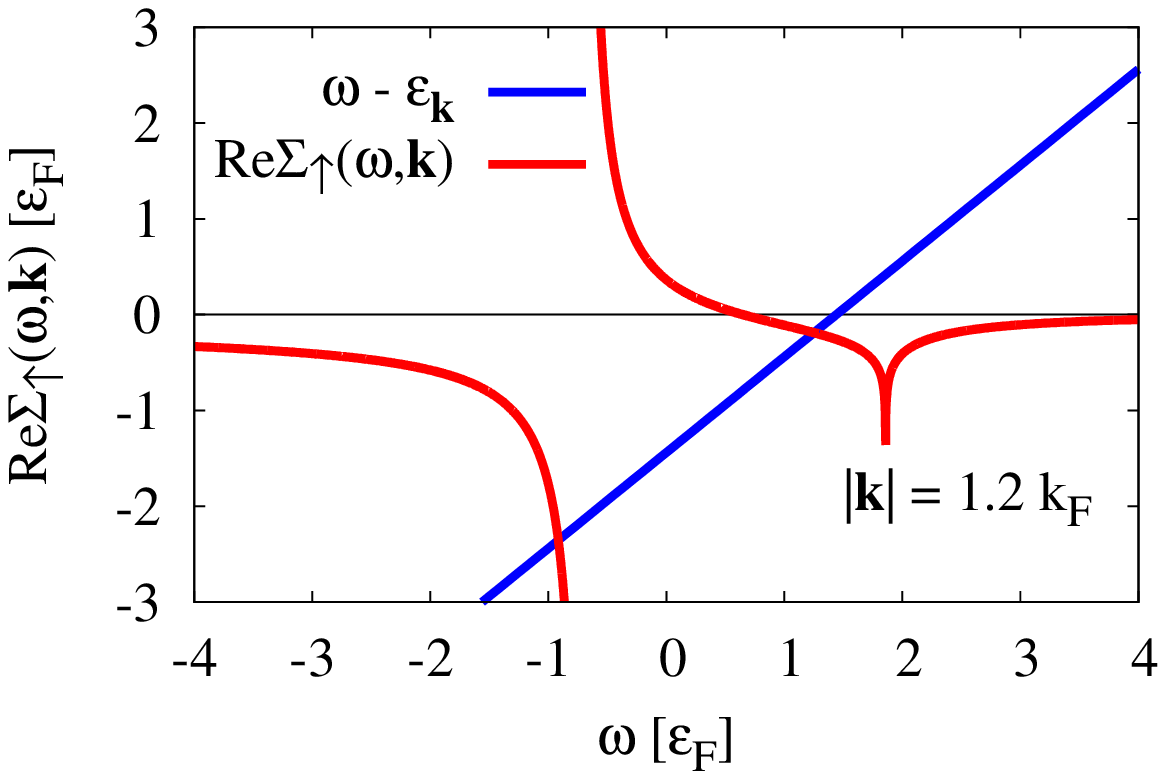} 
\includegraphics[width=7.5cm]{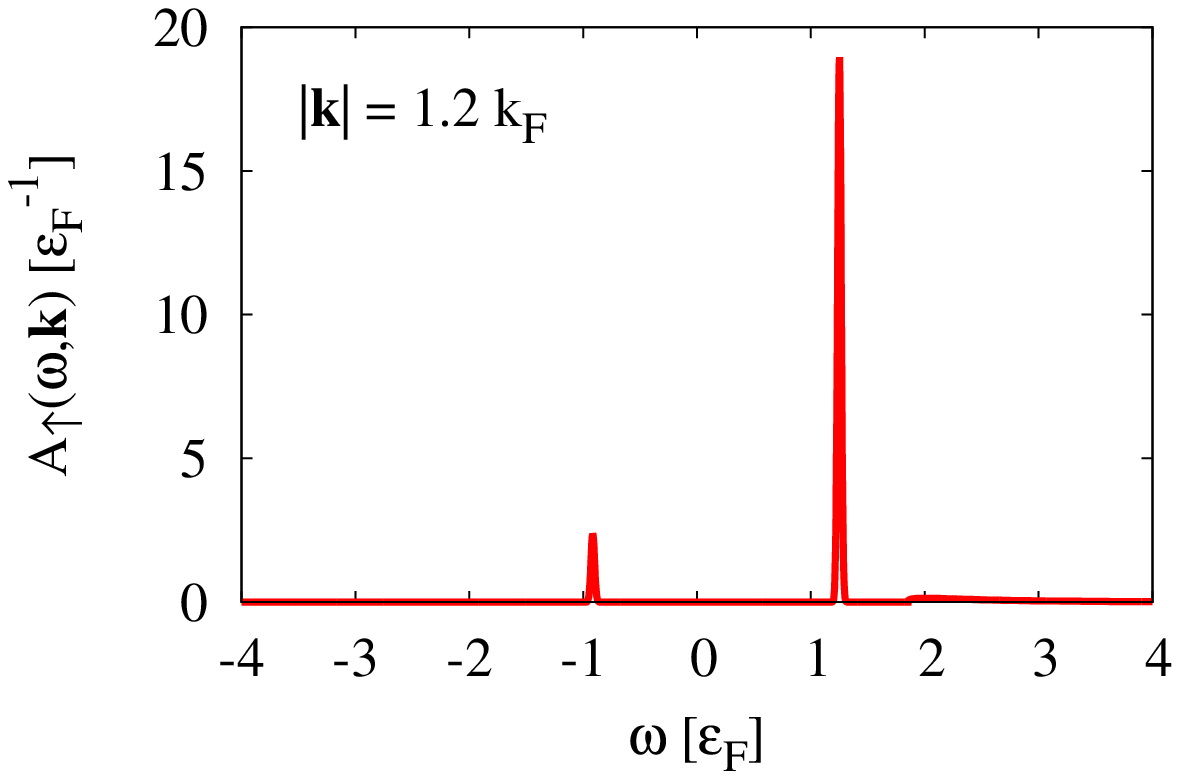}
\vspace{-1.2cm}
\end{center}
\caption{Left column: Real parts of the self-energy $\Sigma_{\uparrow}(\omega, \bm{k})$ [see Eq.(\ref{eq:Sigma})]. 
The intersections with 
the red line (${\rm Re} \Sigma=\omega-\ek$) represent the peak
positions of single-particle spectral densities [see Eq.(\ref{eq:peak})]. 
Right column: Single-particle spectral densities ${\cal A}_{\uparrow}(\omega, \bm{k})$ 
[see Eq.(\ref{eq:A})]. For better visibility, the delta functions $\delta(\omega-\omega_n)$ 
are approximated by $\sqrt{t/\pi}e^{-t(\omega-\omega_n)^2}$ 
with $t=2000$. 
Each row from top to bottom corresponds to $|\bm{k}|/k_F = 0.0$, $0.6$ and $1.2$.}
\label{fig:self}
\end{figure}
Here the pole(s) $\omega=\omega_n(y)$ ($n=1,2,\cdots$) are 
the solution(s) of
\beq
\label{eq:peak}
\omega - \ek - \Sigma_{\uparrow} (\omega+i0^{+},\bm{k}) =0,
\eeq
and the residue(s) $F_n(y)$ are given by 
\beq
F_n^{-1} = 1 - \left. \frac{\d \Sigma_{\uparrow} (\omega+i0^{+},\bm{k}) }
{\d \omega}\right|_{\omega=\omega_n}.
\eeq

Figure \ref{fig:self} shows the plots of 
${\rm Re} \Sigma_{\uparrow} (\omega+i0^{+},\bm{k}) $ and 
${\cal A}_{\uparrow}(\omega, \bm{k})$, respectively, 
for $\xi=0.372$ and $\zeta=3.40$. 
\begin{figure}
\begin{center}
\includegraphics[width=10.5cm]{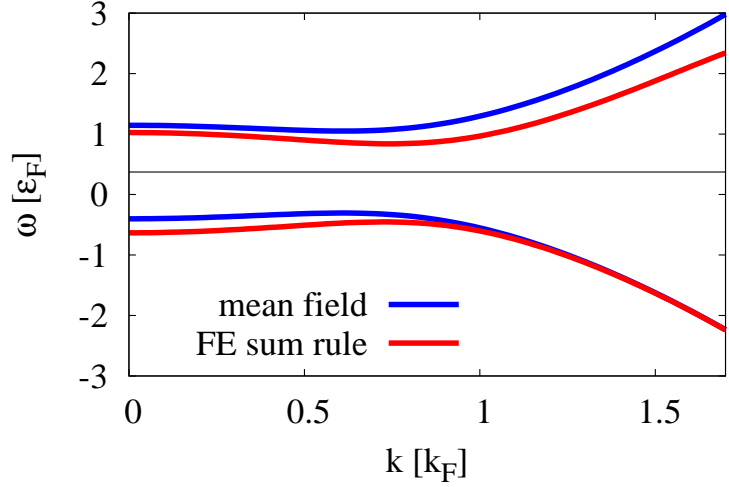}
\vspace{-1.3cm}
\end{center}
\caption{Peak positions of single-particle spectral densities
in the $(|\bm{k}|, \omega)$ plane (red lines). For comparison, mean-field 
dispersion relations, $\omega= \xi \pm\sqrt{(\ek-\xi)^2+C_4}$, 
are shown by the blue lines. The thin black line shows the position of $\omega=\xi$.}
\label{fig:zeros}
\end{figure}
The peak positions of the single-particle spectral densities
as functions of $|\bm{k}|$ 
are shown in Fig. \ref{fig:zeros}, where, for comparison, we 
also show the mean-field dispersion relations, $\omega= \xi \pm\sqrt{(\ek-\xi)^2+C_4}$. 
By investigating the point, at which the particle- and hole-branches approach each other most closely, 
we obtain a pairing gap value of $0.65\,\epsilon_F$, which is not much different from 
the mean-field result, $\sqrt{C_4} \simeq 0.68\,\epsilon_F$. 

\section{\label{MEM} The maximum entropy method} 
Let us here briefly recapitulate the basic steps of MEM and especially explain the differences of our analysis to 
the application of MEM to statistical Monte-Carlo data. For more details, 
consult for instance \cite{Gubler,Gubler2,Jarrel,Asakawa}. 

The problem to be solved with the help of MEM is given in 
Eqs.(\ref{eq:sum.rule3.1}) and (\ref{eq:sum.rule3.2}). 
As, however, the OPE on the right-hand side of these equations is only known  
with limited accuracy and is moreover only valid in a finite range of the Borel mass $M$, the problem 
of obtaining $\mathrm{Im} \Sigma_{\up}(\omega, \bm{k})$ from the OPE is ill-posed and cannot be solved analytically. 

MEM now uses Bayes' theorem, by which additional information 
about $\mathrm{Im} \Sigma_{\up}(\omega, \bm{k})$ such as positivity and its asymptotic behavior at 
large energies can be incorporated into 
the analysis and by which one then can extract the most probable from 
of $\mathrm{Im} \Sigma_{\up}(\omega, \bm{k})$. Bayes' theorem can be expressed as 
\begin{equation}
P[\mathrm{Im} \Sigma|D H] = \frac{P[D|\mathrm{Im} \Sigma H] P[\mathrm{Im} \Sigma|H]}{P[D|H]}, 
\label{eq:bayes}
\end{equation} 
where $H$ denotes prior knowledge of $\mathrm{Im} \Sigma_{\up}(\omega, \bm{k})$ and 
$P[\mathrm{Im} \Sigma|D H]$ represents 
the conditional probability of $\mathrm{Im} \Sigma_{\up}(\omega, \bm{k})$ for given $D^{\mathrm{OPE}}_{\up}(M,\bm{k})$ and $H$. 
Maximizing the above functional with respect to $\mathrm{Im} \Sigma_{\up}(\omega, \bm{k})$ will provide the most probable spectral function. 
$P[D |\mathrm{Im} \Sigma H]$ is called the ``likelihood 
function'' and is obtained as 
\begin{equation}
\begin{split}
P[D|\mathrm{Im} \Sigma H] &= e^{-L[\mathrm{Im} \Sigma]}, \\
L[\mathrm{Im} \Sigma] &= \frac{1}{2(M_{\mathrm{max}} -M_{\mathrm{min}})}
\displaystyle \int_{M_{\mathrm{min}}}^{M_{\mathrm{max}}} 
dM \frac{ \bigl[D^{\mathrm{OPE}}_{n,\,\up}(M,\bm{k}) - D^{\mathrm{Im} \Sigma}_{n,\,\up}(M,\bm{k}) \bigr]^2}{\sigma_{n,\,\up}^2(M,\bm{k})},  
\end{split}
\label{eq:likelihood}
\end{equation} 
with $n=0$ or $1$. 
Here, $D^{\mathrm{OPE}}_{n,\,\up}(M,\bm{k})$ is given on the right-hand sides of Eqs.(\ref{eq:sum.rule3.1}) or 
(\ref{eq:sum.rule3.2}), while $D^{\mathrm{Im} \Sigma}_{n,\,\up}(M,\bm{k})$ is defined as 
\begin{equation}
D^{\mathrm{Im} \Sigma}_{n,\,\up}(M,\bm{k}) = \int_{-\infty}^{\infty} d \omega \K_n(\omega,M) \mathrm{Im} \Sigma_{\up}(\omega,\bm{k}), 
\label{eq:sigma.rho}
\end{equation} 
and hence implicitly depends on $\mathrm{Im} \Sigma_{\up}(\omega, \bm{k})$. 
The error function $\sigma_{n,\,\up}(M,\bm{k})$ stands for the uncertainty of $D^{\mathrm{OPE}}_{n,\,\up}(M,\bm{k})$ at Borel mass $M$ and momentum $|\k|$, which 
we determine from the uncertainties of the parameters $\xi$ and $\zeta$ (e.g. the Bertsch parameter and the contact density) appearing in the OPE. 

$P[\mathrm{Im} \Sigma|H]$ on the other hand is called the ``prior probability'' and can be written down as follows: 
\begin{equation}
\begin{split}
P[\mathrm{Im} \Sigma|H] &= e^{\alpha S[\mathrm{Im} \Sigma]}, \\
S[\mathrm{Im} \Sigma] &= \displaystyle \int_{-\infty}^{\infty} d\omega \Bigr[ \mathrm{Im} \Sigma_{\up}(\omega,\bm{k}) - m(\omega) 
- \mathrm{Im} \Sigma_{\up}(\omega,\bm{k})\log \Bigl( \frac{\mathrm{Im} \Sigma_{\up}(\omega,\bm{k})}{m(\omega)} \Bigr) \Bigl]. 
\end{split}
\label{eq:prior}
\end{equation}
$S[\mathrm{Im} \Sigma]$ is known as the Shannon-Jaynes entropy and 
the function $m(\omega)$ is the so-called ``default model''. 
In case of no available data $D^{\mathrm{OPE}}_{n,\,\up}(M,\bm{k})$ or infinitely large error $\sigma_{n,\,\up}(M,\k)$, the MEM procedure will just 
give $m(\omega)$ for $\mathrm{Im} \Sigma_{\up}(\omega,\bm{k})$ because 
this function maximizes $P[\mathrm{Im} \Sigma|H]$. 
The default model can thus be utilized to incorporate already known information about $\mathrm{Im} \Sigma_{\up}(\omega,\k)$ into the 
analysis. 

Collecting all the terms discussed above, we reach the final form of the probability $P[\mathrm{Im} \Sigma|D H]$:
\begin{equation}
\begin{split}
P[\mathrm{Im} \Sigma|D H] & \propto P[D|\mathrm{Im} \Sigma H] P[\mathrm{Im} \Sigma|H]  \\
&= e^{Q[\mathrm{Im} \Sigma]}, \\
Q[\mathrm{Im} \Sigma] &\equiv \alpha S[\mathrm{Im} \Sigma] - L[\mathrm{Im} \Sigma]. 
\end{split}
\label{eq:finalprob}
\end{equation}
It is now merely a numerical problem to obtain the form of $\mathrm{Im} \Sigma_{\up}(\omega,\k)$ that 
maximizes $Q[\mathrm{Im} \Sigma]$ and is therefore the 
most probable $\mathrm{Im} \Sigma_{\up}(\omega,\k)$ for given $D^{\mathrm{OPE}}_{n,\,\up}(M,\bm{k})$ and $H$. For this task, we will use the 
Bryan algorithm \cite{Bryan}. 

Once $\mathrm{Im} \Sigma_{\alpha,\,\up}(\omega,\k)$ maximizing $Q[\mathrm{Im} \Sigma]$ for a fixed value of $\alpha$ is found, it is integrated out by 
averaging $\mathrm{Im} \Sigma_{\alpha,\,\up}(\omega,\k)$ over a certain range of $\alpha$, which then leads to our final result. 
Explicit formulae for this step and all other practical details specific to the application of MEM to QCD sum rules are discussed in 
\cite{Gubler,Gubler2}. 

As a final point, let us mention here that Eqs.(\ref{eq:sum.rule3.1}) and (\ref{eq:sum.rule3.2}) give two independent sum rules, 
which have to be combined in the analysis of this work. How this can be done is explained in \cite{Gubler2}.


\begin{thebibliography}{99}
\bibitem{Bloch}
I. Bloch, J. Dalibard, and W. Zwerger, 
Rev. Mod. Phys. \textbf{80}, 885 (2008). 
\bibitem{Giorgini}
S. Giorgini, L.P. Pitaevskii, and S. Stringari, 
Rev. Mod. Phys. \textbf{80}, 1215 (2008). 
\bibitem{Zwerger}
W. Zwerger (Editor), 
\textit{The BCS-BEC Crossover and the Unitary Fermi Gas},  Lecture Notes in Physics, Springer (2011). 
\bibitem{Stewart}
J.T. Stewart, J.P. Gaebler and D.S. Jin, 
Nature (London) \textbf{454}, 744 (2008). 
\bibitem{Dao}
T.L. Dao, A. Georges, J. Dalibard, C. Salomon and I. Carusotto, 
Phys. Rev. Lett. \textbf{98}, 240402 (2007). 
\bibitem{Haussmann}
R. Haussmann, M. Punk and W. Zwerger, 
Phys. Rev. A \textbf{80}, 063612 (2009). 
\bibitem{Magierski}
P. Magierski, G. Wlazlowski and A. Bulgac, 
Phys. Rev. Lett. \textbf{107}, 145304 (2011). 
\bibitem{Carlson2}
J. Carlson and S. Reddy, 
Phys. Rev. Lett. \textbf{95}, 060401 (2005). 
\bibitem{Wilson}
K.G. Wilson, 
Phys. Rev. \textbf{179}, 1499 (1969). 
\bibitem{Kadanoff}
L.P. Kadanoff, 
Phys. Rev. Lett. \textbf{23}, 1430 (1969). 
\bibitem{Polyakov}
A.M. Polyakov, 
Zh. Eksp. Teor. Fiz. \textbf{57}, 271 (1969). 
\bibitem{Muta}
T. Muta, 
\textit{Foundations of quantum chromodynamics} (World Scientific, Singapore, 1998). 
\bibitem{Shifman1}
M.A. Shifman MA, A.I. Vainshtein and V.I. Zakharov, 
Nucl. Phys. \textbf{B147}, 385 (1979). 
\bibitem{Shifman2}
M.A. Shifman MA, A.I. Vainshtein and V.I. Zakharov, 
Nucl. Phys. \textbf{B147}, 448 (1979). 
\bibitem{Braaten1}
E. Braaten and L. Platter, 
Phys. Rev. Lett. \textbf{100}, 205301 (2008). 
\bibitem{Braaten2}
E. Braaten, D. Kang and L. Platter, 
Phys. Rev. A \textbf{78}, 053606 (2008). 
\bibitem{Braaten3}
E. Braaten, D. Kang and L. Platter, 
Phys. Rev. Lett. \textbf{104}, 223004 (2010). 
\bibitem{Braaten4}
E. Braaten, D. Kang and L. Platter, 
Phys. Rev. Lett. \textbf{106}, 153005 (2011). 
\bibitem{Braaten5}
E. Braaten, 
in \textit{The BCS-BEC Crossover and the Unitary Fermi Gas}, 
Lecture Notes in Physics, edited by W. Zwerger, 
Chap. 6 (Springer, Berlin, 2012). 
\bibitem{Son}
D.T. Son and E.G. Thompson, 
Phys. Rev. A \textbf{81}, 063634 (2010). 
\bibitem{Barth}
M. Barth and W. Zwerger, 
Ann. Phys. (NY) \textbf{326}, 2544 (2011). 
\bibitem{Hofmann}
J. Hofmann, 
Phys. Rev. A \textbf{84}, 043603 (2011). 
\bibitem{Goldberger}
W.D. Goldberger and I.Z. Rothstein, 
Phys. Rev. A \textbf{85}, 013613 (2012). 
\bibitem{Goldberger2}
W.D. Goldberger and Z.U. Khandker, 
Phys. Rev. A \textbf{85}, 013624 (2012). 
\bibitem{Nishida}
Y. Nishida, 
Phys. Rev. A \textbf{85}, 053643 (2012). 
\bibitem{Golkar}
S. Golkar and D.T. Son, 
JHEP \textbf {1412}, 063 (2014). 
\bibitem{Goldberger3}
W.D. Goldberger, Z.U. Khandker and S. Prabhu, 
arXiv:1412.8507 [hep-th]. 
\bibitem{Tan1}
S. Tan, 
Ann. Phys. \textbf{323}, 2952 (2008). 
\bibitem{Tan2}
S. Tan, 
Ann. Phys. \textbf{323}, 2971 (2008). 
\bibitem{Tan3}
S. Tan, 
Ann. Phys. \textbf{323}, 2987 (2008). 
\bibitem{Ku}
M.J.H. Ku, A.T. Sommer, L.W. Cheuk and M.W. Zwierlein, 
Science \textbf{335}, 563 (2012). 
\bibitem{Zurn}
G. Z$\mathrm{\ddot{u}}$rn \textit{et al.}, 
Phys. Rev. Lett. \textbf{110}, 135301 (2013). 
\bibitem{Hoinka}
S. Hoinka \textit{et al.}, 
Phys. Rev. Lett. \textbf{110}, 055305 (2013). 
\bibitem{Carlson}
J. Carlson, S. Gandolfi, K.E. Schmidt and S. Zhang, 
Phys. Rev. A \textbf{84}, 061602(R) (2011). 
\bibitem{Gandolfi}
S. Gandolfi, K.E. Schmidt and J. Carlson, 
Phys. Rev. A \textbf{83}, 041601(R) (2011). 
\bibitem{Gubler}
P. Gubler and M. Oka, 
Prog. Theor. Phys. \textbf{124}, 995 (2010). 
\bibitem{Krasnikov}
N.V. Krasnikov, A.A. Pivovarov and N.N. Tavkhelidze, 
Z. Phys. C \textbf{19}, 301 (1983). 
\bibitem{Bertlmann}
R.A. Bertlmann, G. Launer and E. de Rafael, 
Nucl. Phys. \textbf{B250}, 61 (1985). 
\bibitem{Orlandini}
G. Orlandini, T.G. Steele and D. Harnett, 
Nucl. Phys. \textbf{A686}, 261 (2001). 
\bibitem{Ohtani}
K. Ohtani, P. Gubler and M. Oka, 
Eur. Phys. J. A \textbf{47}, 114 (2011). 
\bibitem{Ioffe}
B.L. Ioffe and K.N. Zyablyuk, 
Nucl. Phys. \textbf{A687}, 437 (2001). 
\bibitem{Araki}
K.J. Araki, K. Ohtani, P. Gubler and M. Oka, 
Prog. Theor. Exp. Phys. \textbf{2014} 073B03 (2014). 
\bibitem{Gubler3}
P. Gubler, K. Morita and M. Oka, 
Phys. Rev. Lett. \textbf{107}, 092003 (2011). 
\bibitem{Gubler2}
P. Gubler, 
\textit{A Bayesian Analysis of QCD Sum Rules}, 
Springer Theses, Springer Japan, 2013. 
\bibitem{Suzuki}
K. Suzuki, P. Gubler, K. Morita and M. Oka, 
Nucl. Phys. \textbf{A897}, 28 (2013). 
\bibitem{Ohtani2}
K. Ohtani, P. Gubler and M. Oka, 
Phys. Rev. D \textbf{87}, 034027 (2013). 
\bibitem{Gubler4}
P. Gubler and K. Ohtani, 
Phys. Rev. D \textbf{90} 094002 (2014). 
\bibitem{Drut}
J.E. Drut, T.A. L$\mathrm{\ddot{a}}$hde and T. Ten, 
Phys. Rev. Lett. \textbf{106}, 205302 (2011). 
\bibitem{Jarrel}
M. Jarrell and J.E. Gubernatis, 
Phys. Rep. \textbf{269}, 133 (1996).
\bibitem{Asakawa}
M. Asakawa, T. Hatsuda, and Y. Nakahara, 
Prog. Part. Nucl. Phys. \textbf{46}, 459 (2001). 
\bibitem{Bryan}
R.K. Bryan, 
Eur. Biophys. J. \textbf{18}, 165 (1990). 
\end{thebibliography}
\end{document}